\RequirePackage{fix-cm}
\documentclass[twocolumn]{revtex4}

\RequirePackage{graphicx}
\RequirePackage{amsmath}
\RequirePackage{amssymb}
\RequirePackage{bm}
\RequirePackage{color}
\RequirePackage[pdfencoding=auto, psdextra]{hyperref}
\RequirePackage{verbatim}
\newcommand{\ba}{\begin{eqnarray}}
\newcommand{\ea}{\end{eqnarray}}

\begin{document}

\title{Decay widths and mass spectra of single bottom baryons}

\author{H. Garc{\'i}a-Tecocoatzi}

\affiliation{INFN, Sezione di Genova, Via Dodecaneso 33, 16146 Genova, Italy}

\author{A. Giachino} 
\affiliation{Institute of Nuclear Physics Polish Academy of Sciences Radzikowskiego 152, 31-342 Cracow, Poland}


\author{A. Ramirez-Morales}
\affiliation{Center for High Energy Physics, Kyungpook National University, 80 Daehak-ro, Daegu 41566, Korea}

\author{Ailier Rivero-Acosta} 
\affiliation{Departamento de F\'isica, DCI, Campus Le\'on, Universidad de Guanajuato, Loma del Bosque 103, Lomas del Campestre, C.P. 37150, Le\'on, Guanajuato, Mexico}
\affiliation{Dipartimento di Fisica, Universit\`a di Genova, Via Dodecaneso 33, 16146 Genova, Italy}
\affiliation{INFN, Sezione di Genova, Via Dodecaneso 33, 16146 Genova, Italy}

\author{E. Santopinto}\email[]{elena.santopinto@ge.infn.it}
\affiliation{INFN, Sezione di Genova, Via Dodecaneso 33, 16146 Genova, Italy }

\author{Carlos Alberto Vaquera-Araujo} 
\affiliation{Consejo Nacional de Humanidades, Ciencias y Tecnolog\'ias, Av. Insurgentes Sur 1582. Colonia Cr\'edito Constructor, Del. Benito Ju\'arez, C.P. 03940, Ciudad de M\'exico, M\'exico}
\affiliation{Departamento de F\'isica, DCI, Campus Le\'on, Universidad de
  Guanajuato, Loma del Bosque 103, Lomas del Campestre C.P. 37150, Le\'on, Guanajuato, M\'exico}
\affiliation{Dual CP Institute of High Energy Physics, C.P. 28045, Colima, M\'exico}




\begin{abstract}

We  develop a Hamiltonian model that incorporates the spin, spin-orbit, and isospin interactions to determine the masses of the ground states of single-bottom baryons and their excitations up to the $D$-wave.
Furthermore, we calculate the strong decay widths of single-bottom baryons using the $^3P_0$ model. Our calculations consider final states comprising bottom baryon-(vector/pseudoscalar) meson pairs and (octet/decuplet) baryon-(pseudoscalar/vector) bottom meson pairs within a constituent quark model. In that respect, this is the most complete investigation which has ever been performed in the single bottom
baryon sector so far. Additionally, we compute the electromagnetic decay widths from $P$-wave states to ground states. The electromagnetic decays become dominant in cases where the strong decays are suppressed.
The experimental uncertainties are  propagated to the model parameters  using a Monte Carlo bootstrap method. Our  quantum number assignments, as well as our  mass and strong decay width predictions, are in reasonable agreement with the available data.  We also provide the partial decay widths for each open flavor channel. Our predictions of mass spectra and decay widths provide valuable information for the experiments seeking to identify new bottom baryons and knowledge of possible decay channels can aid in their identification in the data. Therefore, our results will be able to guide future searches for the undiscovered single bottom baryons at LHCb, ATLAS, and CMS.
\end{abstract}

\maketitle

\section{Introduction }  
The study of the heavy baryon mass spectra is one of the most relevant access doors toward the understanding of non-perturbative Quantum Chromodynamics (QCD) since it provides essential information regarding the way the quarks interact with each other in the strong-coupling regime of QCD.
For this reason, establishing and improving hadron spectroscopy is a key subject in hadron physics.  The search for bottom states is quite difficult for the experiment since higher energy and higher beam luminosity are required to produce them. Nevertheless, in the last years, many new excited bottom baryon states have been discovered  and many efforts have been made to identify their quantum numbers and to understand their properties.  

In 2012,   two narrow $P$-wave $\Lambda_b^0$ baryons,  denoted as $\Lambda_b(5912)^0$ and $\Lambda_b(5920)^0$, were first discovered  by the LHCb Collaboration~\cite{LHCb:2012kxf}, with 5.2 $\sigma$ and 10.2 $\sigma$ statistical significance, respectively. They were confirmed by the CDF Collaboration~\cite{CDF:2013pvu} one year later. In 2018 the LHCb Collaboration reported the discovery of one excited $\Xi_b$  state, $\Xi_b(6227)^-$  \cite{LHCb:2018vuc}, with a statistical  significance of about 7.9 $\sigma$,  and one excited $\Sigma_b$ state,  $\Sigma_b(6097)^{\pm}$ \cite{LHCb:2018haf} with a significance of 12.6 $\sigma$. 
In 2020, two $D$-wave $\Lambda_b^0$ candidates, $\Lambda_b(6146)^0$ and $\Lambda_b(6152)^0$, were discovered by  LHCb in the $\Lambda_b^0\pi^+\pi^-$ spectrum~\cite{LHCb:2019soc},  both states, were discovered with statistical significance exceeding 6 standard deviations. In the same year  the LHCb collaboration \cite{LHCb:2020tqd} reported the observation of four narrow peaks in the $\Xi_b^{0}K^{-}$ invariant mass spectrum, $\Omega_b(6316)$, $\Omega_b(6330)$, $\Omega_b(6340)$ and $\Omega_b(6350)$,  with significances of 2.1 $\sigma$, 2.6 $\sigma$, 6.7 $\sigma$, and 6.2 $\sigma$, respectively. 

More recently, in 2021, the LHCb collaboration reported the discovery of two  new $\Xi_b$ states, namely $\Xi_b(6327)^{0}$ and $\Xi_b(6333)$, in $\Lambda_b^0 K^{-} \pi^{+}$ channel with a statistical significance larger than nine standard deviations \cite{LHCb:2021ssn}.
It is also worth mentioning the recent discovery of  $\Lambda_b(6072)$ by LHCb \cite{LHCb:2020lzx} with a statistical significance exceeding 7 standard deviations in the  $\Lambda_b^0\pi^+\pi^-$ channel and CMS \cite{CMS:2020zzv} experiments. 

\indent
The first predictions of  $\Lambda_b$ and   $\Sigma_b$ baryon mass spectra  were presented by Capstick and Isgur in their pioneering work in 1968 \cite{Capstick:1986bm}.  Over the last few years, the interest in heavy hadron spectroscopy has increased more and more. Some examples of the recent wide literature on theoretical investigations into the heavy baryon spectroscopy are: the QCD-inspired relativistic quark-diquark picture \cite{Ebert:2007nw,Ebert:2011kk},
the non-relativistic quark model \cite{Roberts:2007ni,Yoshida:2015tia},  the QCD sum rules in the
framework of the Heavy Quark Effective Theory (HQET) \cite{Chen:2016phw,Bagan:1992tp} and the symmetry-preserving Schwinger-Dyson equation approach \cite{Gutierrez-Guerrero:2019uwa}. Alternative discussions employing other models can be found in Refs. \cite{Garcilazo:2007eh,Hasenfratz:1980ka,Kim:2020imk,Kim:2021ywp}, and lattice QCD studies in Ref.  \cite{Vijande:2014uma}.
For more references, see the review articles \cite{Korner:1994nh,Chen:2016spr,Crede:2013kia,Amhis:2019ckw}. 
The Particle Data Group \cite{Workman:2022ynf} lists  24 bottom baryons. For most of them none of isospin $I$, parity $P$ and angular momentum $J$ 
  have actually been measured but they are based on quark model expectations. 

Besides the mass spectrum, decay properties  are one of the important features for a consistent classification of the hadrons.  The matching between the experimental data and the predicted mass spectra and  decay widths provide indeed one of the clearest ways for classifying these states. 
There are not many studies on the heavy baryon strong decays.  Moreover, a systematic investigation that includes the strong decay calculations for ground and excited states up to the $D$-wave shell within the same model has never been performed so far.

In Refs. \cite{Wang:2017kfr} and \cite{Yao:2018jmc} the strong decays of the low-lying 
$S$-  $P$- and $D$-wave singly heavy baryons with emission of one light pseudoscalar meson are studied within ChQM. However, these studies did not include 
the decays into the charmed baryon-vector meson channels or the charmed meson-octet/decuplet baryon channels. Moreover, both these works  \cite{Wang:2017kfr}, \cite{Yao:2018jmc} were restricted only to the low-lying  $\lambda$-mode excitations.
 In Ref.~\cite{Nagahiro:2016nsx} the authors calculate the strong decay widths of $\Lambda_{b}^{}$,  $\Sigma_{b}^{}$ and $\Xi_{b}^{}$ states adopting an interaction   inspired by chiral symmetry but this study  focuses only on the ground states, the first  $\lambda$-mode $P$-wave excitations and the roper excitations and moreover, in this work, only the ground state bottom baryon plus pion channels are considered. 
 
 In \cite{Liang:2020hbo} the $^3P_0$ model was applied to calculate the strong decay widths of the $\Omega_b$  ground states and the first excited states. In a subsequent work, \cite{He:2021xrh}, this analysis was further extended to strong decays of the low-lying bottom strange baryons. Nevertheless, these studies were restricted to the low-lying $\lambda$-mode excitations and only the strong decay widths into ground state bottom baryon plus pion or $K$ mesons have been calculated.
  For the radiative transitions, the literature contains various discussions on the radiative decays of singly heavy baryons, as evidenced by numerous references~\cite{Wang:2017kfr,Cheng:1992xi,Wang:2009ic,Wang:2009cd,Jiang:2015xqa,Zhu:1998ih,Tawfiq:1999cf,Bernotas:2013eia,Gamermann:2010ga,Aliev:2014bma,Aliev:2009jt,Aliev:2016xvq,Aliev:2011bm,Chow:1995nw,Ivanov:1998wj,Banuls:1999br}. However, there is no experimental data for the single bottom baryons to compare with the theoretical predictions.
  
In a previous work \cite{Santopinto:2018ljf}, motivated by the discovery of the five $\Omega_c$ baryons by LHCb \cite{LHCb:2017uwr}, we conducted calculations of the mass spectra for $\Omega_c$ baryons developing a mass formula that incorporates spin, spin-orbit, isospin, and flavor interactions. Additionally, in Ref.~\cite{Santopinto:2018ljf}, we investigated the decay widths in the $\Xi_c^{+} K^{-}$ and $\Xi_c^{'+} K^{-}$ channels within the framework of the $^3P_0$ model. These calculations were also extended to the $\Omega_b$ states. Subsequently, in Ref.~\cite{Bijker:2020tns}, we extend our calculations to the $\Xi_c^{'}$ and the $\Xi_b^{'}$ states. We computed the mass spectra and strong partial decay widths for the ground states and $P$-wave excitations of the $\Xi_c^{'}$ baryons into $^2\Sigma_c \bar{K}$, $^2\Xi_c^{'} \pi$, $^4\Sigma_c \bar{K}$, $^4\Xi_c^{'} \pi$, $\Lambda_c^{} \bar{K}$, $\Xi_c^{} \pi$ and $\Xi_c^{} \eta$  and of the $\Xi_b^{'}$-ground states  and $P$-wave excitations into $^2\Sigma_b \bar{K}$, $^2\Xi_b^{'} \pi$, $^4\Sigma_b \bar{K}$, $^4\Xi_b^{'} \pi$, $\Lambda_b^{} \bar{K}$, $\Xi_b^{} \pi$ and $\Xi_b^{} \eta$, within both the Elementary Emission Model (EEM) and the $^3P_0$ model. In that work, we also calculated the electromagnetic decay widths of the $\Xi'_{c/b}$ and $\Xi_{c/b}$ states.

Recently, in Ref. \cite{Garcia-Tecocoatzi:2022zrf}, we  made our investigation more systematic and we further extended our model to the whole sector of the charmed baryon states ($cqq,cqs$, and $css$ systems). In that study we calculated the mass spectra of the charmed baryons by employing the same mass formula originally introduced in Ref. \cite{Santopinto:2018ljf} and we calculated the decay widths of the ground and excited charmed baryon states ($\boldsymbol \rho$- and $\boldsymbol \lambda$-mode excitations  up to the $D$-wave shell) into the charmed baryon-(vector/pseudoscalar) meson pairs and the (octet/ decuplet) baryon-(pseudoscalar/vector) charmed meson pairs. 

The aim of this article is to  extend the same model of Refs. \cite{Santopinto:2018ljf,Bijker:2020tns,Garcia-Tecocoatzi:2022zrf} to calculate the bottom baryon mass spectra including the calculation of the strong and radiative decay widths. In particular, we calculate the decay widths of the ground and excited bottom baryon states ($\boldsymbol{\rho}$- and $\boldsymbol{\lambda}$-mode excitations  up to the $D$-wave shell) into the bottom baryon-(vector/pseudoscalar) meson pairs and the (octet/ decuplet) baryon-(pseudoscalar/vector) bottom meson pairs. 
The experimental uncertainties are also propagated to the model parameters by means of the Monte Carlo bootstrap method \cite{Efron1994}, which is a specialized method used to properly estimate the error propagation by obtaining probability density functions for the fitted parameters. With this method, we are able to use 
 the same set of parameters to predict the bottom baryon masses, strong partial decay widths, and radiative decays.
 In that respect, this is the most complete investigation which has ever been performed  in the single bottom baryon sector so far.
\section{Methodology} 
\label{methodology}

\subsection{Mass spectra of the bottom baryons }  
\label{secIIA}

The masses of the bottom baryon states are calculated as the eigenvalues of the Hamiltonian  of Ref. \cite{Santopinto:2018ljf}, which we report  below for convenience
\begin{eqnarray}
\label{eq:mass}
	H &=& H_{\rm h.o.}+a_{\rm S}\; {\bf S}^2_{\rm tot}
 + a_{\rm SL} \; {\bf S}_{\rm tot} \cdot {\bf L}_{\rm tot}+a_{\rm I}  \;  \bm{{\rm I}}^2+a_{\rm F}\; {\bf \hat{C}}_2.
 \nonumber
 \\
	\label{MassFormula}
\end{eqnarray}
In Eq.~\ref{MassFormula} the symbols ${\bf S}_{\rm tot}, {\bf L}_{\rm tot}, \bm{{\rm I}}$ and  ${\bf \hat{C}}_2$ denote respectively the spin, orbital angular momentum, isospin, and the $SU_f(3)$ Casimir operators, and are weighted with the model parameters $a_{\rm S}, a_{\rm SL}, a_{\rm I}$, and $a_{\rm F}$.

\begin{figure}[h]
     \centering
     \includegraphics[scale=.35]{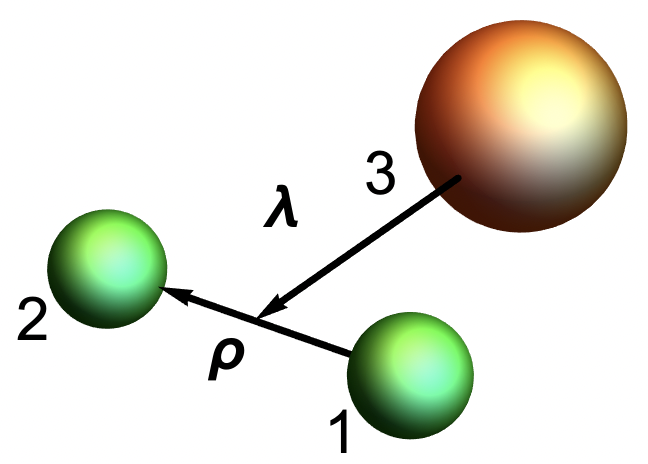}
     \caption{The $\boldsymbol \rho$ and $\boldsymbol{ \lambda}$ coordinates in the single bottom baryons.}
     \label{fig:jacobi}
 \end{figure}

In the case in which the baryons are modeled as three-quark systems, the three-dimensional h.o. Hamiltonian can be written as:
\begin{eqnarray}
 H_{\rm h.o.}^{3q} =\sum_{i=1}^3m_i + \frac{\mathbf{p}_{\rho}^2}{2 m_{\rho}} 
+ \frac{\mathbf{p}_{\lambda}^2}{2 m_{\lambda}} 
+\frac{1}{2} m_{\rho} \omega^2_{\rho} \boldsymbol{\rho}^2   
+\frac{1}{2}  m_{\lambda} \omega^2_{\lambda} \boldsymbol{\lambda}^2
	\nonumber \\
\label{eq:Hho}
\end{eqnarray}
where $ \boldsymbol{\rho} = (\boldsymbol{r}_1-\boldsymbol{r}_2)/\sqrt{2}$ and $\boldsymbol{\lambda}=(\boldsymbol{r}_1+\boldsymbol{r}_2-2\boldsymbol{r}_3)/\sqrt{6}$ denote the Jacobi coordinates, where $\boldsymbol{r}_{i}$ with $i=1,2$ are the light quark positions while  $\boldsymbol{r}_3$ is the bottom quark position, and    $\mathbf{p}_{\rho}$ and $ \mathbf{p}_{\lambda}$ denote the conjugated momenta. Thus, the $\boldsymbol \rho$ coordinate describes the excitations within the light quark pair while the $\boldsymbol \lambda$ coordinate describes the excitations between the light quark pair  and the bottom quark $b$ as depicted in  Fig. \ref{fig:jacobi}.
 
The $H_{\mathrm{h.o}}$ eigenvalues are 
\begin{eqnarray}
    \label{eq:freq}
E_{\rm h.o.}^{3q} &=&   \sum_{i=1}^3m_i + \omega_{\rho} \; n_{\rho} + \omega_{\lambda}\; n_{\lambda}, 
\end{eqnarray}
where $m_{i}$ with $i=1,2$ are the light quark masses while  $m_3$ is the bottom quark mass;
$m_\rho=(m_1+m_2)/2$, 
and $m_\lambda=3m_\rho m_3/(2m_\rho+m_3)$.
We use the usual definitions for
$ n_{\rho(\lambda)}= 2 k_{\rho(\lambda)}+l_{\rho(\lambda)}$, 
$k_{\rho(\lambda)}=0,1,...$,   and $l_{\rho(\lambda)}=0,1,...$; where, $l_{\rho(\lambda)}$ is the orbital angular momentum of the $\rho$($\lambda$) oscillator,  and $k_{\rho(\lambda)}$ is the number of nodes (radial excitations) in the $\rho$($\lambda$) oscillators. The $\rho$- and  $\lambda$-oscillator frequencies are $\omega_{\rho(\lambda)}=\sqrt{\frac{3K_b}{m_{\rho(\lambda)}}}$ where $K_b$ is the spring constant. 

The eigenvalues of the Hamiltonian \ref{MassFormula}, proposed in Ref. \cite{Santopinto:2018ljf}, are given by
\begin{eqnarray}
E^{3q}  &=& \sum_{i=1}^3m_i +  \omega_{\rho} n_{\rho} 
+ \omega_{\lambda} n_{\lambda} + a_{\rm S} \left[ S_{\rm tot}(S_{\rm tot}+1) - \frac{3}{4} \right]
\nonumber\\
&& + a_{\rm SL} \frac{1}{2} \Big[ J_{}(J_{}+1) - L_{\rm tot}(L_{\rm tot}+1) \nonumber\\
&& - S_{\rm tot}(S_{\rm tot}+1) \Big] 
+a_{\rm I}\left[ I(I+1) - \frac{3}{4} \right]\nonumber \\
&& + a_{\rm F}\frac{1}{3} \left[ p(p+3)+q(q+3)+pq \right] ~.
\label{MassFormula2}
\end{eqnarray} 
The spin-dependent term splits the states with different $\bf S_{\rm tot}$. The spin-orbit interaction, which is small in light baryons \cite{Ebert:2007nw,Capstick:1986bm}, turns out to be fundamental to describe the heavy-light baryon mass patterns \cite{Santopinto:2018ljf}. The effect of the spin-orbit term is to split the states with different $\bf J_{}$. 
Finally, the flavor-dependent term splits the baryons belonging to the flavor sextet, $ \bf  6$$\rm _F$ with $(p,q)=(2,0)$, from the baryons of the anti-triplet, $ \bf  \bar 3$$\rm _F$ with $(p,q)=(0,1)$.

\textcolor{black}{
Additionally, we present a simplification of the three-quark system based on only one  relative coordinate $\boldsymbol{r}$ and momentum $ \mathbf{p}_{r}$, namely, the quark-diquark system ~\cite{Santopinto:2004hw}. 
In this picture, the two light quarks are regarded as a single diquark object interacting with a heavy quark}. The quark-diquark Hamiltonian can be written as
\begin{eqnarray}
 H_{\rm h.o.}^{qD} =m_D + m_b 
+ \frac{\mathbf{p}_{r}^2}{2 \mu }
+\frac{1}{2} \mu \omega^2_{r} \boldsymbol{r}^2,
\label{eq:Hhodi}
\end{eqnarray}
where $\mathbf{p}_{r}=(m_b\mathbf{p}_{D}-m_D\mathbf{p}_{b})/(m_b +m_D)$. The $H_{\rm h.o}$ eigenvalues are
\begin{eqnarray}
    \label{eq:freq_di}
E_{\rm h.o.}^{qD} =\,        m_D + m_b + \omega_{r}\; n_{r};\; \mathrm{with}\;
\omega_{r}=\sqrt{\frac{3K_b}{ \mu}},
\end{eqnarray}
where $m_D$ and $m_b$  are the diquark  and bottom quark masses, respectively, $\mu=m_b m_D/(m_b +m_D)$  is the reduced mass of the system.  $ n_{r}= 2 k_{r}+l_{r}$  where $k_{r}=0,1,...$  is the number of nodes, $l_{r}=0,1,...$ is the orbital angular momenta of the $r$ oscillator, and  $K_b$ is the spring constant.

\subsection{Bottom baryon states} 
\label{secIIIA}

We first construct the single bottom baryon states in both the three-quark and the quark-diquark models.

In the three-quark model, the bottom baryons are described as three quark states made up of one $b$ quark and two light quarks  ($u, d$, or $s)$. 
In this model, the spatial degrees of freedom of the bottom states are expressed by the $\boldsymbol \rho$ coordinate, which describes the excitations within the light quark pair, and the $\boldsymbol \lambda$ coordinate, which describes the excitations between the light quark pair  and the bottom quark $b$ (see Fig. \ref{fig:jacobi}).

The total angular momentum, $\bf J_{}= {\bf L}_{\rm tot} + {\bf S}_{\rm tot} $, is the sum of the orbital angular momentum, $\bf{\rm{\bf L_{\rm tot}}}=\bf l_\rho+\bf l_\lambda$, and the internal spin,  $\bf S_{\rm tot}=\bf S_{\rm 12}+\bf 1/2$, which is the sum of the light quark spin, $\bf S_{\rm 12}=\bf S_{\rm 1}+\bf S_{\rm 2}$, and the $b$ quark spin, $\bf 1/2$.

It is important to note that the color part of a baryon wave function is fully antisymmetric, representing an $SU_c(3)$ singlet of the three colors.
In our model the light quarks are considered to be identical particles; hence, their wave function should be antisymmetric in order to satisfy the Pauli Principle.
Since the two light quarks are in the antisymmetric $ \bf  \bar 3$$\rm _c$ color state, the product of their  spin-, flavor-, and orbital-wave functions has to be symmetric. 
Let us apply this principle to construct the single-bottom baryon ground and excited states up to the second energy band, $N = n_{\rho} + n_{\lambda}$ ($N = n_r$ in the case of the quark-diquark system), of the harmonic oscillator.

\begin{figure}[h]
    \centering
    \includegraphics[width=0.5\textwidth]{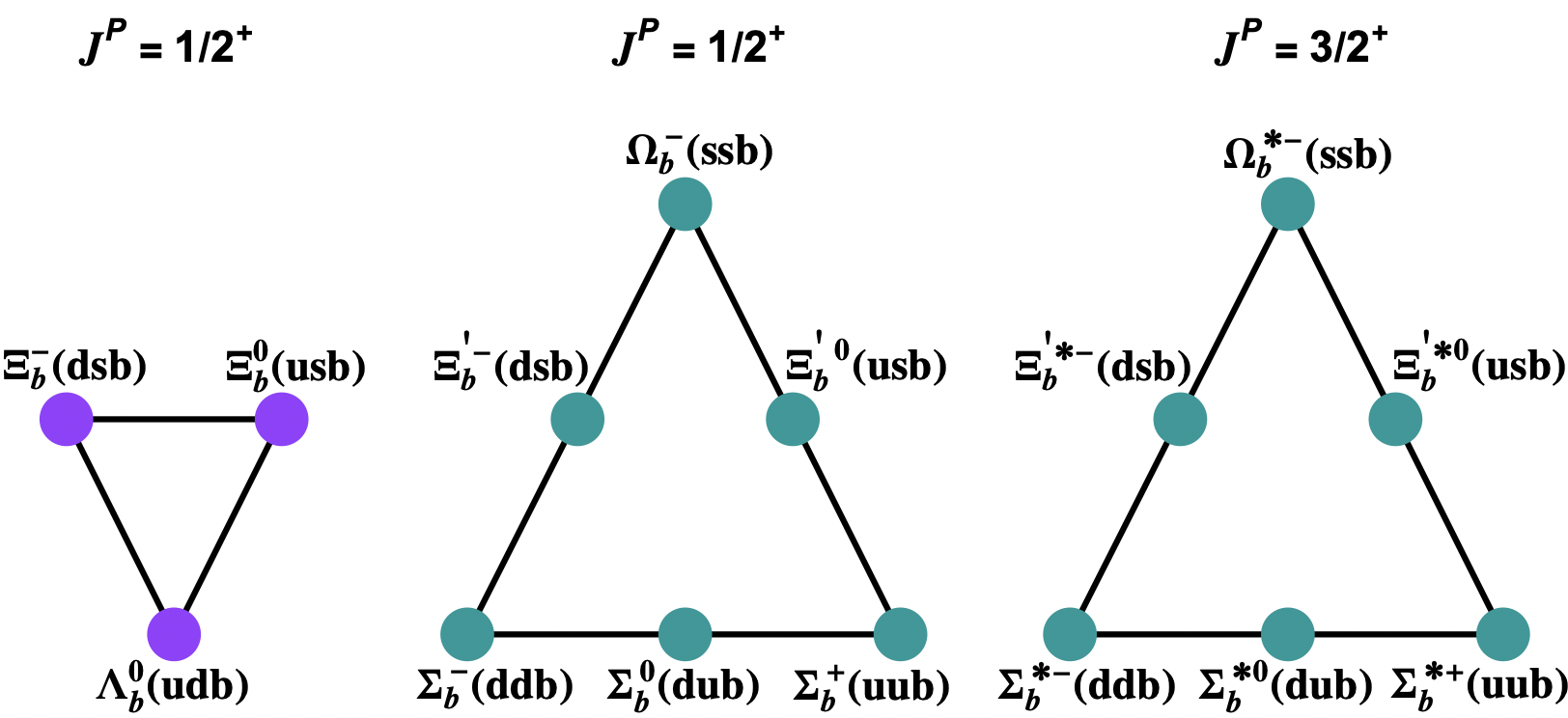}
	\caption{The $SU_f(3)$ flavor multiplets of the ground-state single bottom baryons:  the flavor anti-triplet  $ \bf  \bar 3$$\rm _F$ with spin-parity ${\bf J}^P = {\bf \frac{1}{2}}^+$  (left side), the flavor sextet  $ \bf  6$$\rm _F$ with ${\bf J}^P = {\bf \frac{1}{2}}^+$  (centre), and  the  flavor sextet  $ \bf 6$$\rm _F$ with ${\bf J}^P = {\bf \frac{3}{2}}^+$   (right side).}    
    \label{fig:multiplets}
\end{figure}

 - In the energy band $N=0$, in which $\bf l_\rho=l_\lambda=0$, the spatial wave function of the two light quarks is symmetric implying that their spin-flavor wave function is symmetric. Therefore, we can only combine the antisymmetric $ \bf  \bar 3$$\rm _F$-plet  with  antisymmetric-spin configuration $\bf S_{\rm 12}= 0$ and  the symmetric $ \bf  6$$\rm _F$-plet with spin symmetric configuration $\bf S_{\rm 12}= 1$. This means that the ground state baryons made up of a light quark pair with  antisymmetric-spin configuration $\bf S_{\rm 12}=0$ fill an antisymmetric $ \bf  \bar 3$$\rm _F$-plet  with total spin $\bf J = S_{\rm tot}=\frac{1}{2}$ (displayed on the left-hand side of Fig.~\ref{fig:multiplets}), while the ones made up of a light quark pair with spin symmetric configuration $\bf S_{\rm 12}=1$, fill one $ \bf  6$$\rm _F$-plet  with total spin $\bf J= S_{\rm tot}=\frac{1}{2}$, and one $ \bf  6$$\rm _F$-plet with total spin $\bf J = S_{\rm tot}=\frac{3}{2}$ (displayed on the center and on the right-hand side of Fig.~\ref{fig:multiplets}).

The $ \bf  \bar 3$$\rm _F$-plet and the $ \bf  6$$\rm _F$-plet with  spin-parity ${\bf J}^P = {\bf \frac{1}{2}}^+$  lie on the first floor of the $SU_f(4)$  ${\bf 20}$$\rm _F$-plet with the light octet baryons at the ground level, while the $ \bf  6$$\rm _F$-plet with  spin-parity ${\bf J}^P = {\bf \frac{3}{2}}^+$  lies on the first floor of the $SU_f(4)$ ${\bf 20}$$\rm _F$-plet  with the light decuplet baryons at the ground level.
The $ \bf  \bar 3$$\rm _F$-plet with  spin-parity ${\bf J}^P = {\bf \frac{1}{2}}^+$   contains one isosinglet state, $\Lambda_{b}$, and two isospin $\bf \frac{1}{2}$ states, $\Xi_b^{-}$ and $\Xi_b^{0}$. 

The $ \bf  6$$\rm _F$-plet with  spin-parity ${\bf J}^P = {\bf \frac{1}{2}}^+$  contains one isosinglet, $\Omega_b^{-}$, two isospin $\bf \frac{1}{2}$ states, $\Xi_b^{'-}$ and $\Xi_b^{'0}$, and three states with total isospin $\bf I=1$, $\Sigma_{b}^{+},\Sigma_{b}^{0}$ and $\Sigma_{b}^{-}$. 
The $ \bf  6$$\rm _F$-plet with  spin-parity ${\bf J}^P = {\bf \frac{3}{2}}^+$  contains the 
one isosinglet, $\Omega_b^{*-}$, two isospin $\bf \frac{1}{2}$ states, $\Xi_b^{'*-}$ and $\Xi_b^{'*0}$, and three states with total isospin $\bf I=1$, $\Sigma_{b}^{*+},\Sigma_{b}^{*0}$ and $\Sigma_{b}^{*-}$, where the upper symbol $*$ denotes that the total spin of these states is  $\bf \frac{3}{2}$.
The total spin and the $SU_f(3)$ flavor multiplets of ground state single bottom baryons are reported in Fig. \ref{fig:multiplets}.

 - For the energy band $N=1$, there are two different possibilities. 
 If $\bf l_\rho=0$ and $\bf l_\lambda=1$, the spatial wave function is symmetric under the interchange  of light quarks implying that their spin-flavor wave function is also symmetric. Thus, in  the case of the $ \bf  \bar 3$$\rm _F$-plet baryons, the angular momentum  $\bf{\rm{\bf L_{\rm tot}}}=1$ is coupled with one only spin configuration, $\bf S_{\rm tot}=\frac{1}{2}$ that comes from the light quark spin configuration $\bf S_{\rm 12}=0$, yielding two {$P_\lambda$}-wave excitations, while,  in  the case of the $ \bf  6$$\rm _F$-plet baryons, $\bf{\rm{\bf L_{\rm tot}}}=1$ is coupled with two possible spin configurations, $\bf S_{\rm tot}=\frac{1}{2},\frac{3}{2}$  that come from the light quark spin configuration $\bf S_{\rm 12}=1$, yielding five {$P_\lambda$}-wave excitations.
 
When  $\bf l_\rho=1$ and $\bf l_\lambda=0$, the spatial wave function is antisymmetric under the interchange  of light quarks implying that the two light quark spin-flavor wave function are also antisymmetric, hence the situation is reversed: in  the case of the $ \bf  \bar 3$$\rm _F$-plet baryons,   $\bf{\rm{\bf L_{\rm tot}}}=1$ is coupled with $\bf S_{\rm tot}=\frac{1}{2},\frac{3}{2}$ yielding to five $P_\rho$-wave states,  while,  in  the case of the $ \bf  6$$\rm _F$-plet baryons,  $\bf{\rm{\bf L_{\rm tot}}}=1$ is coupled with $\bf S_{\rm tot}=\frac{1}{2}$, yielding to  two $P_\rho$-wave states. 

  - In the energy band $N=2$, there are three possibilities: the pure $\lambda$-excitations, $\bf l_\rho=0,\,l_\lambda=2$, the pure $\rho$-excitations, $\bf l_\rho=2,\,l_\lambda=0$, and the mixed case $\bf l_\rho=1,\,l_\lambda=1$.  In both the $\bf l_\rho=0,\,l_\lambda=2$ and $\bf l_\rho=2,\,l_\lambda=0$ cases  the total spatial wave function is symmetric under the interchange  of  light quarks implying that their spin-flavor wave function is  also symmetric. 
  
  If  $\bf l_\rho=0$ and $\bf l_\lambda=2$, in  the case of the $ \bf  \bar 3$$\rm _F$-plet baryons, $\bf{\rm{\bf L_{\rm tot}}}=2$ is coupled with $\bf S_{\rm tot}=\frac{1}{2}$  giving two {$D_\lambda$}-wave excitations, while, in  the case of the $ \bf  6$$\rm _F$-plet baryons, $\bf{\rm{\bf L_{\rm tot}}}=2$ is coupled  with two possible spin configurations, $\bf S_{\rm tot}=\frac{1}{2},\frac{3}{2}$, giving six  $D_\lambda$-wave excitations.  If  $\bf l_\rho=2$ and $\bf l_\lambda=0$, in a similar way we have two {$D_\rho$}-wave excitations for the $ \bf  \bar 3$$\rm _F$-plet and six   $D_\rho$-wave excitations for the $ \bf  6$$\rm _F$-plet.
 
When $\bf l_\rho=1$ and $\bf l_\lambda=1$ there are three possible values of the angular momentum ${\bf L_{\rm tot}=0, 1, 2}$. In the case of the $ \bf  \bar 3$$\rm _F$-plet baryons, they  are combined with $\bf{S}_{\rm tot}=\frac{1}{2},\frac{3}{2}$ that come from the light quark spin configuration $\bf S_{\rm 12}=1$, producing  thirteen mixed excited states: six $D$-wave states, five $P$-wave states, and two $S$- wave states. In the case of the $ \bf  6$$\rm _F$-plet baryons, they are combined with $\bf{S}_{\rm tot}=\frac{1}{2}$ that come from the light quark spin configuration $\bf S_{\rm 12}=0$, thus producing five possible states: two $D$-wave states, two $P$-wave states, and one $S$-wave state.

Additionally, there are two possible radial excitation modes in this energy band, $k_\rho=0,k_\lambda=1$, and $k_\rho=1,k_\lambda=0$, both corresponding to a symmetric light quark wave function since $\bf L_{\rm tot}=l_\rho+l_\lambda=0$. 
If $k_\rho=0$ and $k_\lambda=1$, in the case of the $ \bf  \bar 3$$\rm _F$-plet baryons, $\bf L_{\rm tot}=0 $ is combined with $\bf S_{\rm tot}=\frac{1}{2}$ producing one $\lambda$-radial excitation  while, in the case of the $ \bf  6$$\rm _F$-plet baryons, $\bf L_{\rm tot}=0 $ is combined with $\bf J_{}={\bf S}_{\rm tot}=\frac{1}{2},\frac{3}{2}$ producing two $\lambda$-radial excitations. In a similar way, if $k_\rho=1$ and $k_\lambda=1$, in the case of the $ \bf  \bar 3$$\rm _F$-plet baryons, we have one  $\rho$-radial excitation  while, in the case of the $ \bf  6$$\rm _F$-plet baryons, we have two $\rho$-radial excitations.

Finally, when the bottom baryons are seen as quark-diquark systems, the two constituent light quarks of the diquark are considered to be correlated, with no internal spatial excitations ($\bf l_\rho=0$); $i.e$., it is hypothesized that we are within the limit where the diquark internal spatial excitations are higher in energy than the scale of the resonances studied. As a result, the quark-diquark states are a subset of the previously discussed three quark states and they can be obtained by freezing the $\rho$ coordinate. The validity of this scheme for single-bottom systems will ultimately be determined by experimental data. Further investigations and analysis are necessary to confirm its applicability. With the completion of the construction of states in both the three-quark and quark-diquark models, we have established a framework for understanding the properties of single-bottom baryons.  
 
 We make one last remark on the notation used throughout  the paper. The three-quark quantum state is written as $\left| l_{\lambda},l_{\rho}, k_{\lambda},k_{\rho}\right\rangle$, with total angular momentum ${\bf J_{}=} {\bf L}_{\rm tot} + {\bf S}_{\rm tot} $, where ${\bf L_{\rm tot} =l}_{\rho}+{\bf l}_{\lambda}$ and ${\bf S}_{\rm tot} = {\bf S}_{\rm 12}+{\bf \frac{1}{2}}$, ${\bf S}_{\rm 12}$ is the coupled spin of the light quarks. The number of nodes is $k_{\lambda, \rho}$. The quark-diquark quantum state is written as $\left| l_{r},k_r \right\rangle$ where ${\bf L_{\rm tot} =l}_{r}$ and ${\bf S}_{\rm tot} = {\bf S}_{\rm 12}+{\bf \frac{1}{2}}$, and the number of nodes is $k_r$. For each state we also report the information on the total spin, orbital angular momentum and total angular momentum  using the compact spectroscopic notation $^{2S+1}L_{J}\equiv\, ^{2S_{\rm tot}+1}L_{{\rm tot}\, J}$.


 



\section{Decay widths}

\subsection{Strong decay widths}  
\label{secIIC}
In the $^{3}P_0$ model the transition operator is given by \cite{Micu:1968mk,LeYaouanc:1972vsx,Bijker:2015gyk,Garcia-Tecocoatzi:2022zrf}.
\begin{eqnarray}
T^{\dagger} &=& -3 \gamma_0 \, \int d \mathbf{p}_4 \, d \mathbf{p}_5 \, 
\delta(\mathbf{p}_4 + \mathbf{p}_5) \, C_{45} \, F_{45} 
\nonumber\\
&& \hspace{0.5cm}  \left[ \chi_{45} \, \times \, {\cal Y}_{1}(\mathbf{p}_4 -\mathbf{p}_5) \right]^{(0)}_0 \, 
b_4^{\dagger}(\mathbf{p}_4) \, d_5^{\dagger}(\mathbf{p}_5)    \mbox{ }.
\label{3p0}
\end{eqnarray}

Here,  $\gamma_0$ is  the pair-creation strength, and $b_4^{\dagger}(\mathbf{p}_4)$ and $d_5^{\dagger}(\mathbf{p}_5)$ are the creation operators for a quark and an antiquark with momenta $\mathbf{p}_4$ and $\mathbf{p}_5$, respectively. The pair-creation strength,  $\gamma_0=21\pm3$, is fitted to reproduce the experimental $\Sigma^*_b\rightarrow \Lambda_b\pi$ strong decay width.

The $q \bar q$ pair is characterized by a color-singlet wave function $C_{45}$, a flavor-singlet wave function $F_{45}$, a spin-triplet wave function $\chi_{45}$ with spin $\bf S_{\rm 45}=1$ and a solid spherical harmonic ${\cal Y}_{1}(\mathbf{p}_4 - \mathbf{p}_5)$, since the quark and antiquark are in a relative $P$-wave. 

\begin{figure}[h]
    \centering a)
    \includegraphics[scale=0.33]{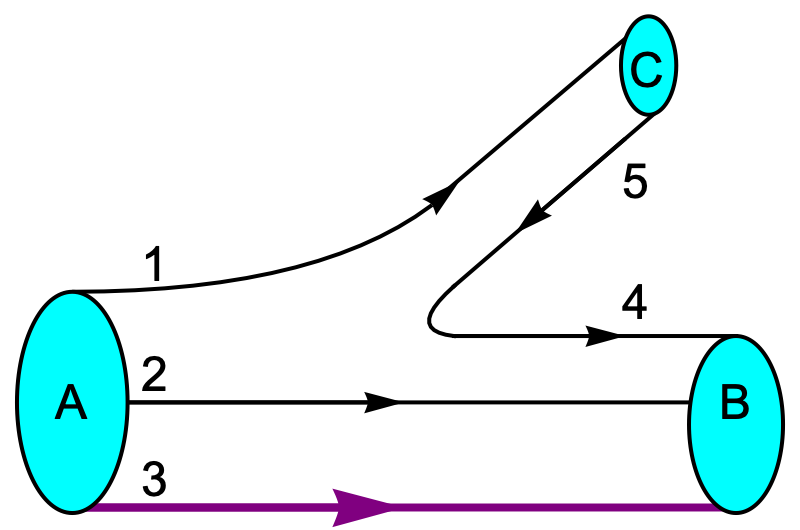}
    \\ \centering b)
    \includegraphics[scale=0.33]{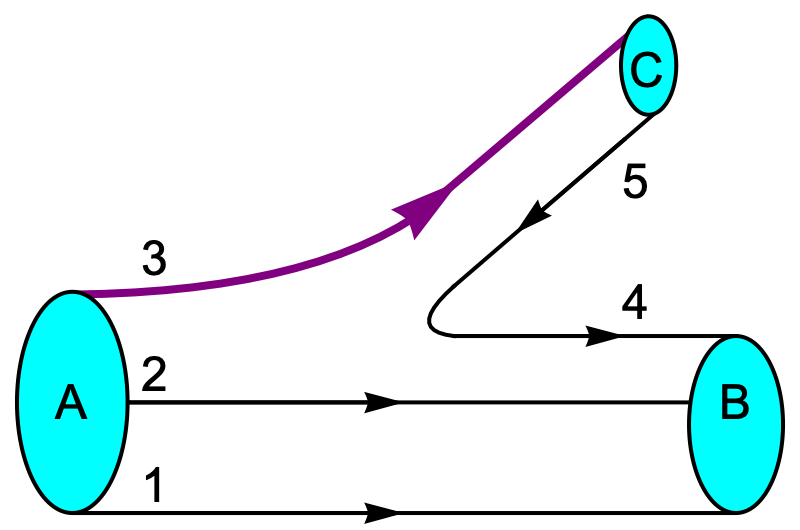}
    \caption{The $^3P_0$ pair-creation model (color online). The violet line 3 denotes a
bottom quark, while the remaining black lines denote light quarks. In diagram  a) the bottom baryon $A$ decays to a bottom baryon $B$ and a light meson $C$. In diagram b) the bottom baryon $A$ decays to a light baryon $B$ and a bottom meson $C$. }
    \label{fig:3P0}
\end{figure}

 According to the $^{3}P_0$ model, the decay of the baryon $A$ proceeds via the creation from the vacuum of the $q \bar q$ pair, the latter recombine into an outgoing  baryon  $B$ and a meson $C$, as depicted in Fig. \ref{fig:3P0}.
\\
The total strong decay width  $\Gamma_{\rm Strong}$ is the sum of the partial width of the single-bottom baryon $A$ decaying to the open-flavor channels $BC$. That is, $\Gamma_{\rm Strong}=\sum_{BC}\Gamma_{\rm Strong}(A \rightarrow BC)$, where 
the  strong partial decay widths $\Gamma_{\rm Strong}(A \rightarrow BC)$, are calculated using
\begin{eqnarray}
	\Gamma_{\rm Strong}(A \rightarrow BC)& = & \frac{2 \pi \gamma_0^2}{2J_{A}+1} \mbox{ } \Phi_{A \rightarrow BC} \nonumber \\
	&\times& \sum_{M_{J_A},M_{J_B}} \big|\mathcal{M}^{M_{J_A},M_{J_B}}\big|^2 \mbox{ },  
        \label{gamma}
\end{eqnarray}
where $\mathcal{M}= \langle BC|T^\dagger|A\rangle$ corresponds to the $^3P_0$ amplitude. The sum runs over the third components $M_{J_A,J_B}$ of the total angular momenta $J_{A,B}$ of $A$ and $B$.
The coefficient $\Phi_{A \rightarrow BC}$ is the relativistic phase space factor~\cite{Bijker:2015gyk,Ferretti:2015rsa}.

We follow the procedure of Ref. \cite{Garcia-Tecocoatzi:2022zrf,Bijker:2015gyk,Ferretti:2015rsa}, where the parameters of the  spatial wave functions, $\alpha_{\rho(\lambda)}$, are related to the baryon $\rho$- and $\lambda$-mode frequencies with $\alpha^2_{\rho(\lambda)}=\omega_{\rho(\lambda)}m_{\rho(\lambda)}$. Thus, $\alpha_{\rho(\lambda)}$ depend on the fit parameter $K_{b}$  and the quark masses. The convention of the h.o. wave functions and coordinate system conventions used in our decay width calculations are given in Ref. \cite{Garcia-Tecocoatzi:2022zrf}.

 The decay widths are calculated for all bottom-baryon up to $D$-wave; the available open-flavor channels include all the pseudoscalar and vector mesons.  In the calculation of the strong-decay width, there is an extra parameter $R$ related to the meson size, we use $R=2.1$ GeV$^{-1}$ \cite{Chen:2007xf,Blundell:1995ev,Garcia-Tecocoatzi:2022zrf}. 

The flavor-meson-wave functions are given in \ref{appme}. All the possible flavor couplings, $\mathcal{F}_{A\rightarrow BC}=\langle\phi_B \phi_C|\phi_0 \phi_A  \rangle$ are given in \ref{flavor}. The masses of the decay products are listed in Table~\ref{tab:exp_dec} in \ref{app2}.

\subsection{Electromagnetic decay widths }  
\label{secIID}
 The calculation of  the radiative-decay width of bottom baryons is done within the constituent quark model. First, we consider  the case of
the emission of a left-handed photon from a single-bottom baryon $A$ to another single-bottom baryon $A'$, i.e. $A\rightarrow A'\gamma$.
  Hence, the partial decay widths of the electromagnetic  transitions are
given by 
\begin{eqnarray}
 \Gamma_{\rm em}(A\rightarrow A'\gamma)=  \Phi_{A\rightarrow A'\gamma}\frac{1}{(2\pi)^2}\frac{2}{2J_A+1}\sum_{\nu>0}
\left|A_{\nu} \right|^2, \label{gammaEM}
\end{eqnarray}
where $J_A$ is the initial single-bottom baryon total angular momentum. The sum runs over the helicities $\nu$ of the initial baryon $A$, and
$ \Phi_{A\rightarrow A'\gamma}$ is the phase space factor, which in the rest frame of the initial baryon is expressed as
\begin{eqnarray}
  \Phi_{A\rightarrow A'\gamma}=4\pi\frac{E_{A'}}{M_{A}}k^2,
\end{eqnarray}
here  the energy of the final state is given by $E_{A'}=\sqrt{M^2_{A'}+k^2}$, $M_A$ and $M_A'$ are the masses of the initial and final baryon, respectively, and  the photon energy $k$ is given by\begin{eqnarray}
 k=\frac{M^2_{A}-M^2_{A'}}{2M_{A}}\ .
\end{eqnarray}
  The transition amplitude for a given helicity $\nu$ is defined as
\begin{eqnarray}
A_{\nu}=\langle J_{A'}, \nu-1| \mathcal{H}_{\rm em} |J_A, \nu\rangle,
\end{eqnarray}
where $\mathcal{H}_{\rm em}$ is the nonrelativistic Hamiltonian that models the electromagnetic transitions given by
\begin{eqnarray}
\mathcal{H}_{\rm em}=2\sqrt{\frac{\pi}{k}}\sum^3_{j=1}\mu_j\Big [k\mathbf{ s}_{j,-}e^{-i\mathbf{k} \cdot \mathbf{ r}_j}-  \nonumber \\ \frac{1}{2}\left ( \mathbf{ p}_{j,-}e^{-i\mathbf{k} \cdot \mathbf{ r}_j}+e^{-i\mathbf{k} \cdot \mathbf{ r}_j}\mathbf{ p}_{j,-} \right) \Big]\label{Hem}
\end{eqnarray}
where $\mathbf{ r}_j$, $\mathbf{ p}_j$,  $\mathbf{ s}_j$, and $\mu_j$ stand for  the coordinate, momentum, spin and magnetic moment of the $j$-th quark, respectively, and $\mathbf{ k}=k \hat{\mathbf z}$  corresponds to the momentum of a photon emitted in the $\hat{z}$ direction.

As described in the strong decay section (Section \ref{secIIC}), the harmonic oscillator wave functions depend on the parameters $\alpha_{\rho(\lambda)}$. Therefore, they only rely on the fit parameter $K_b$ and the constituent quark masses. Consequently, the calculation of electromagnetic decay widths does not introduce any additional parameters.
\begin{table}[htbp]
\caption{Fitted parameters for the three-quark model (second column) and  the quark-diquark model (third column). The $\dagger$ indicates that the parameter is absent in that model.}
\begin{tabular}{c | c c}\hline \hline
 Parameter  &  Three-quark value       & Diquark value    \\ \hline
 $m_{b}$ & $4930^{+12}_{-12}$ MeV & $4677^{+100}_{-64}$ MeV \\
 $m_{s}$ & $464^{+6}_{-6}$ MeV & $\dagger$ \\
 $m_{n}$ & $299^{+10}_{-10}$ MeV & $\dagger$ \\
 $m_{D_{\Omega}}$          & $\dagger$ & $1331^{+59}_{-92}$ MeV \\
 $m_{D_{\Xi}}$             & $\dagger$ & $1185^{+61}_{-92}$ MeV \\
 $m_{D_{\Sigma,\Lambda}}$ & $\dagger$ & $1045^{+55}_{-94}$ MeV \\
 $K_b$   & $0.0254^{+0.0012}_{-0.0012}$ GeV$^{3}$ & $0.0245^{+0.0023}_{-0.0023}$ GeV$^{3}$ \\
 $a_{\rm S}$     & $10^{+2}_{-3}$ MeV & $8^{+2}_{-3}$ MeV \\
 $a_{\rm SL}$     & $4^{+2}_{-2}$ MeV & $6^{+3}_{-1}$ MeV \\
 $a_{\rm I}$     & $36^{+7}_{-7}$ MeV & $19^{+3}_{-2}$ MeV \\
 $a_{\rm F}$     & $60^{+6}_{-5}$ MeV & $13^{+6}_{-3}$ MeV \\
\hline\hline
\end{tabular}

\label{tab:comb_fit}
\end{table}

\section{Parameter determination and uncertainties}
\label{Parameter determination}

We performed a fit to describe the observed masses of single bottom baryons, namely $\Lambda_{b}$, $\Sigma_{b}$, $\Xi_{b}$, $\Sigma_{b}$, and $\Omega_{b}$, with the masses predicted by Eq.\ref{eq:Hho} and Eq.\ref{eq:Hhodi}. This fitting procedure enabled us to determine the masses of the constituent quarks and diquarks ($m_{b}$, $m_s$, $m_{u,d}$, $m_{D_{\Omega}}$, $m_{D_{\Xi}}$, and $m_{D_{\Sigma,\Lambda}}$), and the model parameters  ($a_{\rm S}, a_{\rm SL}, a_{\rm I}, a_{\rm F},$ and $K_{b}$). The goal is that the parameters minimize the sum of the squared differences between the theory-predicted baryon masses and their corresponding experimental values (least-squares method).

Experimental measurements of baryon masses are associated with both statistical and systematic uncertainties. Moreover, the models presented in Eq.\ref{eq:Hho} and Eq.\ref{eq:Hhodi} provide approximate descriptions of bottom baryons. To account for possible deviations between these models and experimental data, we assigned a model uncertainty to each model. The calculation of the model uncertainty, denoted as $\sigma_{mod}$, followed the procedure outlined in Ref.~\cite{Workman:2022ynf}, ensuring that the $\chi^2/NDF$ value approached 1. The computation of $\chi^2$ involved the equation:

\begin{equation}
\label{eq:opt_chi}
\chi^2=\sum_{i} \dfrac{(M_{mod,i}-M_{exp,i})^2} {\sigma_{mod}^2 + \sigma_{exp,i}^2},
\end{equation}
where, $M_{mod,i}$ represents the predicted masses of single bottom baryons, $M_{exp,i}$ denotes the experimental masses of single bottom baryons included in the fitting process, along with their uncertainties $\sigma_{exp,i}$, and $NDF$ refers to the number of degrees of freedom. For the three-quark model, we obtained a value of $\sigma_{mod} = 12$ MeV, while for the quark-diquark model, the value was $\sigma_{mod} = 20$ MeV. The fitted parameters for the bottom baryon masses are shown in Table~\ref{tab:comb_fit}.

To incorporate both experimental and model uncertainties in the fitting process, we performed a statistical simulation using error propagation. This involved randomly sampling the experimental masses from Gaussian distributions with means equal to the central mass values and widths equal to the squared sum of the uncertainties. The fitting procedure was repeated $10^4$ times, with each iteration utilizing a sampled mass corresponding to an experimentally observed state included in the fit. Through this simulation, Gaussian distributions were obtained for each constituent quark mass, model parameter, and baryon mass.
Therefore, the  parameter value is the mean of the distribution, and the parameter uncertainty is its difference from the distribution quantiles at the 68\% confidence level to obtain the confidence interval (C.I.).
This methodology is known as Monte Carlo bootstrap uncertainty propagation \cite{Efron1994, Molina:2020zao}. 

The masses and their uncertainties used in the fit are provided in the PDG~\cite{Workman:2022ynf} and are marked with (*) in Tables~\ref{tab:All_mass_Lambda}-\ref{tab:All_mass_Omega}. Similarly, the procedure of error propagation was carried out for the strong decay widths of the three-quark system. When we calculate the strong decay widths, we took into account the uncertainties associated with the mass model parameters, the decay product masses, and the pair-creation strength $\gamma_0$. The experimental values for the decay product masses, along with their corresponding uncertainties, are given in Table~\ref{app2}. The uncertainty in $\gamma_0$ reported in Sec.~\ref{secIIC}, is computed as the sum in quadrature of a model uncertainty $\sigma_{mod}=2.4$ and an experimental uncertainty $\sigma_{exp}=1.0$.  The fitting and error propagation processes were conducted using \texttt{MINUIT}~\cite{JAMES1975343} and \texttt{NUMPY}~\cite{Harris2020}.

\begin{figure}
    \centering
    \includegraphics[width=0.5\textwidth]{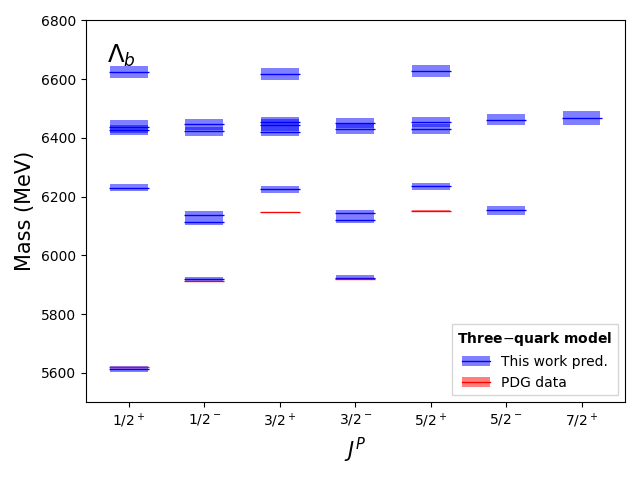}
    \caption{$\Lambda_b$ mass spectra and tentative quantum number assignments based on the three-quark model Hamiltonian of Eqs. \ref{MassFormula} and \ref{eq:Hho}. The theoretical predictions and their uncertainties (red lines and bands) are compared with the experimental results (red lines and bands) given in the PDG \cite{Workman:2022ynf}. The experimental errors are too small to be  reported on this energy scale.}
    \label{fig:lambdas}
\end{figure}

\begin{figure}
    \centering
    \label{fig:cascades}
    \includegraphics[width=0.5\textwidth]{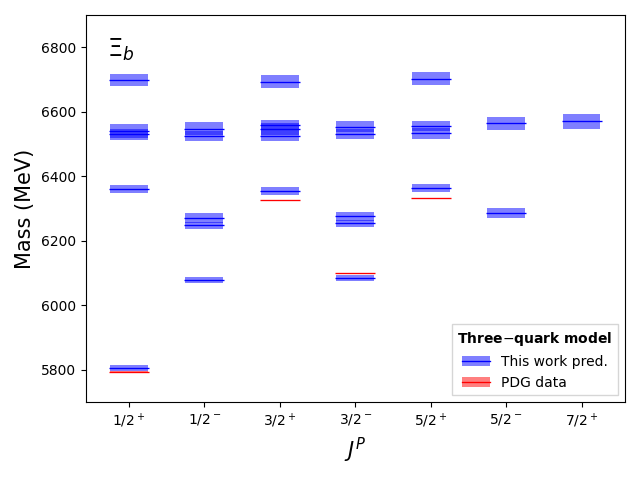}
    \caption{Same as Figure \ref{fig:lambdas}, but for $\Xi_b$ states.}
\end{figure}

\begin{figure}
    \centering
    \includegraphics[width=0.5\textwidth]{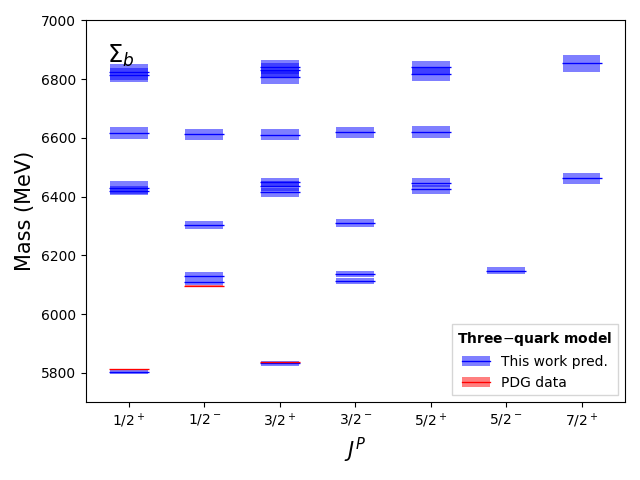}
    \caption{Same as Figure \ref{fig:lambdas}, but for $\Sigma_b$ states.}
    \label{fig:sigmas}
\end{figure}

\begin{figure}
    \centering
    \includegraphics[width=0.5\textwidth]{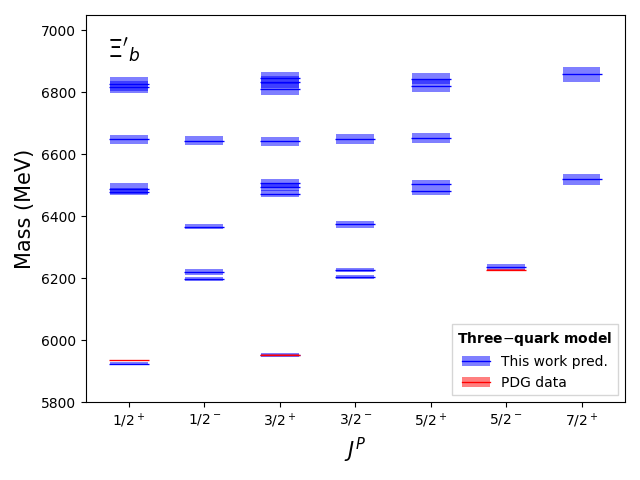}
    \caption{Same as Figure \ref{fig:lambdas}, but for $\Xi'_b$ states.}
    \label{fig:cascades}
\end{figure}

\begin{figure}
    \centering
    \includegraphics[width=0.5\textwidth]{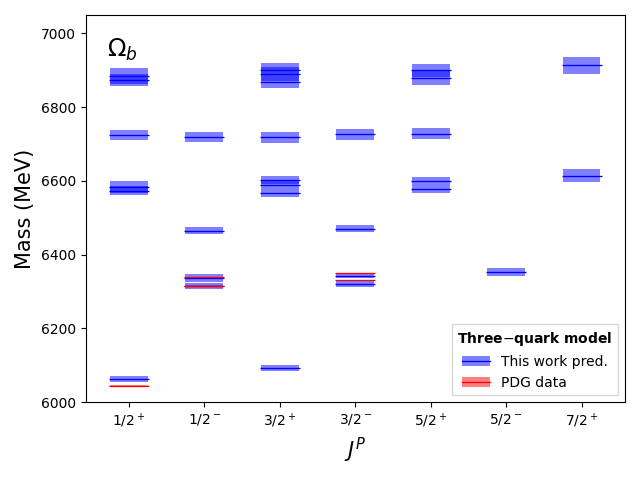}
    \caption{Same as  Figure \ref{fig:lambdas}, but for $\Omega_b$ states.}
    \label{fig:omegas}
\end{figure}

\begin{figure}
    \centering
    \includegraphics[width=0.5\textwidth]{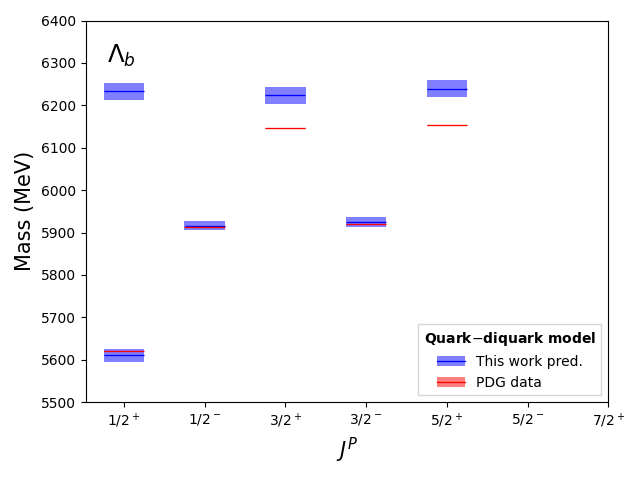}
    \caption{$\Lambda_b$ mass spectra and tentative quantum number assignments based on the quark-diquark model Hamiltonian of Eqs. \ref{MassFormula} and \ref{eq:Hhodi}. The theoretical predictions and their uncertainties (red lines and bands) are compared with the experimental results (red lines and bands) given in the PDG \cite{Workman:2022ynf}. The experimental errors are too small to be  reported on this energy scale.}
    \label{fig:lambdasD}
\end{figure}

\begin{figure}
    \centering
    \label{fig:cascades_anti3D}
    \includegraphics[width=0.5\textwidth]{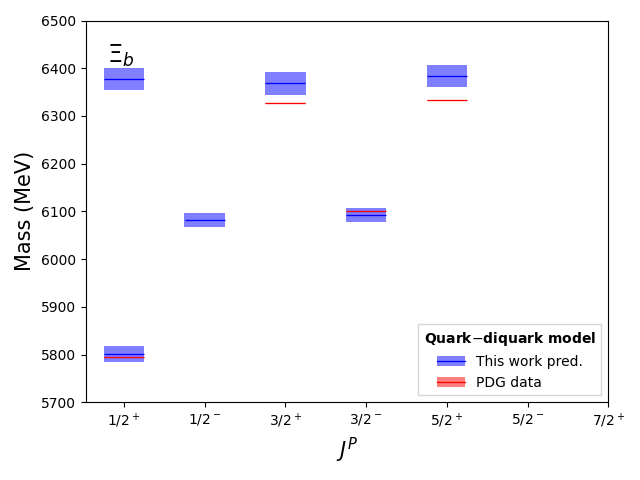}
    \caption{Same as Figure \ref{fig:lambdasD}, but for $\Xi_b$ states.}
\end{figure}

\begin{figure}
    \centering
    \includegraphics[width=0.5\textwidth]{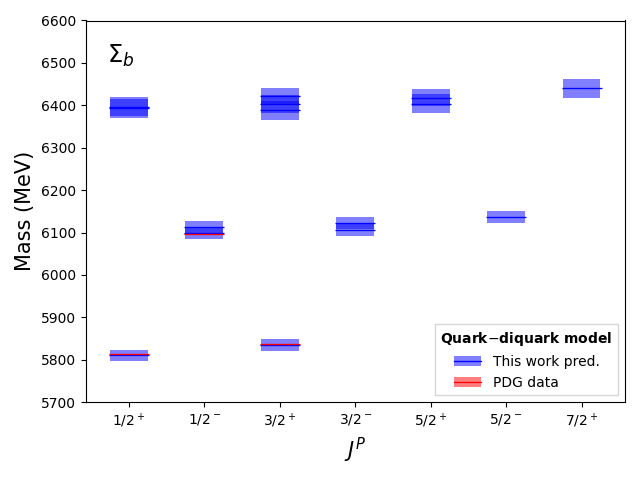}
    \caption{Same as  Figure \ref{fig:lambdasD}, but for $\Sigma_b$ states.}
    \label{fig:sigmasD}
\end{figure}

\begin{figure}
    \centering
    \includegraphics[width=0.5\textwidth]{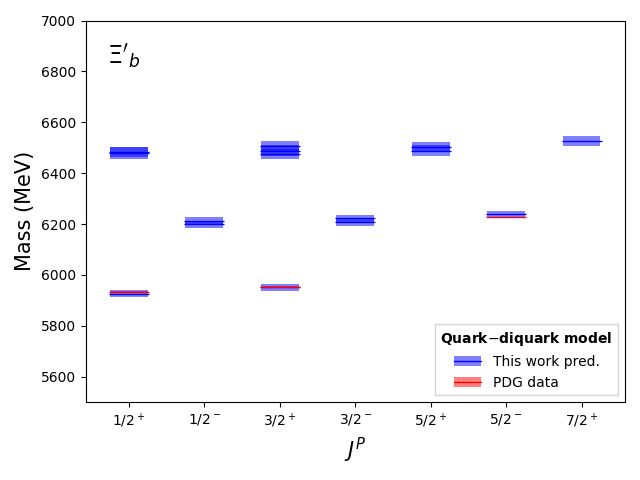}
    \caption{Same as  Figure \ref{fig:lambdasD}, but for $\Xi'_b$ states.}
    \label{fig:cascades_anti3D}
\end{figure}

\begin{figure}
    \centering
    \includegraphics[width=0.5\textwidth]{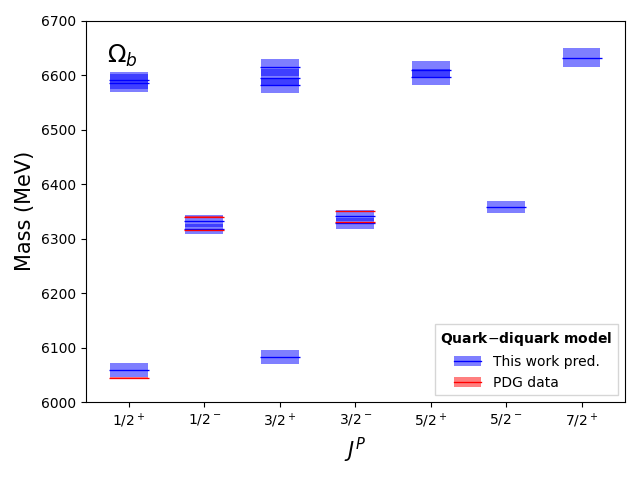}
    \caption{Same as  Figure \ref{fig:lambdasD}, but for $\Lambda_b$ states.}
    \label{fig:omegasD}
\end{figure}

\begin{table*}[htbp]
\caption{$\Lambda_b(nnb)$ states. The flavor multiplet is specified with the symbol $\mathcal{F}$. The first column contains the  three-quark model state, $\left| l_{\lambda},l_{\rho}, k_{\lambda},k_{\rho}\right\rangle$, where $l_{\lambda,\rho}$ are the orbital angular momenta and $k_{\lambda,\rho}$ the number of nodes of the $\lambda$ and $\rho$ oscillators.
The second column displays the spectroscopic notation $^{2S+1}L_J$ for each state. The third column contains the total angular momentum and parity ${\bf J}^P $. In the fourth column the predicted masses, computed within the three-quark model, are shown, whereas the  total strong decay widths  $\Gamma_{\rm Strong}$, computed within the $^3P_0$ model, are shown in the eighth column.
The fifth column reports the quark-diquark model state, $\left| l_{r}, k_{r} \right\rangle$,  where ${ l}_{r}$ is the orbital angular momentum and $k_{r}$ denotes the number of nodes.
 The predicted masses, computed within the quark-diquark framework, are presented in the sixth column. Furthermore, $N=n_\rho+n_\lambda$ in the three-quark model,
 and $N=n_r$, in the quark-diquark model separate the $N=0,1,2$ energy bands. Our theoretical results are compared with the experimental masses of the seventh column and the experimental decay widths of the ninth column as from PDG  \cite{ParticleDataGroup:2018ovx}. The (*) indicates the experimental mass and decay width values included in the fits. The $\dagger$ indicates that there is no reported experimental mass or decay width for that state up to now. The $\dagger\dagger$ indicates that there is no quark-diquark prediction for that state.}

\begingroup
\setlength{\tabcolsep}{1.75pt} 
\renewcommand{\arraystretch}{1.35} 

\begin{tabular}{c c c c |c c | c c c}\hline \hline
${\mathcal{F} = \bf {\bar{3}}}_{\rm F}$&   \multicolumn{2}{c}{Three-quark} & &   \multicolumn{2}{c|}{Quark-diquark} & &    Three-quark  &\\
$\Lambda_{b}(nnb)$ & & &  Predicted   &  &   Predicted &    Experimental &  Predicted            & Experimental \\
  $\vert l_{\lambda}, l_{\rho}, k_{\lambda}, k_{\rho} \rangle$ & $^{2S+1}L_{J}$ & ${\bf J}^P$  & Mass (MeV) & $\vert l_{r}, k_{r} \rangle$  &  Mass (MeV)   &  Mass (MeV)   &  $\Gamma_{\rm Strong}$ (MeV) & $\Gamma$ (MeV) \\ \hline
\hline
 $N=0$  &  &  &  &  &  \\
$\vert \,0\,,\,0\,,\,0\,,\,0 \,\rangle $ & $^{2}S_{1/2}$ & ${\bf \frac{1}{2}}^+$ & $5613^{+9}_{-9}$ & $\vert \,0\,,\,0 \,\rangle$ & $5611^{+15}_{-16}$ & $5619.60\pm 0.17$(*) & $0$ & $\approx 0$ \\
\hline
 $N=1$  &  &  &  &  &  \\
$\vert \,1\,,\,0\,,\,0\,,\,0 \,\rangle $ & $^{2}P_{1/2}$ & ${\bf \frac{1}{2}}^-$ & $5918^{+8}_{-8}$ & $\vert \,1\,,\,0 \,\rangle$ & $5916^{+11}_{-12}$ & $5912.19\pm 0.17$(*) & $0$ & $<0.25$ \\
$\vert \,1\,,\,0\,,\,0\,,\,0 \,\rangle $ & $^{2}P_{3/2}$ & ${\bf \frac{3}{2}}^-$ & $5924^{+8}_{-8}$ & $\vert \,1\,,\,0 \,\rangle$ & $5925^{+12}_{-12}$ & $5920.09\pm 0.17$(*) & $0$ & $<0.19$ \\
$\vert \,0\,,\,1\,,\,0\,,\,0 \,\rangle $ & $^{2}P_{1/2}$ & ${\bf \frac{1}{2}}^-$ & $6114^{+10}_{-10}$ & $\dagger\dagger$ & $\dagger\dagger$ & $\dagger$ & $67^{+16}_{-16}$ & $\dagger$ \\
$\vert \,0\,,\,1\,,\,0\,,\,0 \,\rangle $ & $^{4}P_{1/2}$ & ${\bf \frac{1}{2}}^-$ & $6137^{+14}_{-14}$ & $\dagger\dagger$ & $\dagger\dagger$ & $\dagger$ & $36^{+8}_{-8}$ & $\dagger$ \\
$\vert \,0\,,\,1\,,\,0\,,\,0 \,\rangle $ & $^{2}P_{3/2}$ & ${\bf \frac{3}{2}}^-$ & $6121^{+10}_{-10}$ & $\dagger\dagger$ & $\dagger\dagger$ & $\dagger$ & $85^{+21}_{-21}$ & $\dagger$ \\
$\vert \,0\,,\,1\,,\,0\,,\,0 \,\rangle $ & $^{4}P_{3/2}$ & ${\bf \frac{3}{2}}^-$ & $6143^{+12}_{-12}$ & $\dagger\dagger$ & $\dagger\dagger$ & $\dagger$ & $128^{+31}_{-31}$ & $\dagger$ \\
$\vert \,0\,,\,1\,,\,0\,,\,0 \,\rangle $ & $^{4}P_{5/2}$ & ${\bf \frac{5}{2}}^-$ & $6153^{+14}_{-14}$ & $\dagger\dagger$ & $\dagger\dagger$ & $\dagger$ & $74^{+19}_{-19}$ & $\dagger$ \\
\hline
 $N=2$  &  &  &  &  &  \\
$\vert \,2\,,\,0\,,\,0\,,\,0 \,\rangle $ & $^{2}D_{3/2}$ & ${\bf \frac{3}{2}}^+$ & $6225^{+13}_{-13}$ & $\vert \,2\,,\,0 \,\rangle$ & $6224^{+20}_{-20}$ & $6146.2\pm 0.4$ & $13^{+5}_{-5}$ & $2.9\pm 1.3$ \\
$\vert \,2\,,\,0\,,\,0\,,\,0 \,\rangle $ & $^{2}D_{5/2}$ & ${\bf \frac{5}{2}}^+$ & $6235^{+13}_{-13}$ & $\vert \,2\,,\,0 \,\rangle$ & $6239^{+20}_{-20}$ & $6152.5\pm 0.4$ & $18^{+12}_{-13}$ & $2.1\pm 0.9$ \\
$\vert \,0\,,\,0\,,\,1\,,\,0 \,\rangle $ & $^{2}S_{1/2}$ & ${\bf \frac{1}{2}}^+$ & $6231^{+12}_{-12}$ & $\vert \,0\,,\,1 \,\rangle$ & $6233^{+20}_{-20}$ & $\dagger$ & $1.1^{+0.6}_{-0.6}$ & $\dagger$ \\
$\vert \,0\,,\,0\,,\,0\,,\,1 \,\rangle $ & $^{2}S_{1/2}$ & ${\bf \frac{1}{2}}^+$ & $6624^{+21}_{-21}$ & $\dagger\dagger$ & $\dagger\dagger$ & $\dagger$ & $4^{+1}_{-1}$ & $\dagger$ \\
$\vert \,1\,,\,1\,,\,0\,,\,0 \,\rangle $ & $^{2}D_{3/2}$ & ${\bf \frac{3}{2}}^+$ & $6421^{+16}_{-16}$ & $\dagger\dagger$ & $\dagger\dagger$ & $\dagger$ & $67^{+17}_{-17}$ & $\dagger$ \\
$\vert \,1\,,\,1\,,\,0\,,\,0 \,\rangle $ & $^{2}D_{5/2}$ & ${\bf \frac{5}{2}}^+$ & $6431^{+17}_{-17}$ & $\dagger\dagger$ & $\dagger\dagger$ & $\dagger$ & $108^{+28}_{-28}$ & $\dagger$ \\
$\vert \,1\,,\,1\,,\,0\,,\,0 \,\rangle $ & $^{4}D_{1/2}$ & ${\bf \frac{1}{2}}^+$ & $6438^{+22}_{-22}$ & $\dagger\dagger$ & $\dagger\dagger$ & $\dagger$ & $34^{+9}_{-9}$ & $\dagger$ \\
$\vert \,1\,,\,1\,,\,0\,,\,0 \,\rangle $ & $^{4}D_{3/2}$ & ${\bf \frac{3}{2}}^+$ & $6444^{+19}_{-19}$ & $\dagger\dagger$ & $\dagger\dagger$ & $\dagger$ & $95^{+25}_{-25}$ & $\dagger$ \\
$\vert \,1\,,\,1\,,\,0\,,\,0 \,\rangle $ & $^{4}D_{5/2}$ & ${\bf \frac{5}{2}}^+$ & $6454^{+17}_{-17}$ & $\dagger\dagger$ & $\dagger\dagger$ & $\dagger$ & $128^{+34}_{-33}$ & $\dagger$ \\
$\vert \,1\,,\,1\,,\,0\,,\,0 \,\rangle $ & $^{4}D_{7/2}$ & ${\bf \frac{7}{2}}^+$ & $6468^{+23}_{-22}$ & $\dagger\dagger$ & $\dagger\dagger$ & $\dagger$ & $122^{+34}_{-34}$ & $\dagger$ \\
$\vert \,1\,,\,1\,,\,0\,,\,0 \,\rangle $ & $^{2}P_{1/2}$ & ${\bf \frac{1}{2}}^-$ & $6423^{+16}_{-16}$ & $\dagger\dagger$ & $\dagger\dagger$ & $\dagger$ & $0.5^{+0.1}_{-0.1}$ & $\dagger$ \\
$\vert \,1\,,\,1\,,\,0\,,\,0 \,\rangle $ & $^{2}P_{3/2}$ & ${\bf \frac{3}{2}}^-$ & $6429^{+17}_{-17}$ & $\dagger\dagger$ & $\dagger\dagger$ & $\dagger$ & $1.7^{+0.5}_{-0.5}$ & $\dagger$ \\
$\vert \,1\,,\,1\,,\,0\,,\,0 \,\rangle $ & $^{4}P_{1/2}$ & ${\bf \frac{1}{2}}^-$ & $6446^{+19}_{-18}$ & $\dagger\dagger$ & $\dagger\dagger$ & $\dagger$ & $0.3^{+0.1}_{-0.1}$ & $\dagger$ \\
$\vert \,1\,,\,1\,,\,0\,,\,0 \,\rangle $ & $^{4}P_{3/2}$ & ${\bf \frac{3}{2}}^-$ & $6452^{+17}_{-17}$ & $\dagger\dagger$ & $\dagger\dagger$ & $\dagger$ & $1.2^{+0.3}_{-0.3}$ & $\dagger$ \\
$\vert \,1\,,\,1\,,\,0\,,\,0 \,\rangle $ & $^{4}P_{5/2}$ & ${\bf \frac{5}{2}}^-$ & $6462^{+19}_{-19}$ & $\dagger\dagger$ & $\dagger\dagger$ & $\dagger$ & $2^{+1}_{-1}$ & $\dagger$ \\
$\vert \,1\,,\,1\,,\,0\,,\,0 \,\rangle $ & $^{4}S_{3/2}$ & ${\bf \frac{3}{2}}^+$ & $6456^{+17}_{-18}$ & $\dagger\dagger$ & $\dagger\dagger$ & $\dagger$ & $32^{+8}_{-8}$ & $\dagger$ \\
$\vert \,1\,,\,1\,,\,0\,,\,0 \,\rangle $ & $^{2}S_{1/2}$ & ${\bf \frac{1}{2}}^+$ & $6427^{+16}_{-16}$ & $\dagger\dagger$ & $\dagger\dagger$ & $\dagger$ & $29^{+7}_{-7}$ & $\dagger$ \\
$\vert \,0\,,\,2\,,\,0\,,\,0 \,\rangle $ & $^{2}D_{3/2}$ & ${\bf \frac{3}{2}}^+$ & $6618^{+20}_{-20}$ & $\dagger\dagger$ & $\dagger\dagger$ & $\dagger$ & $131^{+33}_{-32}$ & $\dagger$ \\
$\vert \,0\,,\,2\,,\,0\,,\,0 \,\rangle $ & $^{2}D_{5/2}$ & ${\bf \frac{5}{2}}^+$ & $6628^{+21}_{-22}$ & $\dagger\dagger$ & $\dagger\dagger$ & $\dagger$ & $185^{+49}_{-49}$ & $\dagger$ \\
\hline \hline
\end{tabular}

\endgroup

\label{tab:All_mass_Lambda}
\end{table*}

\begin{table*}[htbp]
\caption{Same as Table \ref{tab:All_mass_Lambda}, but for  $ \Xi_b(snb) $ states.}

\begingroup
\setlength{\tabcolsep}{1.75pt} 
\renewcommand{\arraystretch}{1.35} 

\begin{tabular}{c c c c |c c | c c c}\hline \hline
${\mathcal{F} = \bf {\bar{3}}}_{\rm F}$&   \multicolumn{2}{c}{Three-quark} & &    \multicolumn{2}{c|}{Quark-diquark}  & &    Three-quark  &\\
$\Xi_{b}(snb)$ & & &  Predicted   &  &   Predicted &    Experimental &  Predicted            & Experimental \\
  $\vert l_{\lambda}, l_{\rho}, k_{\lambda}, k_{\rho} \rangle$ & $^{2S+1}L_{J}$ & ${\bf J}^P$  & Mass (MeV) & $\vert l_{r}, k_{r} \rangle$  &  Mass (MeV)   &  Mass (MeV)   &  $\Gamma_{\rm Strong}$ (MeV) & $\Gamma$ (MeV) \\ \hline
\hline
 $N=0$  &  &  &  &  &  \\
$\vert \,0\,,\,0\,,\,0\,,\,0 \,\rangle $ & $^{2}S_{1/2}$ & ${\bf \frac{1}{2}}^+$ & $5806^{+9}_{-9}$ & $\vert \,0\,,\,0 \,\rangle$ & $5801^{+16}_{-16}$ & $5794.5\pm 0.6$(*) & $0$ & $\approx 0$ \\
\hline
 $N=1$  &  &  &  &  &  \\
$\vert \,1\,,\,0\,,\,0\,,\,0 \,\rangle $ & $^{2}P_{1/2}$ & ${\bf \frac{1}{2}}^-$ & $6079^{+9}_{-9}$ & $\vert \,1\,,\,0 \,\rangle$ & $6082^{+15}_{-16}$ & $\dagger$ & $0.2^{+0.1}_{-0.2}$ & $\dagger$ \\
$\vert \,1\,,\,0\,,\,0\,,\,0 \,\rangle $ & $^{2}P_{3/2}$ & ${\bf \frac{3}{2}}^-$ & $6085^{+9}_{-9}$ & $\vert \,1\,,\,0 \,\rangle$ & $6092^{+15}_{-15}$ & $6100.3\pm 0.6$(*) & $1.1^{+0.6}_{-0.6}$ & $<1.9$ \\
$\vert \,0\,,\,1\,,\,0\,,\,0 \,\rangle $ & $^{2}P_{1/2}$ & ${\bf \frac{1}{2}}^-$ & $6248^{+11}_{-11}$ & $\dagger\dagger$ & $\dagger\dagger$ & $\dagger$ & $9^{+2}_{-2}$ & $\dagger$ \\
$\vert \,0\,,\,1\,,\,0\,,\,0 \,\rangle $ & $^{4}P_{1/2}$ & ${\bf \frac{1}{2}}^-$ & $6271^{+15}_{-15}$ & $\dagger\dagger$ & $\dagger\dagger$ & $\dagger$ & $6^{+1}_{-2}$ & $\dagger$ \\
$\vert \,0\,,\,1\,,\,0\,,\,0 \,\rangle $ & $^{2}P_{3/2}$ & ${\bf \frac{3}{2}}^-$ & $6255^{+11}_{-11}$ & $\dagger\dagger$ & $\dagger\dagger$ & $\dagger$ & $66^{+16}_{-16}$ & $\dagger$ \\
$\vert \,0\,,\,1\,,\,0\,,\,0 \,\rangle $ & $^{4}P_{3/2}$ & ${\bf \frac{3}{2}}^-$ & $6277^{+14}_{-14}$ & $\dagger\dagger$ & $\dagger\dagger$ & $\dagger$ & $26^{+7}_{-7}$ & $\dagger$ \\
$\vert \,0\,,\,1\,,\,0\,,\,0 \,\rangle $ & $^{4}P_{5/2}$ & ${\bf \frac{5}{2}}^-$ & $6287^{+15}_{-15}$ & $\dagger\dagger$ & $\dagger\dagger$ & $\dagger$ & $68^{+16}_{-16}$ & $\dagger$ \\
\hline
 $N=2$  &  &  &  &  &  \\
$\vert \,2\,,\,0\,,\,0\,,\,0 \,\rangle $ & $^{2}D_{3/2}$ & ${\bf \frac{3}{2}}^+$ & $6354^{+13}_{-13}$ & $\vert \,2\,,\,0 \,\rangle$ & $6368^{+24}_{-21}$ & $6327.3\pm 2.5$ & $1.9^{+0.8}_{-0.8}$ & $<2.2$ \\
$\vert \,2\,,\,0\,,\,0\,,\,0 \,\rangle $ & $^{2}D_{5/2}$ & ${\bf \frac{5}{2}}^+$ & $6364^{+13}_{-13}$ & $\vert \,2\,,\,0 \,\rangle$ & $6383^{+23}_{-21}$ & $6332.7\pm 2.5$ & $1.5^{+0.5}_{-0.5}$ & $<1.6$ \\
$\vert \,0\,,\,0\,,\,1\,,\,0 \,\rangle $ & $^{2}S_{1/2}$ & ${\bf \frac{1}{2}}^+$ & $6360^{+12}_{-13}$ & $\vert \,0\,,\,1 \,\rangle$ & $6377^{+23}_{-21}$ & $\dagger$ & $0.2^{+0.1}_{-0.1}$ & $\dagger$ \\
$\vert \,0\,,\,0\,,\,0\,,\,1 \,\rangle $ & $^{2}S_{1/2}$ & ${\bf \frac{1}{2}}^+$ & $6699^{+19}_{-19}$ & $\dagger\dagger$ & $\dagger\dagger$ & $\dagger$ & $6^{+1}_{-1}$ & $\dagger$ \\
$\vert \,1\,,\,1\,,\,0\,,\,0 \,\rangle $ & $^{2}D_{3/2}$ & ${\bf \frac{3}{2}}^+$ & $6524^{+16}_{-16}$ & $\dagger\dagger$ & $\dagger\dagger$ & $\dagger$ & $46^{+12}_{-12}$ & $\dagger$ \\
$\vert \,1\,,\,1\,,\,0\,,\,0 \,\rangle $ & $^{2}D_{5/2}$ & ${\bf \frac{5}{2}}^+$ & $6534^{+17}_{-17}$ & $\dagger\dagger$ & $\dagger\dagger$ & $\dagger$ & $108^{+27}_{-27}$ & $\dagger$ \\
$\vert \,1\,,\,1\,,\,0\,,\,0 \,\rangle $ & $^{4}D_{1/2}$ & ${\bf \frac{1}{2}}^+$ & $6540^{+22}_{-22}$ & $\dagger\dagger$ & $\dagger\dagger$ & $\dagger$ & $20^{+5}_{-5}$ & $\dagger$ \\
$\vert \,1\,,\,1\,,\,0\,,\,0 \,\rangle $ & $^{4}D_{3/2}$ & ${\bf \frac{3}{2}}^+$ & $6546^{+19}_{-19}$ & $\dagger\dagger$ & $\dagger\dagger$ & $\dagger$ & $67^{+18}_{-18}$ & $\dagger$ \\
$\vert \,1\,,\,1\,,\,0\,,\,0 \,\rangle $ & $^{4}D_{5/2}$ & ${\bf \frac{5}{2}}^+$ & $6556^{+17}_{-18}$ & $\dagger\dagger$ & $\dagger\dagger$ & $\dagger$ & $100^{+26}_{-26}$ & $\dagger$ \\
$\vert \,1\,,\,1\,,\,0\,,\,0 \,\rangle $ & $^{4}D_{7/2}$ & ${\bf \frac{7}{2}}^+$ & $6570^{+22}_{-22}$ & $\dagger\dagger$ & $\dagger\dagger$ & $\dagger$ & $114^{+30}_{-30}$ & $\dagger$ \\
$\vert \,1\,,\,1\,,\,0\,,\,0 \,\rangle $ & $^{2}P_{1/2}$ & ${\bf \frac{1}{2}}^-$ & $6526^{+16}_{-16}$ & $\dagger\dagger$ & $\dagger\dagger$ & $\dagger$ & $0.3^{+0.1}_{-0.1}$ & $\dagger$ \\
$\vert \,1\,,\,1\,,\,0\,,\,0 \,\rangle $ & $^{2}P_{3/2}$ & ${\bf \frac{3}{2}}^-$ & $6532^{+16}_{-16}$ & $\dagger\dagger$ & $\dagger\dagger$ & $\dagger$ & $2^{+1}_{-1}$ & $\dagger$ \\
$\vert \,1\,,\,1\,,\,0\,,\,0 \,\rangle $ & $^{4}P_{1/2}$ & ${\bf \frac{1}{2}}^-$ & $6548^{+19}_{-19}$ & $\dagger\dagger$ & $\dagger\dagger$ & $\dagger$ & $0.2^{+0.1}_{-0.1}$ & $\dagger$ \\
$\vert \,1\,,\,1\,,\,0\,,\,0 \,\rangle $ & $^{4}P_{3/2}$ & ${\bf \frac{3}{2}}^-$ & $6554^{+18}_{-17}$ & $\dagger\dagger$ & $\dagger\dagger$ & $\dagger$ & $0.9^{+0.3}_{-0.3}$ & $\dagger$ \\
$\vert \,1\,,\,1\,,\,0\,,\,0 \,\rangle $ & $^{4}P_{5/2}$ & ${\bf \frac{5}{2}}^-$ & $6564^{+19}_{-19}$ & $\dagger\dagger$ & $\dagger\dagger$ & $\dagger$ & $3^{+1}_{-1}$ & $\dagger$ \\
$\vert \,1\,,\,1\,,\,0\,,\,0 \,\rangle $ & $^{4}S_{3/2}$ & ${\bf \frac{3}{2}}^+$ & $6558^{+18}_{-18}$ & $\dagger\dagger$ & $\dagger\dagger$ & $\dagger$ & $33^{+8}_{-8}$ & $\dagger$ \\
$\vert \,1\,,\,1\,,\,0\,,\,0 \,\rangle $ & $^{2}S_{1/2}$ & ${\bf \frac{1}{2}}^+$ & $6530^{+16}_{-16}$ & $\dagger\dagger$ & $\dagger\dagger$ & $\dagger$ & $31^{+7}_{-8}$ & $\dagger$ \\
$\vert \,0\,,\,2\,,\,0\,,\,0 \,\rangle $ & $^{2}D_{3/2}$ & ${\bf \frac{3}{2}}^+$ & $6693^{+20}_{-19}$ & $\dagger\dagger$ & $\dagger\dagger$ & $\dagger$ & $127^{+31}_{-31}$ & $\dagger$ \\
$\vert \,0\,,\,2\,,\,0\,,\,0 \,\rangle $ & $^{2}D_{5/2}$ & ${\bf \frac{5}{2}}^+$ & $6703^{+20}_{-20}$ & $\dagger\dagger$ & $\dagger\dagger$ & $\dagger$ & $98^{+25}_{-25}$ & $\dagger$ \\
\hline \hline
\end{tabular}

\endgroup
\label{tab:All_mass_Xi}
\end{table*}

\begin{table*}[htbp]
\caption{Same as  Table \ref{tab:All_mass_Lambda}, but for  $ \Sigma_b(nnb) $ states.}

\begingroup
\setlength{\tabcolsep}{1.75pt} 
\renewcommand{\arraystretch}{1.35} 

\begin{tabular}{c c c c |c c | c c c}\hline \hline
${\mathcal{F} =  \bf {6}}_{\rm F}$&   \multicolumn{2}{c}{Three-quark} & &    \multicolumn{2}{c|}{Quark-diquark}  & &    Three-quark  &\\
$\Sigma_{b}(nnb)$ & & &  Predicted   &  &   Predicted &    Experimental &  Predicted            & Experimental \\
  $\vert l_{\lambda}, l_{\rho}, k_{\lambda}, k_{\rho} \rangle$ & $^{2S+1}L_{J}$ & ${\bf J}^P$  & Mass (MeV) & $\vert l_{r}, k_{r} \rangle$  &  Mass (MeV)   &  Mass (MeV)   &  $\Gamma_{\rm Strong}$ (MeV) & $\Gamma$ (MeV) \\ \hline
\hline
 $N=0$  &  &  &  &  &  \\
$\vert \,0\,,\,0\,,\,0\,,\,0 \,\rangle $ & $^{2}S_{1/2}$ & ${\bf \frac{1}{2}}^+$ & $5804^{+8}_{-8}$ & $\vert \,0\,,\,0 \,\rangle$ & $5811^{+13}_{-14}$ & $5813.1\pm 0.3$(*) & $4^{+2}_{-2}$ & $5.0\pm 0.5$ \\
$\vert \,0\,,\,0\,,\,0\,,\,0 \,\rangle $ & $^{4}S_{3/2}$ & ${\bf \frac{3}{2}}^+$ & $5832^{+8}_{-8}$ & $\vert \,0\,,\,0 \,\rangle$ & $5835^{+14}_{-13}$ & $5832.5\pm 0.5$(*) & $10^{+3}_{-3}$ & $9.9\pm 0.9$(*) \\
\hline
 $N=1$  &  &  &  &  &  \\
$\vert \,1\,,\,0\,,\,0\,,\,0 \,\rangle $ & $^{2}P_{1/2}$ & ${\bf \frac{1}{2}}^-$ & $6108^{+10}_{-10}$ & $\vert \,1\,,\,0 \,\rangle$ & $6098^{+14}_{-13}$ & $6096.9\pm 1.8$(*) & $24^{+6}_{-6}$ & $30\pm 7$ \\
$\vert \,1\,,\,0\,,\,0\,,\,0 \,\rangle $ & $^{4}P_{1/2}$ & ${\bf \frac{1}{2}}^-$ & $6131^{+12}_{-13}$ & $\vert \,1\,,\,0 \,\rangle$ & $6113^{+15}_{-14}$ & $\dagger$ & $13^{+3}_{-3}$ & $\dagger$ \\
$\vert \,1\,,\,0\,,\,0\,,\,0 \,\rangle $ & $^{2}P_{3/2}$ & ${\bf \frac{3}{2}}^-$ & $6114^{+10}_{-10}$ & $\vert \,1\,,\,0 \,\rangle$ & $6107^{+14}_{-14}$ & $\dagger$ & $84^{+20}_{-20}$ & $\dagger$ \\
$\vert \,1\,,\,0\,,\,0\,,\,0 \,\rangle $ & $^{4}P_{3/2}$ & ${\bf \frac{3}{2}}^-$ & $6137^{+10}_{-10}$ & $\vert \,1\,,\,0 \,\rangle$ & $6122^{+14}_{-13}$ & $\dagger$ & $57^{+14}_{-14}$ & $\dagger$ \\
$\vert \,1\,,\,0\,,\,0\,,\,0 \,\rangle $ & $^{4}P_{5/2}$ & ${\bf \frac{5}{2}}^-$ & $6147^{+12}_{-12}$ & $\vert \,1\,,\,0 \,\rangle$ & $6137^{+15}_{-14}$ & $\dagger$ & $96^{+23}_{-23}$ & $\dagger$ \\
$\vert \,0\,,\,1\,,\,0\,,\,0 \,\rangle $ & $^{2}P_{1/2}$ & ${\bf \frac{1}{2}}^-$ & $6304^{+13}_{-13}$ & $\dagger\dagger$ & $\dagger\dagger$ & $\dagger$ & $134^{+31}_{-32}$ & $\dagger$ \\
$\vert \,0\,,\,1\,,\,0\,,\,0 \,\rangle $ & $^{2}P_{3/2}$ & ${\bf \frac{3}{2}}^-$ & $6311^{+13}_{-13}$ & $\dagger\dagger$ & $\dagger\dagger$ & $\dagger$ & $129^{+32}_{-32}$ & $\dagger$ \\
\hline
 $N=2$  &  &  &  &  &  \\
$\vert \,2\,,\,0\,,\,0\,,\,0 \,\rangle $ & $^{2}D_{3/2}$ & ${\bf \frac{3}{2}}^+$ & $6415^{+15}_{-15}$ & $\vert \,2\,,\,0 \,\rangle$ & $6388^{+22}_{-22}$ & $\dagger$ & $58^{+15}_{-15}$ & $\dagger$ \\
$\vert \,2\,,\,0\,,\,0\,,\,0 \,\rangle $ & $^{2}D_{5/2}$ & ${\bf \frac{5}{2}}^+$ & $6425^{+16}_{-16}$ & $\vert \,2\,,\,0 \,\rangle$ & $6404^{+22}_{-22}$ & $\dagger$ & $130^{+33}_{-33}$ & $\dagger$ \\
$\vert \,2\,,\,0\,,\,0\,,\,0 \,\rangle $ & $^{4}D_{1/2}$ & ${\bf \frac{1}{2}}^+$ & $6431^{+21}_{-21}$ & $\vert \,2\,,\,0 \,\rangle$ & $6393^{+22}_{-22}$ & $\dagger$ & $78^{+19}_{-20}$ & $\dagger$ \\
$\vert \,2\,,\,0\,,\,0\,,\,0 \,\rangle $ & $^{4}D_{3/2}$ & ${\bf \frac{3}{2}}^+$ & $6437^{+17}_{-17}$ & $\vert \,2\,,\,0 \,\rangle$ & $6403^{+21}_{-21}$ & $\dagger$ & $106^{+27}_{-27}$ & $\dagger$ \\
$\vert \,2\,,\,0\,,\,0\,,\,0 \,\rangle $ & $^{4}D_{5/2}$ & ${\bf \frac{5}{2}}^+$ & $6448^{+15}_{-15}$ & $\vert \,2\,,\,0 \,\rangle$ & $6418^{+20}_{-21}$ & $\dagger$ & $133^{+33}_{-33}$ & $\dagger$ \\
$\vert \,2\,,\,0\,,\,0\,,\,0 \,\rangle $ & $^{4}D_{7/2}$ & ${\bf \frac{7}{2}}^+$ & $6462^{+20}_{-20}$ & $\vert \,2\,,\,0 \,\rangle$ & $6440^{+23}_{-22}$ & $\dagger$ & $145^{+38}_{-39}$ & $\dagger$ \\
$\vert \,0\,,\,0\,,\,1\,,\,0 \,\rangle $ & $^{2}S_{1/2}$ & ${\bf \frac{1}{2}}^+$ & $6421^{+15}_{-15}$ & $\vert \,0\,,\,1 \,\rangle$ & $6397^{+22}_{-21}$ & $\dagger$ & $4^{+1}_{-1}$ & $\dagger$ \\
$\vert \,0\,,\,0\,,\,1\,,\,0 \,\rangle $ & $^{4}S_{3/2}$ & ${\bf \frac{3}{2}}^+$ & $6450^{+15}_{-15}$ & $\vert \,0\,,\,1 \,\rangle$ & $6421^{+20}_{-21}$ & $\dagger$ & $4^{+1}_{-1}$ & $\dagger$ \\
$\vert \,0\,,\,0\,,\,0\,,\,1 \,\rangle $ & $^{2}S_{1/2}$ & ${\bf \frac{1}{2}}^+$ & $6813^{+24}_{-24}$ & $\dagger\dagger$ & $\dagger\dagger$ & $\dagger$ & $23^{+5}_{-6}$ & $\dagger$ \\
$\vert \,0\,,\,0\,,\,0\,,\,1 \,\rangle $ & $^{4}S_{3/2}$ & ${\bf \frac{3}{2}}^+$ & $6842^{+24}_{-23}$ & $\dagger\dagger$ & $\dagger\dagger$ & $\dagger$ & $32^{+8}_{-8}$ & $\dagger$ \\
$\vert \,1\,,\,1\,,\,0\,,\,0 \,\rangle $ & $^{2}D_{3/2}$ & ${\bf \frac{3}{2}}^+$ & $6611^{+19}_{-19}$ & $\dagger\dagger$ & $\dagger\dagger$ & $\dagger$ & $376^{+93}_{-95}$ & $\dagger$ \\
$\vert \,1\,,\,1\,,\,0\,,\,0 \,\rangle $ & $^{2}D_{5/2}$ & ${\bf \frac{5}{2}}^+$ & $6621^{+20}_{-20}$ & $\dagger\dagger$ & $\dagger\dagger$ & $\dagger$ & $252^{+67}_{-66}$ & $\dagger$ \\
$\vert \,1\,,\,1\,,\,0\,,\,0 \,\rangle $ & $^{2}P_{1/2}$ & ${\bf \frac{1}{2}}^-$ & $6613^{+19}_{-19}$ & $\dagger\dagger$ & $\dagger\dagger$ & $\dagger$ & $4^{+1}_{-1}$ & $\dagger$ \\
$\vert \,1\,,\,1\,,\,0\,,\,0 \,\rangle $ & $^{2}P_{3/2}$ & ${\bf \frac{3}{2}}^-$ & $6619^{+20}_{-20}$ & $\dagger\dagger$ & $\dagger\dagger$ & $\dagger$ & $5^{+1}_{-1}$ & $\dagger$ \\
$\vert \,1\,,\,1\,,\,0\,,\,0 \,\rangle $ & $^{2}S_{1/2}$ & ${\bf \frac{1}{2}}^+$ & $6617^{+19}_{-19}$ & $\dagger\dagger$ & $\dagger\dagger$ & $\dagger$ & $58^{+15}_{-16}$ & $\dagger$ \\
$\vert \,0\,,\,2\,,\,0\,,\,0 \,\rangle $ & $^{2}D_{3/2}$ & ${\bf \frac{3}{2}}^+$ & $6807^{+23}_{-23}$ & $\dagger\dagger$ & $\dagger\dagger$ & $\dagger$ & $549^{+130}_{-130}$ & $\dagger$ \\
$\vert \,0\,,\,2\,,\,0\,,\,0 \,\rangle $ & $^{2}D_{5/2}$ & ${\bf \frac{5}{2}}^+$ & $6817^{+24}_{-25}$ & $\dagger\dagger$ & $\dagger\dagger$ & $\dagger$ & $616^{+172}_{-170}$ & $\dagger$ \\
$\vert \,0\,,\,2\,,\,0\,,\,0 \,\rangle $ & $^{4}D_{1/2}$ & ${\bf \frac{1}{2}}^+$ & $6824^{+27}_{-27}$ & $\dagger\dagger$ & $\dagger\dagger$ & $\dagger$ & $1349^{+320}_{-321}$ & $\dagger$ \\
$\vert \,0\,,\,2\,,\,0\,,\,0 \,\rangle $ & $^{4}D_{3/2}$ & ${\bf \frac{3}{2}}^+$ & $6830^{+24}_{-24}$ & $\dagger\dagger$ & $\dagger\dagger$ & $\dagger$ & $741^{+176}_{-175}$ & $\dagger$ \\
$\vert \,0\,,\,2\,,\,0\,,\,0 \,\rangle $ & $^{4}D_{5/2}$ & ${\bf \frac{5}{2}}^+$ & $6840^{+23}_{-23}$ & $\dagger\dagger$ & $\dagger\dagger$ & $\dagger$ & $376^{+90}_{-89}$ & $\dagger$ \\
$\vert \,0\,,\,2\,,\,0\,,\,0 \,\rangle $ & $^{4}D_{7/2}$ & ${\bf \frac{7}{2}}^+$ & $6854^{+28}_{-28}$ & $\dagger\dagger$ & $\dagger\dagger$ & $\dagger$ & $1178^{+331}_{-334}$ & $\dagger$ \\
\hline \hline
\end{tabular}

\endgroup

\label{tab:All_mass_Sigma}
\end{table*}

\begin{table*}[htbp]
\caption{Same as  Table \ref{tab:All_mass_Lambda}, but for  $ \Xi'_b(snb) $ states.}

\begingroup
\setlength{\tabcolsep}{1.75pt} 
\renewcommand{\arraystretch}{1.35} 

\begin{tabular}{c c c c |c c | c c c}\hline \hline
${\mathcal{F} =  \bf {6}}_{\rm F}$&   \multicolumn{2}{c}{Three-quark} & &    \multicolumn{2}{c|}{Quark-diquark}  & &    Three-quark  &\\
$\Xi'_{b}(snb)$ & & &  Predicted   &  &   Predicted &    Experimental &  Predicted            & Experimental \\
  $\vert l_{\lambda}, l_{\rho}, k_{\lambda}, k_{\rho} \rangle$ & $^{2S+1}L_{J}$ & ${\bf J}^P$  & Mass (MeV) & $\vert l_{r}, k_{r} \rangle$  &  Mass (MeV)   &  Mass (MeV)   &  $\Gamma_{\rm Strong}$ (MeV) & $\Gamma$ (MeV) \\ \hline
\hline
 $N=0$  &  &  &  &  &  \\
$\vert \,0\,,\,0\,,\,0\,,\,0 \,\rangle $ & $^{2}S_{1/2}$ & ${\bf \frac{1}{2}}^+$ & $5925^{+6}_{-6}$ & $\vert \,0\,,\,0 \,\rangle$ & $5927^{+13}_{-13}$ & $5935.02\pm 0.05$(*) & $0$ & $<0.08$ \\
$\vert \,0\,,\,0\,,\,0\,,\,0 \,\rangle $ & $^{4}S_{3/2}$ & ${\bf \frac{3}{2}}^+$ & $5953^{+7}_{-7}$ & $\vert \,0\,,\,0 \,\rangle$ & $5951^{+13}_{-14}$ & $5953.8\pm 0.6$(*) & $0.2^{+0.1}_{-0.1}$ & $0.90\pm 0.18$ \\
\hline
 $N=1$  &  &  &  &  &  \\
$\vert \,1\,,\,0\,,\,0\,,\,0 \,\rangle $ & $^{2}P_{1/2}$ & ${\bf \frac{1}{2}}^-$ & $6198^{+7}_{-7}$ & $\vert \,1\,,\,0 \,\rangle$ & $6199^{+14}_{-15}$ & $\dagger$ & $3^{+1}_{-1}$ & $\dagger$ \\
$\vert \,1\,,\,0\,,\,0\,,\,0 \,\rangle $ & $^{4}P_{1/2}$ & ${\bf \frac{1}{2}}^-$ & $6220^{+10}_{-10}$ & $\vert \,1\,,\,0 \,\rangle$ & $6213^{+14}_{-14}$ & $\dagger$ & $4^{+1}_{-1}$ & $\dagger$ \\
$\vert \,1\,,\,0\,,\,0\,,\,0 \,\rangle $ & $^{2}P_{3/2}$ & ${\bf \frac{3}{2}}^-$ & $6204^{+7}_{-7}$ & $\vert \,1\,,\,0 \,\rangle$ & $6208^{+14}_{-14}$ & $\dagger$ & $29^{+7}_{-7}$ & $\dagger$ \\
$\vert \,1\,,\,0\,,\,0\,,\,0 \,\rangle $ & $^{4}P_{3/2}$ & ${\bf \frac{3}{2}}^-$ & $6226^{+7}_{-7}$ & $\vert \,1\,,\,0 \,\rangle$ & $6223^{+14}_{-13}$ & $\dagger$ & $8^{+2}_{-2}$ & $\dagger$ \\
$\vert \,1\,,\,0\,,\,0\,,\,0 \,\rangle $ & $^{4}P_{5/2}$ & ${\bf \frac{5}{2}}^-$ & $6237^{+10}_{-10}$ & $\vert \,1\,,\,0 \,\rangle$ & $6238^{+14}_{-14}$ & $6227.9\pm 1.6$ & $31^{+8}_{-8}$ & $19.9\pm 2.6$ \\
$\vert \,0\,,\,1\,,\,0\,,\,0 \,\rangle $ & $^{2}P_{1/2}$ & ${\bf \frac{1}{2}}^-$ & $6367^{+9}_{-9}$ & $\dagger\dagger$ & $\dagger\dagger$ & $\dagger$ & $197^{+48}_{-49}$ & $\dagger$ \\
$\vert \,0\,,\,1\,,\,0\,,\,0 \,\rangle $ & $^{2}P_{3/2}$ & ${\bf \frac{3}{2}}^-$ & $6374^{+10}_{-10}$ & $\dagger\dagger$ & $\dagger\dagger$ & $\dagger$ & $97^{+24}_{-24}$ & $\dagger$ \\
\hline
 $N=2$  &  &  &  &  &  \\
$\vert \,2\,,\,0\,,\,0\,,\,0 \,\rangle $ & $^{2}D_{3/2}$ & ${\bf \frac{3}{2}}^+$ & $6473^{+12}_{-12}$ & $\vert \,2\,,\,0 \,\rangle$ & $6474^{+20}_{-22}$ & $\dagger$ & $14^{+5}_{-5}$ & $\dagger$ \\
$\vert \,2\,,\,0\,,\,0\,,\,0 \,\rangle $ & $^{2}D_{5/2}$ & ${\bf \frac{5}{2}}^+$ & $6483^{+13}_{-13}$ & $\vert \,2\,,\,0 \,\rangle$ & $6489^{+21}_{-22}$ & $\dagger$ & $30^{+11}_{-11}$ & $\dagger$ \\
$\vert \,2\,,\,0\,,\,0\,,\,0 \,\rangle $ & $^{4}D_{1/2}$ & ${\bf \frac{1}{2}}^+$ & $6489^{+18}_{-18}$ & $\vert \,2\,,\,0 \,\rangle$ & $6479^{+22}_{-22}$ & $\dagger$ & $25^{+10}_{-10}$ & $\dagger$ \\
$\vert \,2\,,\,0\,,\,0\,,\,0 \,\rangle $ & $^{4}D_{3/2}$ & ${\bf \frac{3}{2}}^+$ & $6495^{+14}_{-14}$ & $\vert \,2\,,\,0 \,\rangle$ & $6488^{+21}_{-21}$ & $\dagger$ & $35^{+12}_{-12}$ & $\dagger$ \\
$\vert \,2\,,\,0\,,\,0\,,\,0 \,\rangle $ & $^{4}D_{5/2}$ & ${\bf \frac{5}{2}}^+$ & $6506^{+11}_{-11}$ & $\vert \,2\,,\,0 \,\rangle$ & $6504^{+20}_{-21}$ & $\dagger$ & $46^{+13}_{-13}$ & $\dagger$ \\
$\vert \,2\,,\,0\,,\,0\,,\,0 \,\rangle $ & $^{4}D_{7/2}$ & ${\bf \frac{7}{2}}^+$ & $6520^{+18}_{-18}$ & $\vert \,2\,,\,0 \,\rangle$ & $6526^{+20}_{-22}$ & $\dagger$ & $47^{+14}_{-14}$ & $\dagger$ \\
$\vert \,0\,,\,0\,,\,1\,,\,0 \,\rangle $ & $^{2}S_{1/2}$ & ${\bf \frac{1}{2}}^+$ & $6479^{+11}_{-12}$ & $\vert \,0\,,\,1 \,\rangle$ & $6483^{+21}_{-22}$ & $\dagger$ & $1.6^{+0.5}_{-0.5}$ & $\dagger$ \\
$\vert \,0\,,\,0\,,\,1\,,\,0 \,\rangle $ & $^{4}S_{3/2}$ & ${\bf \frac{3}{2}}^+$ & $6508^{+12}_{-12}$ & $\vert \,0\,,\,1 \,\rangle$ & $6507^{+20}_{-21}$ & $\dagger$ & $3^{+1}_{-1}$ & $\dagger$ \\
$\vert \,0\,,\,0\,,\,0\,,\,1 \,\rangle $ & $^{2}S_{1/2}$ & ${\bf \frac{1}{2}}^+$ & $6818^{+19}_{-19}$ & $\dagger\dagger$ & $\dagger\dagger$ & $\dagger$ & $20^{+5}_{-5}$ & $\dagger$ \\
$\vert \,0\,,\,0\,,\,0\,,\,1 \,\rangle $ & $^{4}S_{3/2}$ & ${\bf \frac{3}{2}}^+$ & $6847^{+19}_{-19}$ & $\dagger\dagger$ & $\dagger\dagger$ & $\dagger$ & $20^{+5}_{-5}$ & $\dagger$ \\
$\vert \,1\,,\,1\,,\,0\,,\,0 \,\rangle $ & $^{2}D_{3/2}$ & ${\bf \frac{3}{2}}^+$ & $6642^{+15}_{-15}$ & $\dagger\dagger$ & $\dagger\dagger$ & $\dagger$ & $234^{+59}_{-60}$ & $\dagger$ \\
$\vert \,1\,,\,1\,,\,0\,,\,0 \,\rangle $ & $^{2}D_{5/2}$ & ${\bf \frac{5}{2}}^+$ & $6653^{+17}_{-17}$ & $\dagger\dagger$ & $\dagger\dagger$ & $\dagger$ & $116^{+30}_{-29}$ & $\dagger$ \\
$\vert \,1\,,\,1\,,\,0\,,\,0 \,\rangle $ & $^{2}P_{1/2}$ & ${\bf \frac{1}{2}}^-$ & $6644^{+15}_{-15}$ & $\dagger\dagger$ & $\dagger\dagger$ & $\dagger$ & $3^{+1}_{-1}$ & $\dagger$ \\
$\vert \,1\,,\,1\,,\,0\,,\,0 \,\rangle $ & $^{2}P_{3/2}$ & ${\bf \frac{3}{2}}^-$ & $6651^{+16}_{-16}$ & $\dagger\dagger$ & $\dagger\dagger$ & $\dagger$ & $3^{+1}_{-1}$ & $\dagger$ \\
$\vert \,1\,,\,1\,,\,0\,,\,0 \,\rangle $ & $^{2}S_{1/2}$ & ${\bf \frac{1}{2}}^+$ & $6649^{+15}_{-15}$ & $\dagger\dagger$ & $\dagger\dagger$ & $\dagger$ & $59^{+14}_{-14}$ & $\dagger$ \\
$\vert \,0\,,\,2\,,\,0\,,\,0 \,\rangle $ & $^{2}D_{3/2}$ & ${\bf \frac{3}{2}}^+$ & $6812^{+19}_{-19}$ & $\dagger\dagger$ & $\dagger\dagger$ & $\dagger$ & $315^{+81}_{-85}$ & $\dagger$ \\
$\vert \,0\,,\,2\,,\,0\,,\,0 \,\rangle $ & $^{2}D_{5/2}$ & ${\bf \frac{5}{2}}^+$ & $6822^{+20}_{-20}$ & $\dagger\dagger$ & $\dagger\dagger$ & $\dagger$ & $209^{+54}_{-55}$ & $\dagger$ \\
$\vert \,0\,,\,2\,,\,0\,,\,0 \,\rangle $ & $^{4}D_{1/2}$ & ${\bf \frac{1}{2}}^+$ & $6828^{+22}_{-22}$ & $\dagger\dagger$ & $\dagger\dagger$ & $\dagger$ & $529^{+156}_{-151}$ & $\dagger$ \\
$\vert \,0\,,\,2\,,\,0\,,\,0 \,\rangle $ & $^{4}D_{3/2}$ & ${\bf \frac{3}{2}}^+$ & $6834^{+20}_{-20}$ & $\dagger\dagger$ & $\dagger\dagger$ & $\dagger$ & $364^{+94}_{-93}$ & $\dagger$ \\
$\vert \,0\,,\,2\,,\,0\,,\,0 \,\rangle $ & $^{4}D_{5/2}$ & ${\bf \frac{5}{2}}^+$ & $6845^{+19}_{-19}$ & $\dagger\dagger$ & $\dagger\dagger$ & $\dagger$ & $194^{+48}_{-47}$ & $\dagger$ \\
$\vert \,0\,,\,2\,,\,0\,,\,0 \,\rangle $ & $^{4}D_{7/2}$ & ${\bf \frac{7}{2}}^+$ & $6859^{+24}_{-24}$ & $\dagger\dagger$ & $\dagger\dagger$ & $\dagger$ & $349^{+98}_{-98}$ & $\dagger$ \\
\hline \hline
\end{tabular}

\endgroup

\label{tab:All_mass_Xiprime}
\end{table*}

\begin{table*}[htbp]
\caption{Same as  Table \ref{tab:All_mass_Lambda}, but for  $ \Omega_b(ssb) $ states.}

\begingroup
\setlength{\tabcolsep}{1.75pt} 
\renewcommand{\arraystretch}{1.35} 

\begin{tabular}{c c c c |c c | c c c}\hline \hline
${\mathcal{F} =  \bf {6}}_{\rm F}$&   \multicolumn{2}{c}{Three-quark} & &    \multicolumn{2}{c|}{Quark-diquark}  & &    Three-quark  &\\
$\Omega_{b}(ssb)$ & & &  Predicted   &  &   Predicted &    Experimental &  Predicted            & Experimental \\
  $\vert l_{\lambda}, l_{\rho}, k_{\lambda}, k_{\rho} \rangle$ & $^{2S+1}L_{J}$ & ${\bf J}^P$  & Mass (MeV) & $\vert l_{r}, k_{r} \rangle$  &  Mass (MeV)   &  Mass (MeV)   &  $\Gamma_{\rm Strong}$ (MeV) & $\Gamma$ (MeV) \\ \hline
\hline
 $N=0$  &  &  &  &  &  \\
$\vert \,0\,,\,0\,,\,0\,,\,0 \,\rangle $ & $^{2}S_{1/2}$ & ${\bf \frac{1}{2}}^+$ & $6064^{+8}_{-8}$ & $\vert \,0\,,\,0 \,\rangle$ & $6059^{+13}_{-12}$ & $6045.2\pm 1.2$(*) & $0$ & $\approx 0$ \\
$\vert \,0\,,\,0\,,\,0\,,\,0 \,\rangle $ & $^{4}S_{3/2}$ & ${\bf \frac{3}{2}}^+$ & $6093^{+9}_{-8}$ & $\vert \,0\,,\,0 \,\rangle$ & $6083^{+13}_{-14}$ & $\dagger$ & $0$ & $\dagger$ \\
\hline
 $N=1$  &  &  &  &  &  \\
$\vert \,1\,,\,0\,,\,0\,,\,0 \,\rangle $ & $^{2}P_{1/2}$ & ${\bf \frac{1}{2}}^-$ & $6315^{+7}_{-7}$ & $\vert \,1\,,\,0 \,\rangle$ & $6318^{+9}_{-10}$ & $6315.6\pm 0.6$(*) & $5^{+1}_{-1}$ & $<4.2$ \\
$\vert \,1\,,\,0\,,\,0\,,\,0 \,\rangle $ & $^{4}P_{1/2}$ & ${\bf \frac{1}{2}}^-$ & $6337^{+10}_{-10}$ & $\vert \,1\,,\,0 \,\rangle$ & $6333^{+11}_{-11}$ & $6330.3\pm 0.6$(*) & $11^{+3}_{-3}$ & $<4.7$ \\
$\vert \,1\,,\,0\,,\,0\,,\,0 \,\rangle $ & $^{2}P_{3/2}$ & ${\bf \frac{3}{2}}^-$ & $6321^{+8}_{-8}$ & $\vert \,1\,,\,0 \,\rangle$ & $6328^{+10}_{-10}$ & $6339.7\pm 0.6$ & $24^{+6}_{-6}$ & $<1.8$ \\
$\vert \,1\,,\,0\,,\,0\,,\,0 \,\rangle $ & $^{4}P_{3/2}$ & ${\bf \frac{3}{2}}^-$ & $6343^{+7}_{-7}$ & $\vert \,1\,,\,0 \,\rangle$ & $6342^{+10}_{-10}$ & $6349.8\pm 0.6$ & $6^{+2}_{-2}$ & $<3.2$ \\
$\vert \,1\,,\,0\,,\,0\,,\,0 \,\rangle $ & $^{4}P_{5/2}$ & ${\bf \frac{5}{2}}^-$ & $6353^{+11}_{-11}$ & $\vert \,1\,,\,0 \,\rangle$ & $6358^{+11}_{-10}$ & $\dagger$ & $40^{+10}_{-10}$ & $\dagger$ \\
$\vert \,0\,,\,1\,,\,0\,,\,0 \,\rangle $ & $^{2}P_{1/2}$ & ${\bf \frac{1}{2}}^-$ & $6465^{+9}_{-8}$ & $\dagger\dagger$ & $\dagger\dagger$ & $\dagger$ & $10^{+2}_{-2}$ & $\dagger$ \\
$\vert \,0\,,\,1\,,\,0\,,\,0 \,\rangle $ & $^{2}P_{3/2}$ & ${\bf \frac{3}{2}}^-$ & $6471^{+10}_{-10}$ & $\dagger\dagger$ & $\dagger\dagger$ & $\dagger$ & $54^{+14}_{-14}$ & $\dagger$ \\
\hline
 $N=2$  &  &  &  &  &  \\
$\vert \,2\,,\,0\,,\,0\,,\,0 \,\rangle $ & $^{2}D_{3/2}$ & ${\bf \frac{3}{2}}^+$ & $6568^{+11}_{-11}$ & $\vert \,2\,,\,0 \,\rangle$ & $6581^{+14}_{-15}$ & $\dagger$ & $4^{+1}_{-1}$ & $\dagger$ \\
$\vert \,2\,,\,0\,,\,0\,,\,0 \,\rangle $ & $^{2}D_{5/2}$ & ${\bf \frac{5}{2}}^+$ & $6578^{+12}_{-12}$ & $\vert \,2\,,\,0 \,\rangle$ & $6596^{+15}_{-15}$ & $\dagger$ & $10^{+2}_{-2}$ & $\dagger$ \\
$\vert \,2\,,\,0\,,\,0\,,\,0 \,\rangle $ & $^{4}D_{1/2}$ & ${\bf \frac{1}{2}}^+$ & $6584^{+17}_{-17}$ & $\vert \,2\,,\,0 \,\rangle$ & $6585^{+17}_{-17}$ & $\dagger$ & $1.0^{+0.3}_{-0.3}$ & $\dagger$ \\
$\vert \,2\,,\,0\,,\,0\,,\,0 \,\rangle $ & $^{4}D_{3/2}$ & ${\bf \frac{3}{2}}^+$ & $6590^{+13}_{-13}$ & $\vert \,2\,,\,0 \,\rangle$ & $6595^{+16}_{-16}$ & $\dagger$ & $3^{+1}_{-1}$ & $\dagger$ \\
$\vert \,2\,,\,0\,,\,0\,,\,0 \,\rangle $ & $^{4}D_{5/2}$ & ${\bf \frac{5}{2}}^+$ & $6600^{+10}_{-10}$ & $\vert \,2\,,\,0 \,\rangle$ & $6610^{+15}_{-15}$ & $\dagger$ & $8^{+2}_{-2}$ & $\dagger$ \\
$\vert \,2\,,\,0\,,\,0\,,\,0 \,\rangle $ & $^{4}D_{7/2}$ & ${\bf \frac{7}{2}}^+$ & $6614^{+18}_{-18}$ & $\vert \,2\,,\,0 \,\rangle$ & $6632^{+17}_{-16}$ & $\dagger$ & $18^{+10}_{-9}$ & $\dagger$ \\
$\vert \,0\,,\,0\,,\,1\,,\,0 \,\rangle $ & $^{2}S_{1/2}$ & ${\bf \frac{1}{2}}^+$ & $6574^{+11}_{-11}$ & $\vert \,0\,,\,1 \,\rangle$ & $6590^{+15}_{-15}$ & $\dagger$ & $0.7^{+0.2}_{-0.2}$ & $\dagger$ \\
$\vert \,0\,,\,0\,,\,1\,,\,0 \,\rangle $ & $^{4}S_{3/2}$ & ${\bf \frac{3}{2}}^+$ & $6602^{+11}_{-11}$ & $\vert \,0\,,\,1 \,\rangle$ & $6614^{+15}_{-15}$ & $\dagger$ & $0.6^{+0.2}_{-0.2}$ & $\dagger$ \\
$\vert \,0\,,\,0\,,\,0\,,\,1 \,\rangle $ & $^{2}S_{1/2}$ & ${\bf \frac{1}{2}}^+$ & $6874^{+17}_{-17}$ & $\dagger\dagger$ & $\dagger\dagger$ & $\dagger$ & $13^{+4}_{-4}$ & $\dagger$ \\
$\vert \,0\,,\,0\,,\,0\,,\,1 \,\rangle $ & $^{4}S_{3/2}$ & ${\bf \frac{3}{2}}^+$ & $6902^{+17}_{-17}$ & $\dagger\dagger$ & $\dagger\dagger$ & $\dagger$ & $8^{+2}_{-2}$ & $\dagger$ \\
$\vert \,1\,,\,1\,,\,0\,,\,0 \,\rangle $ & $^{2}D_{3/2}$ & ${\bf \frac{3}{2}}^+$ & $6718^{+14}_{-14}$ & $\dagger\dagger$ & $\dagger\dagger$ & $\dagger$ & $116^{+39}_{-39}$ & $\dagger$ \\
$\vert \,1\,,\,1\,,\,0\,,\,0 \,\rangle $ & $^{2}D_{5/2}$ & ${\bf \frac{5}{2}}^+$ & $6728^{+15}_{-15}$ & $\dagger\dagger$ & $\dagger\dagger$ & $\dagger$ & $82^{+22}_{-23}$ & $\dagger$ \\
$\vert \,1\,,\,1\,,\,0\,,\,0 \,\rangle $ & $^{2}P_{1/2}$ & ${\bf \frac{1}{2}}^-$ & $6720^{+14}_{-14}$ & $\dagger\dagger$ & $\dagger\dagger$ & $\dagger$ & $1.1^{+0.4}_{-0.4}$ & $\dagger$ \\
$\vert \,1\,,\,1\,,\,0\,,\,0 \,\rangle $ & $^{2}P_{3/2}$ & ${\bf \frac{3}{2}}^-$ & $6726^{+15}_{-15}$ & $\dagger\dagger$ & $\dagger\dagger$ & $\dagger$ & $2^{+1}_{-1}$ & $\dagger$ \\
$\vert \,1\,,\,1\,,\,0\,,\,0 \,\rangle $ & $^{2}S_{1/2}$ & ${\bf \frac{1}{2}}^+$ & $6724^{+14}_{-14}$ & $\dagger\dagger$ & $\dagger\dagger$ & $\dagger$ & $72^{+21}_{-21}$ & $\dagger$ \\
$\vert \,0\,,\,2\,,\,0\,,\,0 \,\rangle $ & $^{2}D_{3/2}$ & ${\bf \frac{3}{2}}^+$ & $6868^{+17}_{-17}$ & $\dagger\dagger$ & $\dagger\dagger$ & $\dagger$ & $180^{+56}_{-54}$ & $\dagger$ \\
$\vert \,0\,,\,2\,,\,0\,,\,0 \,\rangle $ & $^{2}D_{5/2}$ & ${\bf \frac{5}{2}}^+$ & $6878^{+19}_{-19}$ & $\dagger\dagger$ & $\dagger\dagger$ & $\dagger$ & $157^{+39}_{-39}$ & $\dagger$ \\
$\vert \,0\,,\,2\,,\,0\,,\,0 \,\rangle $ & $^{4}D_{1/2}$ & ${\bf \frac{1}{2}}^+$ & $6884^{+21}_{-21}$ & $\dagger\dagger$ & $\dagger\dagger$ & $\dagger$ & $126^{+31}_{-32}$ & $\dagger$ \\
$\vert \,0\,,\,2\,,\,0\,,\,0 \,\rangle $ & $^{4}D_{3/2}$ & ${\bf \frac{3}{2}}^+$ & $6890^{+18}_{-18}$ & $\dagger\dagger$ & $\dagger\dagger$ & $\dagger$ & $195^{+47}_{-47}$ & $\dagger$ \\
$\vert \,0\,,\,2\,,\,0\,,\,0 \,\rangle $ & $^{4}D_{5/2}$ & ${\bf \frac{5}{2}}^+$ & $6900^{+17}_{-17}$ & $\dagger\dagger$ & $\dagger\dagger$ & $\dagger$ & $172^{+41}_{-41}$ & $\dagger$ \\
$\vert \,0\,,\,2\,,\,0\,,\,0 \,\rangle $ & $^{4}D_{7/2}$ & ${\bf \frac{7}{2}}^+$ & $6914^{+23}_{-23}$ & $\dagger\dagger$ & $\dagger\dagger$ & $\dagger$ & $230^{+62}_{-63}$ & $\dagger$ \\
\hline \hline
\end{tabular}

\endgroup
\label{tab:All_mass_Omega}
\end{table*}

\section{Results and assignments }
\label{Results}
In this section, we present our results for the masses, strong and electromagnetic decays of bottom baryons.  We study the $\Lambda_b$, $\Xi_b$, $\Sigma_b$, $\Xi'_b$, and $\Omega_b$ states simultaneously. 


\subsection{Mass spectra of bottom baryons}

 Our predictions for the $\Lambda_b$, $\Xi_b$, $\Sigma_b$, $\Xi'_b$, and $\Omega_b$ states are reported in Tables~\ref{tab:All_mass_Lambda}-\ref{tab:All_mass_Omega} respectively. In the fourth column of  Tables~\ref{tab:All_mass_Lambda}-\ref{tab:All_mass_Omega}, we provide the theoretical masses, along with their errors, calculated using the three-quark model Hamiltonian given by Eqs.~\ref{MassFormula} and \ref{eq:Hho}.
  In the sixth column, we present our theoretical results for the quark-diquark model description calculated using the  Hamiltonian given by Eqs.~\ref{MassFormula} and \ref{eq:Hhodi}. In the seventh column, we report the experimental masses as  from PDG \cite{ParticleDataGroup:2018ovx}.
  
 Furthermore, we compare our theoretical results with the experimental data \cite{ParticleDataGroup:2018ovx}  in Figs.~\ref{fig:lambdas}-\ref{fig:omegas} for the three-quark model, and in Figs.~\ref{fig:lambdasD}~\ref{fig:omegasD} for the quark-diquark model.
 As one can see, our theoretical mass predictions are in good agreement with the available experimental data.

 In addition, we compare our mass spectra with the previous studies such as NRQM \cite{Yoshida:2015tia}, QCD sum rules \cite{Liu:2007fg, Mao:2015gya, Chen:2016phw}, NRQM \cite{Roberts:2007ni}, $\chi$QM \cite{Kim:2021ywp}, and LQCD \cite{Mohanta:2019mxo}, as shown in Tables  \ref{comparisonlambdas}-\ref{comparisonomegas}. 
Unfortunately, these  studies did not fully complete the calculation up to the $D$-wave.
For instance, LQCD \cite{Mohanta:2019mxo} only provided predictions for the ground-state bottom  baryons. QCD sum rules \cite{Liu:2007fg, Mao:2015gya, Chen:2016phw}  predicted $P$-wave states but did not offer any predictions for $D$-wave states.

Only NRQM \cite{Yoshida:2015tia} and \cite{Roberts:2007ni} made predictions for some $D$-wave states; however, they did not provide predictions for all possible mass states within the $D$-wave.  Due to the lack of data, we cannot make any conclusion about the differences in the prediction for each model. It is crucial to emphasize that the validation of our model requires the identification of multiplets through additional data. However, the difficulty in identifying new bottom baryons within the data remains a challenge.

\subsection{Bottom baryon strong decay widths }

We also investigate the open-flavor strong decay widths of the $\Lambda_b$, $\Xi_b$, $\Sigma_b$, $\Xi'_b$, and $\Omega_b$ states. The strong-decay widths are computed using the $^3P_0$ model by means of Eq. (\ref{gamma}). Our theoretical width predictions are presented in the eighth column of Tables \ref{tab:All_mass_Lambda}-\ref{tab:All_mass_Omega}. The results exhibit good agreement with the experimental widths \cite{ParticleDataGroup:2018ovx}, reported in the ninth column of  Tables  \ref{tab:All_mass_Lambda}-\ref{tab:All_mass_Omega}.

It is worth noting that the experimental widths encompass contributions from strong, electromagnetic, and weak interactions. However, the dominant contribution is typically from the strong decay process. In fact, our electromagnetic results are a small contribution to the width, typically no more than 1.5 MeV. It is important to note that this contribution is relatively minor compared to the uncertainties associated with the strong decay width. 

Additionally, the  partial decay widths of each open flavor channel are given in Tables \ref{tab:part_dec_lambdas}-\ref{tab:part_dec_Omega}. The partial decay widths obtained in this study can serve as valuable information for experimentalists in their efforts to identify bottom baryons. The knowledge of potential decay channels can greatly assist in the identification process by guiding the analysis of experimental data.

Nevertheless, in the single-bottom baryon sector, there are a few cases where the strong decay is suppressed due to the absence of phase space, leading to the dominance of electromagnetic or even weak interactions. Specifically, the ground states, $\Lambda_b$, $\Xi_b$, and $\Omega_b$, can only decay via weak interaction. In the case of  $\Xi'_b$, and $\Omega_b^*$, all the strong decay channels are closed due to the lack of  phase space. In such cases, the decay width is primarily dominated by electromagnetic interaction.

\subsection{Electromagnetic decay widths of bottom baryons}
In this subsection, we compute the electromagnetic decays  of $\Lambda_b$, $\Xi_b$, $\Sigma_b$, $\Xi'_b$, and $\Omega_b$ baryons using Eq. (\ref{gammaEM}) for $P$-wave excited states transitioning to ground states, as well as for ground state to ground state transitions. 

Our results are presented in Tables~\ref{lambdasEM}-\ref{omegasEM}.  These results extend those obtained by several works on the subject. For example, the analysis performed in~\cite{Wang:2017kfr} employs a constituent quark model. Other studies of radiative decays in different frameworks use Heavy Quark Symmetry~\cite{Tawfiq:1999cf}, bound state picture~\cite{Chow:1995nw}, dynamically generated states \cite{Gamermann:2010ga}, relativistic three-quark model \cite{Ivanov:1998wj}, light cone QCD sum rules \cite{Zhu:1998ih, Wang:2009ic, Wang:2009cd,Aliev:2014bma,Aliev:2009jt,Aliev:2016xvq,Aliev:2011bm}, Heavy Hadron Chiral Perturbation Theory \cite{Banuls:1999br,Cheng:1992xi,Jiang:2015xqa}, and modified bag models~\cite{Bernotas:2013eia}. 

The electromagnetic decay widths are particularly valuable in cases where the strong decays are suppressed. One notable example is the spin excitation of the $\Omega_b^-$ state, denoted as $\Omega_b^*$, which has not yet been observed. The $\Omega_b^*\to\Omega_b^-\pi$ strong decay  is prohibited due to lack of phase space and isospin conservation in strong interactions. Therefore, the $\Omega_b^*\to\Omega_b^-\gamma$ decay mode becomes a particularly important channel, as it serves as a "golden channel" for the observation of the $\Omega_b^*$ state.

\subsection{Assignments of bottom baryons} 

First, we will make the assignments of the bottom baryons  reported in PDG~\cite{Workman:2022ynf} using our theoretical results for $\Lambda_b$, $\Xi_b$, $\Sigma_b$, $\Xi'_b$, and $\Omega_b$. Our first criterion is to use the mass spectrum to identify resonances of bottom baryons, while the decay width serves as a secondary criterion. The classification within the quark-diquark model is the same as that of the three-quark model when it comes to describing ground states and $\boldsymbol \lambda$-mode excitations. However, it is important to note that in the quark-diquark model, the $\boldsymbol\rho$-modes which exist in the three-quark model, are absent   (see Tables~\ref{tab:All_mass_Omega}-\ref{tab:All_mass_Lambda}).


\subsubsection{$\Lambda_b$}

We make the assignment of the six $\Lambda_b$ states reported by the PDG \cite{Workman:2022ynf}, using our predictions given in the table  \ref{tab:All_mass_Lambda}. 

The $\Lambda^0_b$ is identified as the ground state with ${\bf J}^P = {\bf \frac{1}{2}}^+$, and its theoretical mass is well reproduced in both the three-quark model and quark-diquark models.

The $\Lambda_b (5912)^0$ and $\Lambda_b (5920)^0$ are identified as the two $P_\lambda$ waves with ${\bf J}^P = {\bf \frac{1}{2}}^-$  and ${\bf J}^P = {\bf \frac{3}{2}}^-$, respectively.
Our theoretical predictions for their mass are in agreement with the experimental value. There are no strong decay channels, so their strong decay widths are zero. 

The $\Lambda_b (6070)^0$ has been recognized as the first radial excitation with ${\bf J}^P = {\bf \frac{1}{2}}^+$; however, its  quantum numbers have not yet been determined experimentally. In our model, if we consider it as a radial excitation, there is a deviation of approximately 3\% in its mass, and its width is underestimated. Alternatively, we can identify it as a $P_\rho$ state with ${\bf J}^P = {\bf \frac{1}{2}}^-$, featuring an internal spin of $\bf S_{\rm tot}=\frac{1}{2}$. In this scenario, the theoretical mass deviates by less than 1\%, and the width is accurately reproduced. 

In the case of $\Lambda_b (6146)^0$ and $\Lambda_b (6152)^0$, they are identified as the two $D_\lambda$ excitations with quantum numbers ${\bf J}^P = {\bf \frac{3}{2}}^+$ and ${\bf J}^P = {\bf \frac{5}{2}}^+$, respectively, but their quantum numbers have not been measured yet. In this case, our theoretical predictions for the mass have a small deviation of 1\%, and their theoretical widths are slightly overestimated.
\subsubsection{$\Xi_b$ and  $\Xi'_b$}

Now we will make the assignments for the  states $\Xi'_b$ and $\Xi_b$ reported in PDG \cite{Workman:2022ynf}. In this case, the assignment is more complicated because there are several theoretically excited states in the same energy range for $\Xi'_b$ and $\Xi_b$. Here we use our results reported in Table \ref{tab:All_mass_Xiprime} for $\Xi'_b$  and \ref{tab:All_mass_Xi} for  $\Xi_b$. 

In the flavor space, the $\Xi'_b$ states belong to the $\bf 6_{\rm F}$ configuration and the $\Xi_b$ states belong to the $\bf \bar 3_{\rm F}$ configuration.
 Invariance of the strong interaction under $SU_I(2)$ isospin
transformations leads to isospin conservation and the
appearance of degenerate isospin multiplets.  In the case of the $\Xi_b$ and $\Xi'_b$ baryons, both belong to isospin doublets.

Both our three-quark and quark-diquark models predict that  the masses of the $\Xi^0_b$ and $\Xi^-_b$ isospin partners are degenerate, which is in  agreement with experimental data \cite{Workman:2022ynf}.
In our model, we assign them the quantum numbers ${\bf J}^P = {\bf \frac{1}{2}}^+$, even though these quantum numbers have not yet been directly measured.

The $\Xi'^-_b(5935)$ is considered as the ground state of the sextuplet. Its predicted mass agrees with the experimental data  in  the three-quark and  the quark-diquark models, its assignment is ${\bf J}^P = {\bf \frac{1}{2}}^+$, but the quantum numbers have not yet been measured, nor its charge partner, the $\Xi'^0_b(5935)$,  has been observed. 

The spin excitations of $\Xi'^-_b(5935)$ are the $\Xi'^-_b(5955)$ and $\Xi'^0_b(5945)$ they are identified as ${\bf J}^P = {\bf \frac{3}{2}}^+$, but their quantum numbers are based on the expectations of the quark model. The $\Xi'^-_b(5955)$ and $\Xi'^0_b(5945)$ predicted masses  are degenerate and in agreement with experimental data. Their width also is well reproduced in our model. 

The $\Xi'^-_b(6100)$ is identified as a $P$-wave state, ${\bf J}^{P} = {\bf \frac{1}{2}}^-$, but its ${\bf J}^P$ remains to be confirmed. It is identified as one of the two $P_\lambda$ excitations of $\Xi_b$ belonging to the $\bf \bar 3_{\rm F}$, with total internal spin $\bf S_{\rm tot}=\frac{1}{2}$. Both its mass and width are well described. 

Finally, the PDG reports $\Xi'^-_b(6227)$ and $\Xi'^0_b(6227)$, which in our model are identified with the fifth $P_\lambda$ excitation of $\Xi'_b$ with ${\bf J}^P = {\bf \frac{5}{2}}^-$ and total internal spin $\bf S_{\rm tot}=\frac{3}{2}$. Its predicted mass is compatible with the experimental value, and its width is well reproduced. However,  it could be identified as ${\bf J}^P = {\bf \frac{3}{2}}^-$ since these states also have similar mass, but, with this assignment, the predicted width has a deviation of 6 MeV in respect to the experimental one. 

Recently, the PDG \cite{Workman:2022ynf}  added the two  $\Xi^0_b(6327)$ and $\Xi^0_b(6333)$ states observed by LHCb \cite{LHCb:2021ssn}. The observed masses of these states are $m(\Xi^0_b(6327)) = 6327.28 ^{+0.23}_{-0.21} \pm 0.12 \pm 0.24$ MeV and $m(\Xi^0_b(6333)) = 6332.69 ^{+0.17}_{- 0.18} \pm 0.03 \pm 0.22$ MeV, respectively. In our model, we identify them as the two $D_\lambda$ excitations with quantum numbers ${\bf J}^P = {\bf \frac{3}{2}}^+$and ${\bf J}^P = {\bf \frac{5}{2}}^+$, respectively, belonging to the $\bf \bar 3_{\rm F}$ configuration. However, one of these states is also compatible with the first radial excitation with ${\bf J}^P = {\bf \frac{1}{2}}^+$. Further data will be necessary to correctly assign to these states the correct quantum numbers.

\subsubsection{$\Sigma_b$}
The PDG \cite{Workman:2022ynf} reports only four $\Sigma_b$ states and their quantum numbers have not yet been measured. We use our results of $\Sigma_b$ shown in Table \ref{tab:All_mass_Sigma} to identify them. The $\Sigma_b$ is identified as ${\bf J}^P = {\bf \frac{1}{2}}^+$, its mass agrees well in both the three-quark model and quark-diquark models, and our theoretical width agrees perfectly with the experimental result. The spin excitation $\Sigma^*_b$ is identified as ${\bf J}^P = {\bf \frac{3}{2}}^+$ and our predictions for mass and width agree well with the experimental data. 

The $\Sigma_b (6097)^-$ and $\Sigma_b (6097)^+$ states are the two of three charge states of $\Sigma_b (6097)$ that are degenerate in our model, since we assume isospin symmetry. Our predicted mass and width for $\Sigma_b (6097)$ agree well with the experimental data. Our calculations indicate that  this is the first $P_\lambda$ excitation of $\Sigma_b$ with ${\bf J}^P = {\bf \frac{1}{2}}^-$ and  internal spin $\bf S_{\rm tot}=1/2$.


\subsubsection{$\Omega_b$}

The predicted mass spectra and strong decay widths for the $\Omega_b$ states are presented in Table \ref{tab:All_mass_Omega}. It is noteworthy that these new findings are in agreement with our earlier calculation \cite{Santopinto:2018ljf}. Furthermore, with respect to our previous study \cite{Santopinto:2018ljf}, we have extended our investigation to include the $D$-wave  excitations. 

The $\Omega^-_b$ is identified as  a ${\bf J}^P = {\bf \frac{1}{2}}^+$ state. Its experimental mass is well reproduced in the quark-diquark description. In the three-quark model, the predicted mass has a slight deviation of 10 MeV. The $\Omega^-_b$ can only decay weakly, so its strong decay width is zero. The spin excitation of $\Omega^-_b$, the $\Omega^{*-}_b$ with ${\bf J}^P = {\bf \frac{3}{2}}^+$, has not been observed yet. We suggest the $\Omega_b^*\to\Omega_b^-\gamma$ decay mode  as a "golden channel" for the observation of this state, whose mass, according to our predictions, is expected in the 6070-6098 MeV energy range. 

The four $\Omega_b$ resonances, namely $\Omega_b(6316) ^-$, $\Omega_b(6330) ^-$, $\Omega_b(6340) ^-$, and $\Omega_b(6350) ^-$, observed and discovered in LHCb \cite{LHCb:2020tqd}, have to be confirmed in other experiments, and their quantum numbers have also not been measured yet. They are identified in our model as four of the five $P_\lambda$ excitations. 

The mass and width of $\Omega_b(6316) ^-$ are well reproduced. This state is  identified as ${\bf J}^P = {\bf \frac{1}{2}}^-$, with  internal spin $\bf S_{\rm tot}=\frac{1}{2}$. The assignment for $\Omega_b(6330) ^-$ is also ${\bf J}^P = {\bf \frac{1}{2}}^-$, but with internal spin $\bf S_{\rm tot}=\frac{3}{2}$. Its mass is well reproduced in the calculations with three-quark model as well as with quark-diquark model, and our theoretical width is compatible with the experimental data. The mass of $\Omega_b(6340) ^-$ is well reproduced in both the three-quark model
and quark-diquark models, but the experimental width is slightly overestimated. Our preferred assignment is ${\bf J}^P = {\bf \frac{3}{2}}^-$, with a internal spin $\bf S_{\rm tot}=\frac{1}{2}$. Finally, the mass and width of $\Omega_b(6350) ^-$ are well reproduced in our calculation, it is described  as a ${\bf J}^P = {\bf \frac{3}{2}}^-$ state with  internal spin $\bf S_{\rm tot}=\frac{3}{2}$. 

The fifth $P_\lambda$ excitation is characterized by a large width, requiring a high statistical significance for its observation at the LHC. According to our predictions, this state is expected to have a mass in the range of 6345-6365 MeV, with a width of approximately 40 MeV.

In the three-quark model, we  predict  two additional $P_\rho$ excitations which do not appear in the quark-diquark description.
They do not couple to the $\Xi^0_bK^-$ channel, which is the channel where LHCb observed the four excited $\Omega^-_b$ states, but they exhibit a strong coupling in the $\Xi'^0_bK^-$ channel. Therefore, the experimental search for these two $P_\rho$ excitations is a fundamental step  
 toward assessing whether the bottom baryons are a three-quark system or a quark-diquark system.

\begin{table*}[htp]
\caption{$\Lambda_b(nnb)$ states. The flavor multiplet is specified with the symbol $\mathcal{F}$.
The first column contains the  three-quark model state, $\left| l_{\lambda},l_{\rho}, k_{\lambda},k_{\rho}\right\rangle$, where $l_{\lambda,\rho}$ are the orbital angular momenta and $k_{\lambda,\rho}$ the number of nodes of the $\lambda$ and $\rho$ oscillators.
The second column displays the spectroscopic notation $^{2S+1}L_J$ for each state. The third column contains the total angular momentum and parity ${\bf J}^P $. In the fourth column our predicted masses, computed within the three-quark model, are shown. Our results are compared with those of  references \cite{Yoshida:2015tia} (fifth column), \cite{Liu:2007fg, Mao:2015gya, Chen:2016phw} (sixth column), \cite{Roberts:2007ni} (seventh column), \cite{Kim:2021ywp} (eighth column), and \cite{Mohanta:2019mxo} (ninth column). Our theoretical results are also compared with the experimental masses as from PDG  \cite{ParticleDataGroup:2018ovx} (tenth column).
The "$...$" indicates that there is no prediction for that state.}
\begin{center}
\scriptsize{

\begingroup
\setlength{\tabcolsep}{1.75pt} 
\renewcommand{\arraystretch}{1.35} 

\begin{tabular}{c c c| c c c c c c c}\hline \hline
$\Lambda_{b}(nnb)$ & & ${\mathcal{F} = \bf {\bar 3}}_{\rm F}$& This work   &   NRQM \cite{Yoshida:2015tia}     &  QCD sum rules \cite{Liu:2007fg, Mao:2015gya, Chen:2016phw}      &  NRQM \cite{Roberts:2007ni}  & $\chi$QM \cite{Kim:2021ywp}        & LQCD \cite{Mohanta:2019mxo}     & Experimental  \\
 $\vert l_{\lambda}, l_{\rho}, k_{\lambda}, k_{\rho} \rangle$ & $^{2S+1}L_{J}$  &  ${\bf J}^P$ & mass (MeV)  &   mass (MeV)  &  mass (MeV)  &  mass (MeV) & mass (MeV) & mass (MeV) &      mass (MeV) \\ \hline
\hline
 $N=0$  &  &  &  &  &  \\
$\vert \,0\,,\,0\,,\,0\,,\,0 \,\rangle $ & $^{2}S_{1/2}$ & ${\bf \frac{1}{2}}^+$ & $5613^{+9}_{-9}$ & $5618$ & $5637$ & $5612$ & $5620$ & $5667$ & $5619.60\pm 0.17$ \\
\hline
 $N=1$  &  &  &  &  &  \\
$\vert \,1\,,\,0\,,\,0\,,\,0 \,\rangle $ & $^{2}P_{1/2}$ & ${\bf \frac{1}{2}}^-$ & $5918^{+8}_{-8}$ & $5938$ & $6010$ & $5939$ & $5914$ & ... & $5912.19\pm 0.17$ \\
$\vert \,1\,,\,0\,,\,0\,,\,0 \,\rangle $ & $^{2}P_{3/2}$ & ${\bf \frac{3}{2}}^-$ & $5924^{+8}_{-8}$ & $5939$ & $6010$ & $5941$ & $5927$ & ... & $5920.09\pm 0.17$ \\
$\vert \,0\,,\,1\,,\,0\,,\,0 \,\rangle $ & $^{2}P_{1/2}$ & ${\bf \frac{1}{2}}^-$ & $6114^{+10}_{-10}$ & $6236$ & ... & $6180$ & $6207$ & ... & $\dagger$ \\
$\vert \,0\,,\,1\,,\,0\,,\,0 \,\rangle $ & $^{4}P_{1/2}$ & ${\bf \frac{1}{2}}^-$ & $6137^{+14}_{-14}$ & $6273$ & $5870$ & ... & $6233$ & ... & $\dagger$ \\
$\vert \,0\,,\,1\,,\,0\,,\,0 \,\rangle $ & $^{2}P_{3/2}$ & ${\bf \frac{3}{2}}^-$ & $6121^{+10}_{-10}$ & $6273$ & ... & $6191$ & ... & ... & $\dagger$ \\
$\vert \,0\,,\,1\,,\,0\,,\,0 \,\rangle $ & $^{4}P_{3/2}$ & ${\bf \frac{3}{2}}^-$ & $6143^{+12}_{-12}$ & $6285$ & $5880$ & ... & ... & ... & $\dagger$ \\
$\vert \,0\,,\,1\,,\,0\,,\,0 \,\rangle $ & $^{4}P_{5/2}$ & ${\bf \frac{5}{2}}^-$ & $6153^{+14}_{-14}$ & $6289$ & ... & $6206$ & ... & ... & $\dagger$ \\
\hline
 $N=2$  &  &  &  &  &  \\
$\vert \,2\,,\,0\,,\,0\,,\,0 \,\rangle $ & $^{2}D_{3/2}$ & ${\bf \frac{3}{2}}^+$ & $6225^{+13}_{-13}$ & $6211$ & $6010$ & $6181$ & $6172$ & ... & $6146.2\pm 0.4$ \\
$\vert \,2\,,\,0\,,\,0\,,\,0 \,\rangle $ & $^{2}D_{5/2}$ & ${\bf \frac{5}{2}}^+$ & $6235^{+13}_{-13}$ & $6212$ & $6010$ & $6183$ & $6178$ & ... & $6152.5\pm 0.4$ \\
$\vert \,0\,,\,0\,,\,1\,,\,0 \,\rangle $ & $^{2}S_{1/2}$ & ${\bf \frac{1}{2}}^+$ & $6231^{+12}_{-12}$ & $6153$ & ... & $6107$ & $6121$ & ... & $\dagger$ \\
$\vert \,0\,,\,0\,,\,0\,,\,1 \,\rangle $ & $^{2}S_{1/2}$ & ${\bf \frac{1}{2}}^+$ & $6624^{+21}_{-21}$ & ... & ... & ... & ... & ... & $\dagger$ \\
$\vert \,1\,,\,1\,,\,0\,,\,0 \,\rangle $ & $^{2}D_{3/2}$ & ${\bf \frac{3}{2}}^+$ & $6421^{+16}_{-16}$ & $6488$ & ... & $6401$ & ... & ... & $\dagger$ \\
$\vert \,1\,,\,1\,,\,0\,,\,0 \,\rangle $ & $^{2}D_{5/2}$ & ${\bf \frac{5}{2}}^+$ & $6431^{+17}_{-17}$ & $6530$ & ... & $6422$ & ... & ... & $\dagger$ \\
$\vert \,1\,,\,1\,,\,0\,,\,0 \,\rangle $ & $^{4}D_{1/2}$ & ${\bf \frac{1}{2}}^+$ & $6438^{+22}_{-22}$ & $6467$ & ... & ... & ... & ... & $\dagger$ \\
$\vert \,1\,,\,1\,,\,0\,,\,0 \,\rangle $ & $^{4}D_{3/2}$ & ${\bf \frac{3}{2}}^+$ & $6444^{+19}_{-19}$ & $6511$ & $6360$ & ... & ... & ... & $\dagger$ \\
$\vert \,1\,,\,1\,,\,0\,,\,0 \,\rangle $ & $^{4}D_{5/2}$ & ${\bf \frac{5}{2}}^+$ & $6454^{+17}_{-17}$ & $6539$ & $6360$ & ... & ... & ... & $\dagger$ \\
$\vert \,1\,,\,1\,,\,0\,,\,0 \,\rangle $ & $^{4}D_{7/2}$ & ${\bf \frac{7}{2}}^+$ & $6468^{+23}_{-22}$ & ... & ... & $6433$ & ... & ... & $\dagger$ \\
$\vert \,1\,,\,1\,,\,0\,,\,0 \,\rangle $ & $^{2}P_{1/2}$ & ${\bf \frac{1}{2}}^-$ & $6423^{+16}_{-16}$ & ... & ... & ... & ... & ... & $\dagger$ \\
$\vert \,1\,,\,1\,,\,0\,,\,0 \,\rangle $ & $^{2}P_{3/2}$ & ${\bf \frac{3}{2}}^-$ & $6429^{+17}_{-17}$ & ... & ... & ... & ... & ... & $\dagger$ \\
$\vert \,1\,,\,1\,,\,0\,,\,0 \,\rangle $ & $^{4}P_{1/2}$ & ${\bf \frac{1}{2}}^-$ & $6446^{+19}_{-18}$ & ... & $6220$ & ... & ... & ... & $\dagger$ \\
$\vert \,1\,,\,1\,,\,0\,,\,0 \,\rangle $ & $^{4}P_{3/2}$ & ${\bf \frac{3}{2}}^-$ & $6452^{+17}_{-17}$ & ... & $6230$ & ... & ... & ... & $\dagger$ \\
$\vert \,1\,,\,1\,,\,0\,,\,0 \,\rangle $ & $^{4}P_{5/2}$ & ${\bf \frac{5}{2}}^-$ & $6462^{+19}_{-19}$ & ... & ... & ... & ... & ... & $\dagger$ \\
$\vert \,1\,,\,1\,,\,0\,,\,0 \,\rangle $ & $^{4}S_{3/2}$ & ${\bf \frac{3}{2}}^+$ & $6456^{+17}_{-18}$ & ... & ... & ... & ... & ... & $\dagger$ \\
$\vert \,1\,,\,1\,,\,0\,,\,0 \,\rangle $ & $^{2}S_{1/2}$ & ${\bf \frac{1}{2}}^+$ & $6427^{+16}_{-16}$ & ... & ... & ... & ... & ... & $\dagger$ \\
$\vert \,0\,,\,2\,,\,0\,,\,0 \,\rangle $ & $^{2}D_{3/2}$ & ${\bf \frac{3}{2}}^+$ & $6618^{+20}_{-20}$ & ... & $6520$ & ... & ... & ... & $\dagger$ \\
$\vert \,0\,,\,2\,,\,0\,,\,0 \,\rangle $ & $^{2}D_{5/2}$ & ${\bf \frac{5}{2}}^+$ & $6628^{+21}_{-22}$ & ... & $6520$ & ... & ... & ... & $\dagger$ \\
\hline \hline
\end{tabular}

\endgroup

}
\end{center}
\label{comparisonlambdas}
\end{table*}

\begin{table*}[htp]
\caption{Same as  Table \ref{comparisonlambdas}, but for  $ \Xi_{b}(snb) $ states.}
\begin{center}
\scriptsize{
\begingroup
\setlength{\tabcolsep}{1.75pt} 
\renewcommand{\arraystretch}{1.35} 

\begin{tabular}{c c c| c c c c c c c}\hline \hline
$\Xi_{b}(snb)$ & & ${\mathcal{F} = \bf {\bar 3}}_{\rm F}$& This work   &   NRQM \cite{Yoshida:2015tia}     &  QCD sum rules \cite{Liu:2007fg, Mao:2015gya, Chen:2016phw}      &  NRQM \cite{Roberts:2007ni}  & $\chi$QM \cite{Kim:2021ywp}    & LQCD \cite{Mohanta:2019mxo}     & Experimental  \\
 $\vert l_{\lambda}, l_{\rho}, k_{\lambda}, k_{\rho} \rangle$ & $^{2S+1}L_{J}$  &  ${\bf J}^P$ & mass (MeV)  &   mass (MeV)  &  mass (MeV)  &  mass (MeV) & mass (MeV) & mass (MeV) &      mass (MeV) \\ \hline
\hline
 $N=0$  &  &  &  &  &  \\
$\vert \,0\,,\,0\,,\,0\,,\,0 \,\rangle $ & $^{2}S_{1/2}$ & ${\bf \frac{1}{2}}^+$ & $5806^{+9}_{-9}$ & $5618$ & $5637$ & $5612$ & $5620$ & $5667$ & $5794.5\pm 0.6$ \\
\hline
 $N=1$  &  &  &  &  &  \\
$\vert \,1\,,\,0\,,\,0\,,\,0 \,\rangle $ & $^{2}P_{1/2}$ & ${\bf \frac{1}{2}}^-$ & $6079^{+9}_{-9}$ & $5938$ & $6010$ & $5939$ & $5914$ & ... & $\dagger$ \\
$\vert \,1\,,\,0\,,\,0\,,\,0 \,\rangle $ & $^{2}P_{3/2}$ & ${\bf \frac{3}{2}}^-$ & $6085^{+9}_{-9}$ & $5939$ & $6010$ & $5941$ & $5927$ & ... & $6100.3\pm 0.6$ \\
$\vert \,0\,,\,1\,,\,0\,,\,0 \,\rangle $ & $^{2}P_{1/2}$ & ${\bf \frac{1}{2}}^-$ & $6248^{+11}_{-11}$ & $6236$ & ... & $6180$ & $6207$ & ... & $\dagger$ \\
$\vert \,0\,,\,1\,,\,0\,,\,0 \,\rangle $ & $^{4}P_{1/2}$ & ${\bf \frac{1}{2}}^-$ & $6271^{+15}_{-15}$ & $6273$ & $5870$ & ... & $6233$ & ... & $\dagger$ \\
$\vert \,0\,,\,1\,,\,0\,,\,0 \,\rangle $ & $^{2}P_{3/2}$ & ${\bf \frac{3}{2}}^-$ & $6255^{+11}_{-11}$ & $6273$ & ... & $6191$ & ... & ... & $\dagger$ \\
$\vert \,0\,,\,1\,,\,0\,,\,0 \,\rangle $ & $^{4}P_{3/2}$ & ${\bf \frac{3}{2}}^-$ & $6277^{+14}_{-14}$ & $6285$ & $5880$ & ... & ... & ... & $\dagger$ \\
$\vert \,0\,,\,1\,,\,0\,,\,0 \,\rangle $ & $^{4}P_{5/2}$ & ${\bf \frac{5}{2}}^-$ & $6287^{+15}_{-15}$ & $6289$ & ... & $6206$ & ... & ... & $\dagger$ \\
\hline
 $N=2$  &  &  &  &  &  \\
$\vert \,2\,,\,0\,,\,0\,,\,0 \,\rangle $ & $^{2}D_{3/2}$ & ${\bf \frac{3}{2}}^+$ & $6354^{+13}_{-13}$ & $6211$ & $6010$ & $6181$ & $6172$ & ... & $6327.3\pm 2.5$ \\
$\vert \,2\,,\,0\,,\,0\,,\,0 \,\rangle $ & $^{2}D_{5/2}$ & ${\bf \frac{5}{2}}^+$ & $6364^{+13}_{-13}$ & $6212$ & $6010$ & $6183$ & $6178$ & ... & $6332.7\pm 2.5$ \\
$\vert \,0\,,\,0\,,\,1\,,\,0 \,\rangle $ & $^{2}S_{1/2}$ & ${\bf \frac{1}{2}}^+$ & $6360^{+12}_{-13}$ & $6153$ & ... & $6107$ & $6121$ & ... & $\dagger$ \\
$\vert \,0\,,\,0\,,\,0\,,\,1 \,\rangle $ & $^{2}S_{1/2}$ & ${\bf \frac{1}{2}}^+$ & $6699^{+19}_{-19}$ & ... & ... & ... & ... & ... & $\dagger$ \\
$\vert \,1\,,\,1\,,\,0\,,\,0 \,\rangle $ & $^{2}D_{3/2}$ & ${\bf \frac{3}{2}}^+$ & $6524^{+16}_{-16}$ & $6488$ & ... & $6401$ & ... & ... & $\dagger$ \\
$\vert \,1\,,\,1\,,\,0\,,\,0 \,\rangle $ & $^{2}D_{5/2}$ & ${\bf \frac{5}{2}}^+$ & $6534^{+17}_{-17}$ & $6530$ & ... & $6422$ & ... & ... & $\dagger$ \\
$\vert \,1\,,\,1\,,\,0\,,\,0 \,\rangle $ & $^{4}D_{1/2}$ & ${\bf \frac{1}{2}}^+$ & $6540^{+22}_{-22}$ & $6467$ & ... & ... & ... & ... & $\dagger$ \\
$\vert \,1\,,\,1\,,\,0\,,\,0 \,\rangle $ & $^{4}D_{3/2}$ & ${\bf \frac{3}{2}}^+$ & $6546^{+19}_{-19}$ & $6511$ & $6360$ & ... & ... & ... & $\dagger$ \\
$\vert \,1\,,\,1\,,\,0\,,\,0 \,\rangle $ & $^{4}D_{5/2}$ & ${\bf \frac{5}{2}}^+$ & $6556^{+17}_{-18}$ & $6539$ & $6360$ & ... & ... & ... & $\dagger$ \\
$\vert \,1\,,\,1\,,\,0\,,\,0 \,\rangle $ & $^{4}D_{7/2}$ & ${\bf \frac{7}{2}}^+$ & $6570^{+22}_{-22}$ & ... & ... & $6433$ & ... & ... & $\dagger$ \\
$\vert \,1\,,\,1\,,\,0\,,\,0 \,\rangle $ & $^{2}P_{1/2}$ & ${\bf \frac{1}{2}}^-$ & $6526^{+16}_{-16}$ & ... & ... & ... & ... & ... & $\dagger$ \\
$\vert \,1\,,\,1\,,\,0\,,\,0 \,\rangle $ & $^{2}P_{3/2}$ & ${\bf \frac{3}{2}}^-$ & $6532^{+16}_{-16}$ & ... & ... & ... & ... & ... & $\dagger$ \\
$\vert \,1\,,\,1\,,\,0\,,\,0 \,\rangle $ & $^{4}P_{1/2}$ & ${\bf \frac{1}{2}}^-$ & $6548^{+19}_{-19}$ & ... & $6220$ & ... & ... & ... & $\dagger$ \\
$\vert \,1\,,\,1\,,\,0\,,\,0 \,\rangle $ & $^{4}P_{3/2}$ & ${\bf \frac{3}{2}}^-$ & $6554^{+18}_{-17}$ & ... & $6230$ & ... & ... & ... & $\dagger$ \\
$\vert \,1\,,\,1\,,\,0\,,\,0 \,\rangle $ & $^{4}P_{5/2}$ & ${\bf \frac{5}{2}}^-$ & $6564^{+19}_{-19}$ & ... & ... & ... & ... & ... & $\dagger$ \\
$\vert \,1\,,\,1\,,\,0\,,\,0 \,\rangle $ & $^{4}S_{3/2}$ & ${\bf \frac{3}{2}}^+$ & $6558^{+18}_{-18}$ & ... & ... & ... & ... & ... & $\dagger$ \\
$\vert \,1\,,\,1\,,\,0\,,\,0 \,\rangle $ & $^{2}S_{1/2}$ & ${\bf \frac{1}{2}}^+$ & $6530^{+16}_{-16}$ & ... & ... & ... & ... & ... & $\dagger$ \\
$\vert \,0\,,\,2\,,\,0\,,\,0 \,\rangle $ & $^{2}D_{3/2}$ & ${\bf \frac{3}{2}}^+$ & $6693^{+20}_{-19}$ & ... & $6520$ & ... & ... & ... & $\dagger$ \\
$\vert \,0\,,\,2\,,\,0\,,\,0 \,\rangle $ & $^{2}D_{5/2}$ & ${\bf \frac{5}{2}}^+$ & $6703^{+20}_{-20}$ & ... & $6520$ & ... & ... & ... & $\dagger$ \\
\hline \hline
\end{tabular}

\endgroup
}
\end{center}
\end{table*}

\begin{table*}[htp]
\caption{Same as  Table \ref{comparisonlambdas}, but for  $ \Sigma_b(nnb) $ states.}
\label{comparisonsigmas}
\begin{center}
\scriptsize{
\begingroup
\setlength{\tabcolsep}{1.75pt} 
\renewcommand{\arraystretch}{1.35} 

\begin{tabular}{c c c| c c c c c c c}\hline \hline
$\Sigma_{b}(nnb)$ & & ${\mathcal{F} =  \bf {6}}_{\rm F}$& This work   &   NRQM \cite{Yoshida:2015tia}     &  QCD sum rules \cite{Liu:2007fg, Mao:2015gya, Chen:2016phw}      &  NRQM \cite{Roberts:2007ni}  & $\chi$QM \cite{Kim:2021ywp}    & LQCD \cite{Mohanta:2019mxo}     & Experimental  \\
 $\vert l_{\lambda}, l_{\rho}, k_{\lambda}, k_{\rho} \rangle$ & $^{2S+1}L_{J}$  &  ${\bf J}^P$ & mass (MeV)  &   mass (MeV)  &  mass (MeV)  &  mass (MeV) & mass (MeV) & mass (MeV) &      mass (MeV) \\ \hline
\hline
 $N=0$  &  &  &  &  &  \\
$\vert \,0\,,\,0\,,\,0\,,\,0 \,\rangle $ & $^{2}S_{1/2}$ & ${\bf \frac{1}{2}}^+$ & $5804^{+8}_{-8}$ & $5823$ & $5809$ & $5833$ & $5810$ & $5820$ & $5813.1\pm 0.3$ \\
$\vert \,0\,,\,0\,,\,0\,,\,0 \,\rangle $ & $^{4}S_{3/2}$ & ${\bf \frac{3}{2}}^+$ & $5832^{+8}_{-8}$ & $5845$ & $5835$ & $5858$ & $5829$ & $5836$ & $5832.5\pm 0.5$ \\
\hline
 $N=1$  &  &  &  &  &  \\
$\vert \,1\,,\,0\,,\,0\,,\,0 \,\rangle $ & $^{2}P_{1/2}$ & ${\bf \frac{1}{2}}^-$ & $6108^{+10}_{-10}$ & $6127$ & ... & $6099$ & $6043$ & ... & $6096.9\pm 1.8$ \\
$\vert \,1\,,\,0\,,\,0\,,\,0 \,\rangle $ & $^{4}P_{1/2}$ & ${\bf \frac{1}{2}}^-$ & $6131^{+12}_{-13}$ & $6135$ & $6020$ & $6106$ & $6065$ & ... & $\dagger$ \\
$\vert \,1\,,\,0\,,\,0\,,\,0 \,\rangle $ & $^{2}P_{3/2}$ & ${\bf \frac{3}{2}}^-$ & $6114^{+10}_{-10}$ & $6132$ & ... & $6101$ & $6079$ & ... & $\dagger$ \\
$\vert \,1\,,\,0\,,\,0\,,\,0 \,\rangle $ & $^{4}P_{3/2}$ & ${\bf \frac{3}{2}}^-$ & $6137^{+10}_{-10}$ & $6141$ & $5960$ & $6105$ & $6117$ & ... & $\dagger$ \\
$\vert \,1\,,\,0\,,\,0\,,\,0 \,\rangle $ & $^{4}P_{5/2}$ & ${\bf \frac{5}{2}}^-$ & $6147^{+12}_{-12}$ & $6144$ & $5980$ & $6172$ & $6129$ & ... & $\dagger$ \\
$\vert \,0\,,\,1\,,\,0\,,\,0 \,\rangle $ & $^{2}P_{1/2}$ & ${\bf \frac{1}{2}}^-$ & $6304^{+13}_{-13}$ & $6246$ & $5910$ & ... & ... & ... & $\dagger$ \\
$\vert \,0\,,\,1\,,\,0\,,\,0 \,\rangle $ & $^{2}P_{3/2}$ & ${\bf \frac{3}{2}}^-$ & $6311^{+13}_{-13}$ & $6246$ & $5920$ & ... & ... & ... & $\dagger$ \\
\hline
 $N=2$  &  &  &  &  &  \\
$\vert \,2\,,\,0\,,\,0\,,\,0 \,\rangle $ & $^{2}D_{3/2}$ & ${\bf \frac{3}{2}}^+$ & $6415^{+15}_{-15}$ & $6356$ & ... & $6308$ & $6316$ & ... & $\dagger$ \\
$\vert \,2\,,\,0\,,\,0\,,\,0 \,\rangle $ & $^{2}D_{5/2}$ & ${\bf \frac{5}{2}}^+$ & $6425^{+16}_{-16}$ & $6397$ & ... & $6325$ & $6341$ & ... & $\dagger$ \\
$\vert \,2\,,\,0\,,\,0\,,\,0 \,\rangle $ & $^{4}D_{1/2}$ & ${\bf \frac{1}{2}}^+$ & $6431^{+21}_{-21}$ & $6343$ & ... & ... & $6304$ & ... & $\dagger$ \\
$\vert \,2\,,\,0\,,\,0\,,\,0 \,\rangle $ & $^{4}D_{3/2}$ & ${\bf \frac{3}{2}}^+$ & $6437^{+17}_{-17}$ & $6393$ & ... & ... & $6330$ & ... & $\dagger$ \\
$\vert \,2\,,\,0\,,\,0\,,\,0 \,\rangle $ & $^{4}D_{5/2}$ & ${\bf \frac{5}{2}}^+$ & $6448^{+15}_{-15}$ & $6402$ & ... & $6328$ & $6365$ & ... & $\dagger$ \\
$\vert \,2\,,\,0\,,\,0\,,\,0 \,\rangle $ & $^{4}D_{7/2}$ & ${\bf \frac{7}{2}}^+$ & $6462^{+20}_{-20}$ & ... & ... & $6333$ & $6373$ & ... & $\dagger$ \\
$\vert \,0\,,\,0\,,\,1\,,\,0 \,\rangle $ & $^{2}S_{1/2}$ & ${\bf \frac{1}{2}}^+$ & $6421^{+15}_{-15}$ & $6395$ & ... & $6294$ & $6274$ & ... & $\dagger$ \\
$\vert \,0\,,\,0\,,\,1\,,\,0 \,\rangle $ & $^{4}S_{3/2}$ & ${\bf \frac{3}{2}}^+$ & $6450^{+15}_{-15}$ & ... & ... & ... & $6286$ & ... & $\dagger$ \\
$\vert \,0\,,\,0\,,\,0\,,\,1 \,\rangle $ & $^{2}S_{1/2}$ & ${\bf \frac{1}{2}}^+$ & $6813^{+24}_{-24}$ & ... & ... & ... & ... & ... & $\dagger$ \\
$\vert \,0\,,\,0\,,\,0\,,\,1 \,\rangle $ & $^{4}S_{3/2}$ & ${\bf \frac{3}{2}}^+$ & $6842^{+24}_{-23}$ & ... & ... & ... & ... & ... & $\dagger$ \\
$\vert \,1\,,\,1\,,\,0\,,\,0 \,\rangle $ & $^{2}D_{3/2}$ & ${\bf \frac{3}{2}}^+$ & $6611^{+19}_{-19}$ & ... & ... & ... & ... & ... & $\dagger$ \\
$\vert \,1\,,\,1\,,\,0\,,\,0 \,\rangle $ & $^{2}D_{5/2}$ & ${\bf \frac{5}{2}}^+$ & $6621^{+20}_{-20}$ & $6505$ & ... & ... & ... & ... & $\dagger$ \\
$\vert \,1\,,\,1\,,\,0\,,\,0 \,\rangle $ & $^{2}P_{1/2}$ & ${\bf \frac{1}{2}}^-$ & $6613^{+19}_{-19}$ & ... & ... & ... & ... & ... & $\dagger$ \\
$\vert \,1\,,\,1\,,\,0\,,\,0 \,\rangle $ & $^{2}P_{3/2}$ & ${\bf \frac{3}{2}}^-$ & $6619^{+20}_{-20}$ & ... & ... & ... & ... & ... & $\dagger$ \\
$\vert \,1\,,\,1\,,\,0\,,\,0 \,\rangle $ & $^{2}S_{1/2}$ & ${\bf \frac{1}{2}}^+$ & $6617^{+19}_{-19}$ & ... & ... & ... & ... & ... & $\dagger$ \\
$\vert \,0\,,\,2\,,\,0\,,\,0 \,\rangle $ & $^{2}D_{3/2}$ & ${\bf \frac{3}{2}}^+$ & $6807^{+23}_{-23}$ & ... & ... & ... & ... & ... & $\dagger$ \\
$\vert \,0\,,\,2\,,\,0\,,\,0 \,\rangle $ & $^{2}D_{5/2}$ & ${\bf \frac{5}{2}}^+$ & $6817^{+24}_{-25}$ & ... & ... & ... & ... & ... & $\dagger$ \\
$\vert \,0\,,\,2\,,\,0\,,\,0 \,\rangle $ & $^{4}D_{1/2}$ & ${\bf \frac{1}{2}}^+$ & $6824^{+27}_{-27}$ & ... & ... & ... & ... & ... & $\dagger$ \\
$\vert \,0\,,\,2\,,\,0\,,\,0 \,\rangle $ & $^{4}D_{3/2}$ & ${\bf \frac{3}{2}}^+$ & $6830^{+24}_{-24}$ & ... & ... & ... & ... & ... & $\dagger$ \\
$\vert \,0\,,\,2\,,\,0\,,\,0 \,\rangle $ & $^{4}D_{5/2}$ & ${\bf \frac{5}{2}}^+$ & $6840^{+23}_{-23}$ & ... & ... & ... & ... & ... & $\dagger$ \\
$\vert \,0\,,\,2\,,\,0\,,\,0 \,\rangle $ & $^{4}D_{7/2}$ & ${\bf \frac{7}{2}}^+$ & $6854^{+28}_{-28}$ & ... & ... & $6554$ & ... & ... & $\dagger$ \\
\hline \hline
\end{tabular}

\endgroup
}
\end{center}
\end{table*}

\begin{table*}[htp]
\caption{Same as  Table \ref{comparisonlambdas}, but for  $ \Xi'_{b}(snb) $ states.}
\begin{center}
\scriptsize{
\begingroup
\setlength{\tabcolsep}{1.75pt} 
\renewcommand{\arraystretch}{1.35} 

\begin{tabular}{c c c| c c c c c c c}\hline \hline
$\Xi'_{b}(snb)$ & & ${\mathcal{F} =  \bf {6}}_{\rm F}$& This work   &   NRQM \cite{Yoshida:2015tia}     &  QCD sum rules \cite{Liu:2007fg, Mao:2015gya, Chen:2016phw}      &  NRQM \cite{Roberts:2007ni}  & $\chi$QM \cite{Kim:2021ywp}    & LQCD \cite{Mohanta:2019mxo}     & Experimental  \\
 $\vert l_{\lambda}, l_{\rho}, k_{\lambda}, k_{\rho} \rangle$ & $^{2S+1}L_{J}$  &  ${\bf J}^P$ & mass (MeV)  &   mass (MeV)  &  mass (MeV)  &  mass (MeV) & mass (MeV) & mass (MeV) &      mass (MeV) \\ \hline
\hline
 $N=0$  &  &  &  &  &  \\
$\vert \,0\,,\,0\,,\,0\,,\,0 \,\rangle $ & $^{2}S_{1/2}$ & ${\bf \frac{1}{2}}^+$ & $5925^{+6}_{-6}$ & ... & $5903$ & ... & $5934$ & $5946$ & $5935.02\pm 0.05$ \\
$\vert \,0\,,\,0\,,\,0\,,\,0 \,\rangle $ & $^{4}S_{3/2}$ & ${\bf \frac{3}{2}}^+$ & $5953^{+7}_{-7}$ & ... & ... & ... & $5952$ & ... & $5953.8\pm 0.6$ \\
\hline
 $N=1$  &  &  &  &  &  \\
$\vert \,1\,,\,0\,,\,0\,,\,0 \,\rangle $ & $^{2}P_{1/2}$ & ${\bf \frac{1}{2}}^-$ & $6198^{+7}_{-7}$ & ... & ... & ... & $6164$ & ... & $\dagger$ \\
$\vert \,1\,,\,0\,,\,0\,,\,0 \,\rangle $ & $^{4}P_{1/2}$ & ${\bf \frac{1}{2}}^-$ & $6220^{+10}_{-10}$ & ... & $6240$ & ... & $6183$ & ... & $\dagger$ \\
$\vert \,1\,,\,0\,,\,0\,,\,0 \,\rangle $ & $^{2}P_{3/2}$ & ${\bf \frac{3}{2}}^-$ & $6204^{+7}_{-7}$ & ... & ... & ... & $6195$ & ... & $\dagger$ \\
$\vert \,1\,,\,0\,,\,0\,,\,0 \,\rangle $ & $^{4}P_{3/2}$ & ${\bf \frac{3}{2}}^-$ & $6226^{+7}_{-7}$ & ... & $6170$ & ... & $6227$ & ... & $\dagger$ \\
$\vert \,1\,,\,0\,,\,0\,,\,0 \,\rangle $ & $^{4}P_{5/2}$ & ${\bf \frac{5}{2}}^-$ & $6237^{+10}_{-10}$ & ... & $6180$ & ... & $6238$ & ... & $6227.9\pm 1.6$ \\
$\vert \,0\,,\,1\,,\,0\,,\,0 \,\rangle $ & $^{2}P_{1/2}$ & ${\bf \frac{1}{2}}^-$ & $6367^{+9}_{-9}$ & ... & $6110$ & ... & ... & ... & $\dagger$ \\
$\vert \,0\,,\,1\,,\,0\,,\,0 \,\rangle $ & $^{2}P_{3/2}$ & ${\bf \frac{3}{2}}^-$ & $6374^{+10}_{-10}$ & ... & $6110$ & ... & ... & ... & $\dagger$ \\
\hline
 $N=2$  &  &  &  &  &  \\
$\vert \,2\,,\,0\,,\,0\,,\,0 \,\rangle $ & $^{2}D_{3/2}$ & ${\bf \frac{3}{2}}^+$ & $6473^{+12}_{-12}$ & ... & ... & ... & $6423$ & ... & $\dagger$ \\
$\vert \,2\,,\,0\,,\,0\,,\,0 \,\rangle $ & $^{2}D_{5/2}$ & ${\bf \frac{5}{2}}^+$ & $6483^{+13}_{-13}$ & ... & ... & ... & $6444$ & ... & $\dagger$ \\
$\vert \,2\,,\,0\,,\,0\,,\,0 \,\rangle $ & $^{4}D_{1/2}$ & ${\bf \frac{1}{2}}^+$ & $6489^{+18}_{-18}$ & ... & ... & ... & $6411$ & ... & $\dagger$ \\
$\vert \,2\,,\,0\,,\,0\,,\,0 \,\rangle $ & $^{4}D_{3/2}$ & ${\bf \frac{3}{2}}^+$ & $6495^{+14}_{-14}$ & ... & ... & ... & $6434$ & ... & $\dagger$ \\
$\vert \,2\,,\,0\,,\,0\,,\,0 \,\rangle $ & $^{4}D_{5/2}$ & ${\bf \frac{5}{2}}^+$ & $6506^{+11}_{-11}$ & ... & ... & ... & $6465$ & ... & $\dagger$ \\
$\vert \,2\,,\,0\,,\,0\,,\,0 \,\rangle $ & $^{4}D_{7/2}$ & ${\bf \frac{7}{2}}^+$ & $6520^{+18}_{-18}$ & ... & ... & ... & $6472$ & ... & $\dagger$ \\
$\vert \,0\,,\,0\,,\,1\,,\,0 \,\rangle $ & $^{2}S_{1/2}$ & ${\bf \frac{1}{2}}^+$ & $6479^{+11}_{-12}$ & ... & ... & ... & $6381$ & ... & $\dagger$ \\
$\vert \,0\,,\,0\,,\,1\,,\,0 \,\rangle $ & $^{4}S_{3/2}$ & ${\bf \frac{3}{2}}^+$ & $6508^{+12}_{-12}$ & ... & ... & ... & $6392$ & ... & $\dagger$ \\
$\vert \,0\,,\,0\,,\,0\,,\,1 \,\rangle $ & $^{2}S_{1/2}$ & ${\bf \frac{1}{2}}^+$ & $6818^{+19}_{-19}$ & ... & ... & ... & ... & ... & $\dagger$ \\
$\vert \,0\,,\,0\,,\,0\,,\,1 \,\rangle $ & $^{4}S_{3/2}$ & ${\bf \frac{3}{2}}^+$ & $6847^{+19}_{-19}$ & ... & ... & ... & ... & ... & $\dagger$ \\
$\vert \,1\,,\,1\,,\,0\,,\,0 \,\rangle $ & $^{2}D_{3/2}$ & ${\bf \frac{3}{2}}^+$ & $6642^{+15}_{-15}$ & ... & ... & ... & ... & ... & $\dagger$ \\
$\vert \,1\,,\,1\,,\,0\,,\,0 \,\rangle $ & $^{2}D_{5/2}$ & ${\bf \frac{5}{2}}^+$ & $6653^{+17}_{-17}$ & ... & ... & ... & ... & ... & $\dagger$ \\
$\vert \,1\,,\,1\,,\,0\,,\,0 \,\rangle $ & $^{2}P_{1/2}$ & ${\bf \frac{1}{2}}^-$ & $6644^{+15}_{-15}$ & ... & ... & ... & ... & ... & $\dagger$ \\
$\vert \,1\,,\,1\,,\,0\,,\,0 \,\rangle $ & $^{2}P_{3/2}$ & ${\bf \frac{3}{2}}^-$ & $6651^{+16}_{-16}$ & ... & ... & ... & ... & ... & $\dagger$ \\
$\vert \,1\,,\,1\,,\,0\,,\,0 \,\rangle $ & $^{2}S_{1/2}$ & ${\bf \frac{1}{2}}^+$ & $6649^{+15}_{-15}$ & ... & ... & ... & ... & ... & $\dagger$ \\
$\vert \,0\,,\,2\,,\,0\,,\,0 \,\rangle $ & $^{2}D_{3/2}$ & ${\bf \frac{3}{2}}^+$ & $6812^{+19}_{-19}$ & ... & ... & ... & ... & ... & $\dagger$ \\
$\vert \,0\,,\,2\,,\,0\,,\,0 \,\rangle $ & $^{2}D_{5/2}$ & ${\bf \frac{5}{2}}^+$ & $6822^{+20}_{-20}$ & ... & ... & ... & ... & ... & $\dagger$ \\
$\vert \,0\,,\,2\,,\,0\,,\,0 \,\rangle $ & $^{4}D_{1/2}$ & ${\bf \frac{1}{2}}^+$ & $6828^{+22}_{-22}$ & ... & ... & ... & ... & ... & $\dagger$ \\
$\vert \,0\,,\,2\,,\,0\,,\,0 \,\rangle $ & $^{4}D_{3/2}$ & ${\bf \frac{3}{2}}^+$ & $6834^{+20}_{-20}$ & ... & ... & ... & ... & ... & $\dagger$ \\
$\vert \,0\,,\,2\,,\,0\,,\,0 \,\rangle $ & $^{4}D_{5/2}$ & ${\bf \frac{5}{2}}^+$ & $6845^{+19}_{-19}$ & ... & ... & ... & ... & ... & $\dagger$ \\
$\vert \,0\,,\,2\,,\,0\,,\,0 \,\rangle $ & $^{4}D_{7/2}$ & ${\bf \frac{7}{2}}^+$ & $6859^{+24}_{-24}$ & ... & ... & ... & ... & ... & $\dagger$ \\
\hline \hline
\end{tabular}

\endgroup
}
\end{center}
\end{table*}

\begin{table*}[htp]
\caption{Same as  Table \ref{comparisonlambdas}, but for  $ \Omega_{b}(ssb) $ states.}
\label{comparisonomegas}
\begin{center}
\scriptsize{

\begingroup
\setlength{\tabcolsep}{1.75pt} 
\renewcommand{\arraystretch}{1.35} 

\begin{tabular}{c c c| c c c c c c c}\hline \hline
$\Omega_{b}(ssb)$ & & ${\mathcal{F} =  \bf {6}}_{\rm F}$& This work   &   NRQM \cite{Yoshida:2015tia}     &  QCD sum rules \cite{Liu:2007fg, Mao:2015gya, Chen:2016phw}      &  NRQM \cite{Roberts:2007ni}  & $\chi$QM \cite{Kim:2021ywp}    & LQCD \cite{Mohanta:2019mxo}     & Experimental  \\
 $\vert l_{\lambda}, l_{\rho}, k_{\lambda}, k_{\rho} \rangle$ & $^{2S+1}L_{J}$  &  ${\bf J}^P$ & mass (MeV)  &   mass (MeV)  &  mass (MeV)  &  mass (MeV) & mass (MeV) & mass (MeV) &      mass (MeV) \\ \hline
\hline
 $N=0$  &  &  &  &  &  \\
$\vert \,0\,,\,0\,,\,0\,,\,0 \,\rangle $ & $^{2}S_{1/2}$ & ${\bf \frac{1}{2}}^+$ & $6064^{+8}_{-8}$ & $6076$ & $6036$ & $6081$ & $6047$ & $6014$ & $6045.2\pm 1.2$ \\
$\vert \,0\,,\,0\,,\,0\,,\,0 \,\rangle $ & $^{4}S_{3/2}$ & ${\bf \frac{3}{2}}^+$ & $6093^{+9}_{-8}$ & $6094$ & $6063$ & $6102$ & $6064$ & $6019$ & $\dagger$ \\
\hline
 $N=1$  &  &  &  &  &  \\
$\vert \,1\,,\,0\,,\,0\,,\,0 \,\rangle $ & $^{2}P_{1/2}$ & ${\bf \frac{1}{2}}^-$ & $6315^{+7}_{-7}$ & $6333$ & ... & $6301$ & $6273$ & ... & $6315.6\pm 0.6$ \\
$\vert \,1\,,\,0\,,\,0\,,\,0 \,\rangle $ & $^{4}P_{1/2}$ & ${\bf \frac{1}{2}}^-$ & $6337^{+10}_{-10}$ & $6340$ & $6500$ & $6312$ & $6290$ & ... & $6330.3\pm 0.6$ \\
$\vert \,1\,,\,0\,,\,0\,,\,0 \,\rangle $ & $^{2}P_{3/2}$ & ${\bf \frac{3}{2}}^-$ & $6321^{+8}_{-8}$ & $6336$ & ... & $6304$ & $6301$ & ... & $6339.7\pm 0.6$ \\
$\vert \,1\,,\,0\,,\,0\,,\,0 \,\rangle $ & $^{4}P_{3/2}$ & ${\bf \frac{3}{2}}^-$ & $6343^{+7}_{-7}$ & $6344$ & $6430$ & $6311$ & $6329$ & ... & $6349.8\pm 0.6$ \\
$\vert \,1\,,\,0\,,\,0\,,\,0 \,\rangle $ & $^{4}P_{5/2}$ & ${\bf \frac{5}{2}}^-$ & $6353^{+11}_{-11}$ & $6345$ & $6430$ & $6311$ & $6339$ & ... & $\dagger$ \\
$\vert \,0\,,\,1\,,\,0\,,\,0 \,\rangle $ & $^{2}P_{1/2}$ & ${\bf \frac{1}{2}}^-$ & $6465^{+9}_{-8}$ & $6437$ & $6340$ & ... & ... & ... & $\dagger$ \\
$\vert \,0\,,\,1\,,\,0\,,\,0 \,\rangle $ & $^{2}P_{3/2}$ & ${\bf \frac{3}{2}}^-$ & $6471^{+10}_{-10}$ & $6438$ & $6340$ & ... & ... & ... & $\dagger$ \\
\hline
 $N=2$  &  &  &  &  &  \\
$\vert \,2\,,\,0\,,\,0\,,\,0 \,\rangle $ & $^{2}D_{3/2}$ & ${\bf \frac{3}{2}}^+$ & $6568^{+11}_{-11}$ & $6528$ & ... & $6478$ & $6522$ & ... & $\dagger$ \\
$\vert \,2\,,\,0\,,\,0\,,\,0 \,\rangle $ & $^{2}D_{5/2}$ & ${\bf \frac{5}{2}}^+$ & $6578^{+12}_{-12}$ & $6561$ & ... & $6492$ & $6541$ & ... & $\dagger$ \\
$\vert \,2\,,\,0\,,\,0\,,\,0 \,\rangle $ & $^{4}D_{1/2}$ & ${\bf \frac{1}{2}}^+$ & $6584^{+17}_{-17}$ & $6517$ & ... & ... & $6511$ & ... & $\dagger$ \\
$\vert \,2\,,\,0\,,\,0\,,\,0 \,\rangle $ & $^{4}D_{3/2}$ & ${\bf \frac{3}{2}}^+$ & $6590^{+13}_{-13}$ & $6559$ & ... & ... & $6532$ & ... & $\dagger$ \\
$\vert \,2\,,\,0\,,\,0\,,\,0 \,\rangle $ & $^{4}D_{5/2}$ & ${\bf \frac{5}{2}}^+$ & $6600^{+10}_{-10}$ & $6566$ & ... & $6494$ & $6559$ & ... & $\dagger$ \\
$\vert \,2\,,\,0\,,\,0\,,\,0 \,\rangle $ & $^{4}D_{7/2}$ & ${\bf \frac{7}{2}}^+$ & $6614^{+18}_{-18}$ & ... & ... & $6497$ & $6567$ & ... & $\dagger$ \\
$\vert \,0\,,\,0\,,\,1\,,\,0 \,\rangle $ & $^{2}S_{1/2}$ & ${\bf \frac{1}{2}}^+$ & $6574^{+11}_{-11}$ & $6561$ & ... & $6472$ & $6480$ & ... & $\dagger$ \\
$\vert \,0\,,\,0\,,\,1\,,\,0 \,\rangle $ & $^{4}S_{3/2}$ & ${\bf \frac{3}{2}}^+$ & $6602^{+11}_{-11}$ & ... & ... & ... & $6491$ & ... & $\dagger$ \\
$\vert \,0\,,\,0\,,\,0\,,\,1 \,\rangle $ & $^{2}S_{1/2}$ & ${\bf \frac{1}{2}}^+$ & $6874^{+17}_{-17}$ & ... & ... & ... & ... & ... & $\dagger$ \\
$\vert \,0\,,\,0\,,\,0\,,\,1 \,\rangle $ & $^{4}S_{3/2}$ & ${\bf \frac{3}{2}}^+$ & $6902^{+17}_{-17}$ & ... & ... & ... & ... & ... & $\dagger$ \\
$\vert \,1\,,\,1\,,\,0\,,\,0 \,\rangle $ & $^{2}D_{3/2}$ & ${\bf \frac{3}{2}}^+$ & $6718^{+14}_{-14}$ & ... & ... & ... & ... & ... & $\dagger$ \\
$\vert \,1\,,\,1\,,\,0\,,\,0 \,\rangle $ & $^{2}D_{5/2}$ & ${\bf \frac{5}{2}}^+$ & $6728^{+15}_{-15}$ & $6657$ & ... & ... & ... & ... & $\dagger$ \\
$\vert \,1\,,\,1\,,\,0\,,\,0 \,\rangle $ & $^{2}P_{1/2}$ & ${\bf \frac{1}{2}}^-$ & $6720^{+14}_{-14}$ & ... & ... & ... & ... & ... & $\dagger$ \\
$\vert \,1\,,\,1\,,\,0\,,\,0 \,\rangle $ & $^{2}P_{3/2}$ & ${\bf \frac{3}{2}}^-$ & $6726^{+15}_{-15}$ & ... & ... & ... & ... & ... & $\dagger$ \\
$\vert \,1\,,\,1\,,\,0\,,\,0 \,\rangle $ & $^{2}S_{1/2}$ & ${\bf \frac{1}{2}}^+$ & $6724^{+14}_{-14}$ & ... & ... & ... & ... & ... & $\dagger$ \\
$\vert \,0\,,\,2\,,\,0\,,\,0 \,\rangle $ & $^{2}D_{3/2}$ & ${\bf \frac{3}{2}}^+$ & $6868^{+17}_{-17}$ & ... & ... & ... & ... & ... & $\dagger$ \\
$\vert \,0\,,\,2\,,\,0\,,\,0 \,\rangle $ & $^{2}D_{5/2}$ & ${\bf \frac{5}{2}}^+$ & $6878^{+19}_{-19}$ & ... & ... & ... & ... & ... & $\dagger$ \\
$\vert \,0\,,\,2\,,\,0\,,\,0 \,\rangle $ & $^{4}D_{1/2}$ & ${\bf \frac{1}{2}}^+$ & $6884^{+21}_{-21}$ & ... & ... & ... & ... & ... & $\dagger$ \\
$\vert \,0\,,\,2\,,\,0\,,\,0 \,\rangle $ & $^{4}D_{3/2}$ & ${\bf \frac{3}{2}}^+$ & $6890^{+18}_{-18}$ & ... & ... & ... & ... & ... & $\dagger$ \\
$\vert \,0\,,\,2\,,\,0\,,\,0 \,\rangle $ & $^{4}D_{5/2}$ & ${\bf \frac{5}{2}}^+$ & $6900^{+17}_{-17}$ & ... & ... & ... & ... & ... & $\dagger$ \\
$\vert \,0\,,\,2\,,\,0\,,\,0 \,\rangle $ & $^{4}D_{7/2}$ & ${\bf \frac{7}{2}}^+$ & $6914^{+23}_{-23}$ & ... & ... & $6667$ & ... & ... & $\dagger$ \\
\hline \hline
\end{tabular}

\endgroup
}
\end{center}
\end{table*}


\begin{table}[htp]
\caption{$\Lambda_b(nnb)$  electromagnetic decay widths within the three-quark model. The first column reports the baryon name with its predicted  mass, calculated using the three-quark model Hamiltonian given by Eqs.~\ref{MassFormula} and \ref{eq:Hho}. The second column displays $\bf J^{\rm P}$, the third column shows the internal configuration of the baryon $\left| l_{\lambda},l_{\rho}, k_{\lambda},k_{\rho}\right\rangle$ within the three-quark model, where $l_{\lambda,\rho}$ represent the orbital angular momenta and $k_{\lambda,\rho}$ denote the number of nodes of the $\lambda$ and $\rho$ oscillators.
The fourth column presents the spectroscopic notation $^{2S+1}L_J$ for each state.  Furthermore, $N=n_\rho+n_\lambda$ separates the  $N=0,1$ energy bands.  Starting from the fifth column, the electromagnetic decay widths, computed by means of Eq.~\ref{gammaEM}, are presented. Each column corresponds to an electromagnetic decay channel,  the decay products are indicated at the top of the column and their masses are shown in Table~\ref{tab:exp_dec}. The electromagnetic widths  are given in KeV. The values denoted by 0 are forbidden by phase space or are too small to be shown in this scale.}
\begin{center}
\scriptsize{
\begingroup
\setlength{\tabcolsep}{1.75pt} 
\renewcommand{\arraystretch}{1.35} 

\begin{tabular}{c c c c|  p{0.65cm}  p{0.65cm}  p{0.65cm}  p{0.65cm}  p{0.65cm}  p{0.65cm}} \hline \hline
$\mathcal{F}={\bf {\bar{3}}}_{\rm f}$  &    &    &    & $\Lambda_{b}^{0} \gamma$  & $\Sigma_{b}^{0} \gamma$  & $\Sigma_{b}^{*} \gamma$  \\
$\Lambda_b(nnb)$  & $ {\bf J}^P $  & $\vert l_{\lambda}, l_{\rho}, k_{\lambda}, k_{\rho} \rangle$  & $^{2S+1}L_{J}$  & KeV  & KeV  & KeV  \\
\hline
 $N=0$  &  &  &  &  &  \\
$\Lambda_b(5613)$  & ${\bf \frac{1}{2}}^+$ & $\vert \,0\,,\,0\,,\,0\,,\,0 \,\rangle $ &$^{2}S_{1/2}$&0  &0  &0 \\
\hline
 $N=1$  &  &  &  &  &  \\
$\Lambda_b(5918)$  & ${\bf \frac{1}{2}}^-$ & $\vert \,1\,,\,0\,,\,0\,,\,0 \,\rangle $ &$^{2}P_{1/2}$&64  &0.4  &0 \\
$\Lambda_b(5924)$  & ${\bf \frac{3}{2}}^-$ & $\vert \,1\,,\,0\,,\,0\,,\,0 \,\rangle $ &$^{2}P_{3/2}$&65  &0.5  &0.1 \\
$\Lambda_b(6114)$  & ${\bf \frac{1}{2}}^-$ & $\vert \,0\,,\,1\,,\,0\,,\,0 \,\rangle $ &$^{2}P_{1/2}$&15  &519  &3 \\
$\Lambda_b(6137)$  & ${\bf \frac{1}{2}}^-$ & $\vert \,0\,,\,1\,,\,0\,,\,0 \,\rangle $ &$^{4}P_{1/2}$&9  &6  &76 \\
$\Lambda_b(6121)$  & ${\bf \frac{3}{2}}^-$ & $\vert \,0\,,\,1\,,\,0\,,\,0 \,\rangle $ &$^{2}P_{3/2}$&16  &1025  &3 \\
$\Lambda_b(6143)$  & ${\bf \frac{3}{2}}^-$ & $\vert \,0\,,\,1\,,\,0\,,\,0 \,\rangle $ &$^{4}P_{3/2}$&25  &17  &382 \\
$\Lambda_b(6153)$  & ${\bf \frac{5}{2}}^-$ & $\vert \,0\,,\,1\,,\,0\,,\,0 \,\rangle $ &$^{4}P_{5/2}$&17  &12  &1023 \\
\hline \hline
\end{tabular}

\endgroup
}
\end{center}
\label{lambdasEM}
\end{table}


\begin{table}[htp]
\caption{Same as Table \ref{lambdasEM}, but for  $ \Xi_b(snb) $ states.}
\begin{center}
\scriptsize{
\begingroup
\setlength{\tabcolsep}{1.75pt} 
\renewcommand{\arraystretch}{1.35} 

\begin{tabular}{c c c c|  p{0.65cm}  p{0.65cm}  p{0.65cm}  p{0.65cm}  p{0.65cm}  p{0.65cm}  p{0.65cm}  p{0.65cm}  p{0.65cm}} \hline \hline
$\mathcal{F}={\bf {\bar{3}}}_{\rm f}$  &    &    &    & $\Xi_{b}^{0} \gamma$  & $\Xi_{b}^{-} \gamma$  & $\Xi'^{0}_{b} \gamma$  & $\Xi'^{-}_{b} \gamma$  & $\Xi^{*0}_{b} \gamma$  & $\Xi^{*-}_{b} \gamma$  \\
$\Xi_b(snb)$  & $ {\bf J}^P $  & $\vert l_{\lambda}, l_{\rho}, k_{\lambda}, k_{\rho} \rangle$  & $^{2S+1}L_{J}$  & KeV  & KeV  & KeV  & KeV  & KeV  & KeV  \\
\hline
 $N=0$  &  &  &  &  &  \\
$\Xi_b(5806)$  & ${\bf \frac{1}{2}}^+$ & $\vert \,0\,,\,0\,,\,0\,,\,0 \,\rangle $ &$^{2}S_{1/2}$&0  &0  &0  &0  &0  &0 \\
\hline
 $N=1$  &  &  &  &  &  \\
$\Xi_b(6079)$  & ${\bf \frac{1}{2}}^-$ & $\vert \,1\,,\,0\,,\,0\,,\,0 \,\rangle $ &$^{2}P_{1/2}$&122  &126  &1.1  &0  &0.2  &0 \\
$\Xi_b(6085)$  & ${\bf \frac{3}{2}}^-$ & $\vert \,1\,,\,0\,,\,0\,,\,0 \,\rangle $ &$^{2}P_{3/2}$&125  &126  &1.3  &0  &0.2  &0 \\
$\Xi_b(6248)$  & ${\bf \frac{1}{2}}^-$ & $\vert \,0\,,\,1\,,\,0\,,\,0 \,\rangle $ &$^{2}P_{1/2}$&19  &28  &494  &9  &2  &0 \\
$\Xi_b(6271)$  & ${\bf \frac{1}{2}}^-$ & $\vert \,0\,,\,1\,,\,0\,,\,0 \,\rangle $ &$^{4}P_{1/2}$&11  &17  &5  &0.1  &75  &1.4 \\
$\Xi_b(6255)$  & ${\bf \frac{3}{2}}^-$ & $\vert \,0\,,\,1\,,\,0\,,\,0 \,\rangle $ &$^{2}P_{3/2}$&20  &29  &950  &17  &3  &0 \\
$\Xi_b(6277)$  & ${\bf \frac{3}{2}}^-$ & $\vert \,0\,,\,1\,,\,0\,,\,0 \,\rangle $ &$^{4}P_{3/2}$&33  &49  &14  &0.3  &363  &7 \\
$\Xi_b(6287)$  & ${\bf \frac{5}{2}}^-$ & $\vert \,0\,,\,1\,,\,0\,,\,0 \,\rangle $ &$^{4}P_{5/2}$&23  &34  &10  &0.2  &945  &17 \\
\hline \hline
\end{tabular}

\endgroup
}
\end{center}
\label{cascadesEM}
\end{table}

\begin{table*}[htp]
\caption{Same as  Table \ref{lambdasEM}, but for  $ \Sigma_b(nnb) $ states.}
\begin{center}
\scriptsize{

\begingroup
\setlength{\tabcolsep}{1.75pt} 
\renewcommand{\arraystretch}{1.35} 

\begin{tabular}{c c c c|  p{0.65cm}  p{0.65cm}  p{0.65cm}  p{0.65cm}  p{0.65cm}  p{0.65cm}  p{0.65cm}  p{0.65cm}  p{0.65cm}  p{0.65cm}} \hline \hline
$\mathcal{F}={\bf {6}}_{\rm f}$   &    &    &    & $\Sigma_{b}^{+} \gamma$  & $\Sigma_{b}^{0} \gamma$  & $\Sigma_{b}^{-} \gamma$  & $\Lambda_{b}^{0} \gamma$  & $\Sigma_{b}^{*+} \gamma$  & $\Sigma_{b}^{*0} \gamma$  & $\Sigma_{b}^{*-} \gamma$  \\
$\Sigma_b(nnb)$  & $ {\bf J}^P $  & $\vert l_{\lambda}, l_{\rho}, k_{\lambda}, k_{\rho} \rangle$  & $^{2S+1}L_{J}$  & KeV  & KeV  & KeV  & KeV  & KeV  & KeV  & KeV  \\
\hline
 $N=0$  &  &  &  &  &  \\
$\Sigma_b(5804)$  & ${\bf \frac{1}{2}}^+$ & $\vert \,0\,,\,0\,,\,0\,,\,0 \,\rangle $ &$^{2}S_{1/2}$&0  &0  &0  &150  &0  &0  &0 \\
$\Sigma_b(5832)$  & ${\bf \frac{3}{2}}^+$ & $\vert \,0\,,\,0\,,\,0\,,\,0 \,\rangle $ &$^{4}S_{3/2}$&0.5  &0  &0.1  &215  &0  &0  &0 \\
\hline
 $N=1$  &  &  &  &  &  \\
$\Sigma_b(6108)$  & ${\bf \frac{1}{2}}^-$ & $\vert \,1\,,\,0\,,\,0\,,\,0 \,\rangle $ &$^{2}P_{1/2}$&407  &34  &73  &195  &7  &0.4  &2 \\
$\Sigma_b(6131)$  & ${\bf \frac{1}{2}}^-$ & $\vert \,1\,,\,0\,,\,0\,,\,0 \,\rangle $ &$^{4}P_{1/2}$&13  &0.8  &3  &111  &36  &4  &5 \\
$\Sigma_b(6114)$  & ${\bf \frac{3}{2}}^-$ & $\vert \,1\,,\,0\,,\,0\,,\,0 \,\rangle $ &$^{2}P_{3/2}$&1202  &89  &252  &202  &7  &0.4  &2 \\
$\Sigma_b(6137)$  & ${\bf \frac{3}{2}}^-$ & $\vert \,1\,,\,0\,,\,0\,,\,0 \,\rangle $ &$^{4}P_{3/2}$&40  &2  &10  &321  &316  &26  &59 \\
$\Sigma_b(6147)$  & ${\bf \frac{5}{2}}^-$ & $\vert \,1\,,\,0\,,\,0\,,\,0 \,\rangle $ &$^{4}P_{5/2}$&29  &2  &7  &217  &1222  &90  &256 \\
$\Sigma_b(6304)$  & ${\bf \frac{1}{2}}^-$ & $\vert \,0\,,\,1\,,\,0\,,\,0 \,\rangle $ &$^{2}P_{1/2}$&247  &15  &62  &424  &103  &6  &26 \\
$\Sigma_b(6311)$  & ${\bf \frac{3}{2}}^-$ & $\vert \,0\,,\,1\,,\,0\,,\,0 \,\rangle $ &$^{2}P_{3/2}$&256  &16  &64  &414  &107  &7  &27 \\
\hline \hline
\end{tabular}

\endgroup
}
\end{center}
\label{sigmasEM}
\end{table*}

\begin{table}[htp]
\caption{Same as  Table \ref{lambdasEM}, but for  $ \Xi'_b(snb) $ states.}
\begin{center}
\scriptsize{

\begingroup
\setlength{\tabcolsep}{1.75pt} 
\renewcommand{\arraystretch}{1.35} 

\begin{tabular}{c c c c|  p{0.65cm}  p{0.65cm}  p{0.65cm}  p{0.65cm}  p{0.65cm}  p{0.65cm}  p{0.65cm}  p{0.65cm}  p{0.65cm}} \hline \hline
$\mathcal{F}={\bf {6}}_{\rm f}$   &    &    &    & $\Xi_{b}^{0} \gamma$  & $\Xi_{b}^{-} \gamma$  & $\Xi'^{0}_{b} \gamma$  & $\Xi'^{-}_{b} \gamma$  & $\Xi^{*0}_{b} \gamma$  & $\Xi^{*-}_{b} \gamma$  \\
$\Xi'_b(snb)$  & $ {\bf J}^P $  & $\vert l_{\lambda}, l_{\rho}, k_{\lambda}, k_{\rho} \rangle$  & $^{2S+1}L_{J}$  & KeV  & KeV  & KeV  & KeV  & KeV  & KeV  \\
\hline
 $N=0$  &  &  &  &  &  \\
$\Xi'_b(5925)$  & ${\bf \frac{1}{2}}^+$ & $\vert \,0\,,\,0\,,\,0\,,\,0 \,\rangle $ &$^{2}S_{1/2}$&33  &0.6  &0  &0  &0  &0 \\
$\Xi'_b(5953)$  & ${\bf \frac{3}{2}}^+$ & $\vert \,0\,,\,0\,,\,0\,,\,0 \,\rangle $ &$^{4}S_{3/2}$&60  &1.1  &0.1  &0.1  &0  &0 \\
\hline
 $N=1$  &  &  &  &  &  \\
$\Xi'_b(6198)$  & ${\bf \frac{1}{2}}^-$ & $\vert \,1\,,\,0\,,\,0\,,\,0 \,\rangle $ &$^{2}P_{1/2}$&65  &1.2  &78.2  &71.1  &0.4  &0.6 \\
$\Xi'_b(6220)$  & ${\bf \frac{1}{2}}^-$ & $\vert \,1\,,\,0\,,\,0\,,\,0 \,\rangle $ &$^{4}P_{1/2}$&40  &0.7  &0.9  &1.4  &11  &9 \\
$\Xi'_b(6204)$  & ${\bf \frac{3}{2}}^-$ & $\vert \,1\,,\,0\,,\,0\,,\,0 \,\rangle $ &$^{2}P_{3/2}$&69  &1.3  &157  &167  &0.4  &0.7 \\
$\Xi'_b(6226)$  & ${\bf \frac{3}{2}}^-$ & $\vert \,1\,,\,0\,,\,0\,,\,0 \,\rangle $ &$^{4}P_{3/2}$&117  &2  &3  &4  &57  &54 \\
$\Xi'_b(6237)$  & ${\bf \frac{5}{2}}^-$ & $\vert \,1\,,\,0\,,\,0\,,\,0 \,\rangle $ &$^{4}P_{5/2}$&83  &1.5  &2  &3  &157  &168 \\
$\Xi'_b(6367)$  & ${\bf \frac{1}{2}}^-$ & $\vert \,0\,,\,1\,,\,0\,,\,0 \,\rangle $ &$^{2}P_{1/2}$&644  &12  &19  &28  &7  &11 \\
$\Xi'_b(6374)$  & ${\bf \frac{3}{2}}^-$ & $\vert \,0\,,\,1\,,\,0\,,\,0 \,\rangle $ &$^{2}P_{3/2}$&637  &12  &20  &29  &8  &12 \\
\hline \hline
\end{tabular}

\endgroup
}
\end{center}
\label{cascadesprimeEM}
\end{table}

\begin{table}[htp]
\caption{Same as Table \ref{lambdasEM}, but for  $ \Omega_b(ssb) $ states.}
\begin{center}
\scriptsize{
\begingroup
\setlength{\tabcolsep}{1.75pt} 
\renewcommand{\arraystretch}{1.35} 

\begin{tabular}{c c c c|  p{0.65cm}  p{0.65cm}  p{0.65cm}  p{0.65cm}  p{0.65cm}} \hline \hline
$\mathcal{F}={\bf {6}}_{\rm f}$   &    &    &    & $\Omega_{b} \gamma$  & $\Omega^{*}_{b} \gamma$  \\
$\Omega_b(ssb)$  & $ {\bf J}^P $  & $\vert l_{\lambda}, l_{\rho}, k_{\lambda}, k_{\rho} \rangle$  & $^{2S+1}L_{J}$  & KeV  & KeV  \\
\hline
 $N=0$  &  &  &  &  &  \\
$\Omega_b(6064)$  & ${\bf \frac{1}{2}}^+$ & $\vert \,0\,,\,0\,,\,0\,,\,0 \,\rangle $ &$^{2}S_{1/2}$&0  &0 \\
$\Omega_b(6093)$  & ${\bf \frac{3}{2}}^+$ & $\vert \,0\,,\,0\,,\,0\,,\,0 \,\rangle $ &$^{4}S_{3/2}$&0.1  &0 \\
\hline
 $N=1$  &  &  &  &  &  \\
$\Omega_b(6315)$  & ${\bf \frac{1}{2}}^-$ & $\vert \,1\,,\,0\,,\,0\,,\,0 \,\rangle $ &$^{2}P_{1/2}$&51  &0.2 \\
$\Omega_b(6337)$  & ${\bf \frac{1}{2}}^-$ & $\vert \,1\,,\,0\,,\,0\,,\,0 \,\rangle $ &$^{4}P_{1/2}$&0.5  &8 \\
$\Omega_b(6321)$  & ${\bf \frac{3}{2}}^-$ & $\vert \,1\,,\,0\,,\,0\,,\,0 \,\rangle $ &$^{2}P_{3/2}$&99  &0.2 \\
$\Omega_b(6343)$  & ${\bf \frac{3}{2}}^-$ & $\vert \,1\,,\,0\,,\,0\,,\,0 \,\rangle $ &$^{4}P_{3/2}$&1.7  &38 \\
$\Omega_b(6353)$  & ${\bf \frac{5}{2}}^-$ & $\vert \,1\,,\,0\,,\,0\,,\,0 \,\rangle $ &$^{4}P_{5/2}$&1.3  &99 \\
$\Omega_b(6465)$  & ${\bf \frac{1}{2}}^-$ & $\vert \,0\,,\,1\,,\,0\,,\,0 \,\rangle $ &$^{2}P_{1/2}$&12  &4 \\
$\Omega_b(6471)$  & ${\bf \frac{3}{2}}^-$ & $\vert \,0\,,\,1\,,\,0\,,\,0 \,\rangle $ &$^{2}P_{3/2}$&12  &5 \\
\hline \hline
\end{tabular}

\endgroup
}
\label{omegasEM}
\end{center}
\end{table}

\begin{figure*}
    \centering
    \includegraphics[scale=.54]{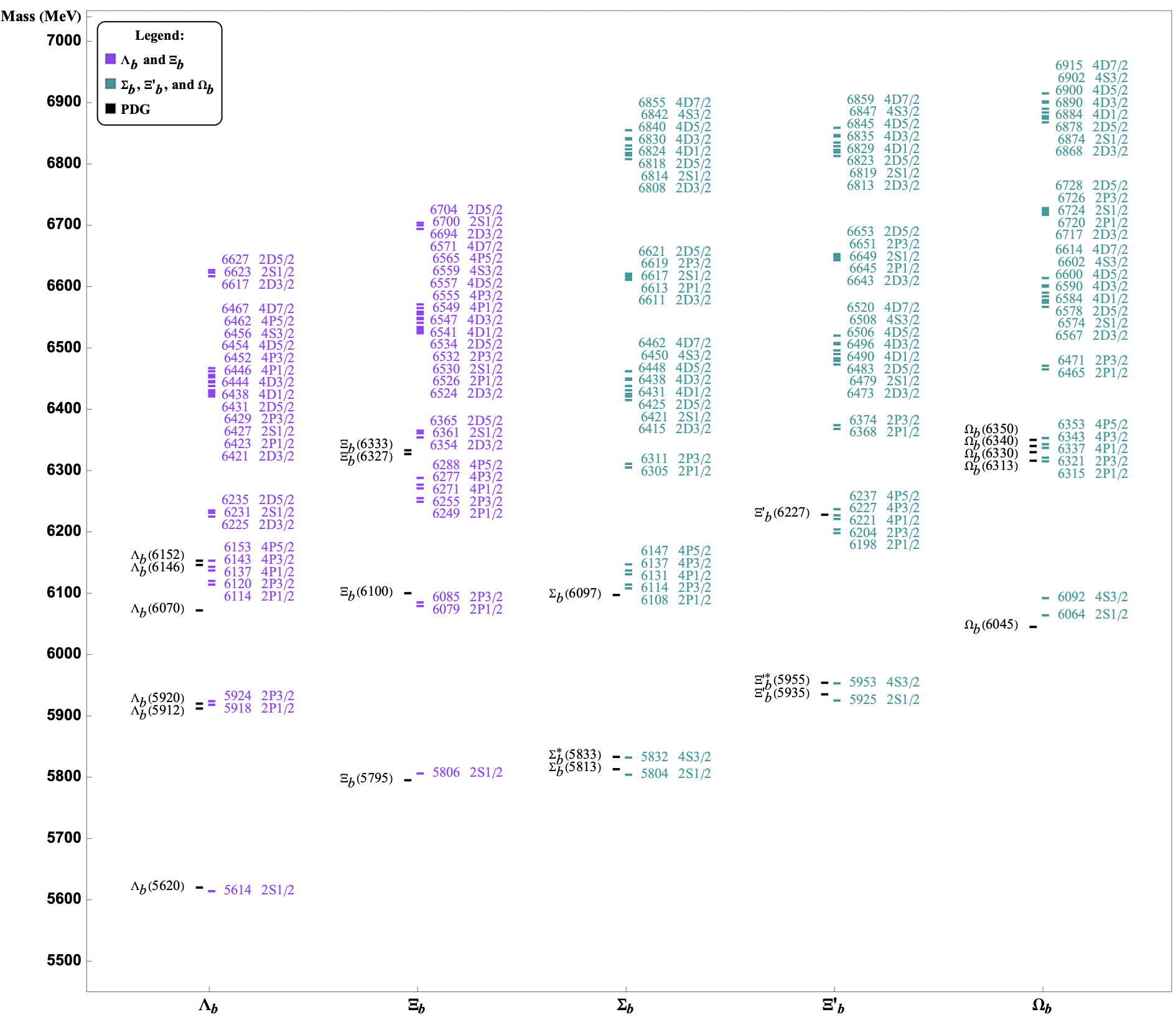}
    \caption{Comparison between the single bottom baryon mass predictions,  computed using the three-quark model Hamiltonian of  Eqs.~\ref{MassFormula} and \ref{eq:Hho}, with the experimental data. The predicted masses for the ${\bf \bar 3}_{\rm F}$ and ${\bf 6}_{\rm F}$ states are displayed  in purple and teal, respectively, while the experimental masses are reported  in black \cite{Workman:2022ynf}. }
    \label{fig:all-states}
\end{figure*}

\section{Discussion and conclusions}
\label{Discussion}

The complexity to identify the bottom baryons arises from the lack of data. However, our results effectively capture the trend observed in the available data reported in PDG \cite{Workman:2022ynf}, see Fig.\ref{fig:all-states}.

Furthermore, we calculate the single-bottom baryon strong decay widths within the $^3P_0$ model. Our calculations consider final states comprising bottom baryon-(vector/pseudoscalar) meson pairs and (octet/decuplet) baryon-(pseudoscalar/vector) bottom meson pairs. In this regard, our investigation represents the most complete study conducted in the single-bottom baryon sector to date. 

To provide further assistance to experimentalists in their search for bottom baryons, we include the partial decay widths for each open flavor channel in Tables \ref{tab:part_dec_lambdas}-\ref{tab:part_dec_Omega}. These partial decay widths can provide valuable information for experimentalists as they aim to identify bottom baryons, and the information regarding possible decay channels can aid in the identification process within the data.

We can observe that electromagnetic decays play a dominant role for the states which cannot decay strongly. 
One notable example is the spin excitation of the $\Omega_b^-$ state, denoted as $\Omega_b^*$, which has not yet been observed. The $\Omega_b^*\to\Omega_b^-\pi$ strong decay  is prohibited due to lack of phase space and isospin conservation in strong interactions.
Given that the $\Omega_b(6093)$ state has not yet been discovered, there exists a fascinating experimental opportunity to simultaneously observe a new electromagnetic decay in the bottom baryon sector and the emergence of a new state, the $\Omega_b(6093)$, by exploring the $\Omega^-_b\gamma$ electromagnetic channel.

 Additionally, we discuss why the presence or absence of the $\rho$-mode excitations in the experimental spectrum is the key to distinguishing between the quark-diquark and three-quark behaviors, as it was originally pointed out in \cite{Santopinto:2018ljf}.
 

In the quark-diquark picture, the effective degrees of freedom are reduced, resulting in fewer predicted states.  It is noteworthy that the quark-diquark model significantly reduces the number of states for $\Lambda_b$ and $\Xi_b$ baryons belonging to the $\bf \bar 3_{\rm F}$ flavor representation. This is evident from Tables \ref{tab:All_mass_Lambda} and \ref{tab:All_mass_Xi}, where the scalar diquark configuration, $\bf S_{\rm tot}=0$, only allows one single ground state with ${\bf J}^P = {\bf \frac{1}{2}}^+$ for $\Lambda_b$ and $\Xi_b$ baryons.

Moreover, in the case of $\Lambda_b$ and $\Xi_b$ baryons, the scalar diquark can only combine with $P_\lambda$ states, $L_{\rm tot}=l_\lambda=1$, leading to two states with ${\bf J}^P = {\bf \frac{1}{2}}^-$ and ${\bf J}^P = {\bf \frac{3}{2}}^-$. A similar constraint applies to the $D$-wave states, where the scalar diquark contributes to the formation of two states with ${\bf J}^P = {\bf \frac{3}{2}}^+$ and ${\bf J}^P = {\bf \frac{5}{2}}^+$, moreover, there is only one radially excited state with ${\bf J}^P = {\bf \frac{1}{2}}^+$. 

It is worth mentioning that the $\Lambda_b (6070)^0$ state has been identified as the first radial excitation with ${\bf J}^P = {\bf \frac{1}{2}}^+$ in the Particle Data Group (PDG) \cite{Workman:2022ynf}. However, its precise quantum numbers have not yet been determined experimentally. In our model, if we consider it as a radial excitation, there is a deviation of approximately 3\% in its mass, and its width is underestimated. Alternatively, we can identify it as a $P_\rho$-wave excited state with ${\bf J}^P = {\bf \frac{1}{2}}^-$, characterized by an internal spin of $\bf S_{\rm tot}=\frac{1}{2}$. In this scenario, the theoretical mass deviates by less than 1\%, and the width is accurately reproduced. Therefore, determining its quantum numbers experimentally would be crucial to ascertain whether it corresponds to a radial excitation or the first $P_\rho$-wave excited state of the $\Lambda_b$ baryon. The latter interpretation would go beyond the diquark picture.

In the case of $\Omega_b$ baryons, we have successfully identified four out of the five $P_\lambda$-wave excited states, matching them with experimental observations. The fifth $P_\lambda$ excitation is characterized by a large width, requiring a high statistical significance for its observation at the LHC. According to our predictions, this state is expected to have a mass in the range of 6345-6365 MeV, with a width of approximately 40 MeV. 

In the three-quark model, we  predict  two additional $P_\rho$ excitations which do not appear in the quark-diquark description.
They do not couple to the $\Xi^0_bK^-$ channel, which is the channel where LHCb observed the four excited $\Omega^-_b$ states, but they exhibit a strong coupling in the $\Xi'^0_bK^-$ channel. 
Therefore, conducting experimental searches for these two $P_\rho$ excitations is crucial in determining whether the bottom baryons can be described as a three-quark system or a quark-diquark system, as these states are not predicted within the quark-diquark framework. This investigation plays a fundamental role in advancing our understanding of the underlying structure of bottom baryons.

In summary, we have performed calculations for the mass spectra, strong partial decay widths, and electromagnetic decay widths of bottom baryons. The mass spectrum predicts all states up  to $D$-wave. Notably, our approach allows for a comprehensive description of all bottom baryons through a global fit, wherein the same set of model parameters predicts the masses, strong partial decay widths, and electromagnetic decay widths of bottom baryons as well. Both the three-quark and quark-diquark schemes are utilized to provide the bottom baryon mass spectra.

 Moreover, we present the spectra of single bottom baryons computed using the three-quark model in Figure \ref{fig:all-states}. The $\bf \bar 3_{\rm F}$ single-bottom baryons are depicted in purple, and their mass predictions are represented by purple lines. The $\bf 6_{\rm F}$ single-bottom baryons are shown in teal, with their mass predictions indicated by teal lines. The experimentally known states and their corresponding experimental values are displayed in black \cite{Workman:2022ynf}.

In contrast to many theoretical papers in this field, we have accounted for the propagation of parameter uncertainties using a Monte Carlo bootstrap method. This inclusion of uncertainties is essential but often missed in related research.

Our predictions for the masses and strong partial decay widths of bottom baryons exhibit good agreement with the available experimental data. Consequently, they can guide future experimental searches at LHCb. Additionally, we have determined the flavor coupling coefficients for all possible decay channels, which can be used in further theoretical investigations.

We calculated the decay widths of the ground and excited bottom baryon states ($\rho$ and $\lambda$ mode excitations  up to the $D$-wave shell) into the bottom baryon-(vector/pseudoscalar) meson pairs and the (octet/ decuplet) baryon-(pseudoscalar/vector) bottom meson pairs.
To the best of our knowledge, our calculations constitute the most comprehensive analysis of strong partial decay widths in the bottom baryon sector to date. It is widely recognized that the predictions for strong decay widths are not highly sensitive to specific models, thus further  enhancing the reliability and accuracy  of our study.

\section*{Acknowledgments}
A. R.-A. acknowledges support by CONAHCyT and by INFN.  C.A. V.-A. is supported by the CONAHCyT Investigadoras e Investigadores por M\'exico project 749 and SNI 58928. A. R.-M. acknowledges the National Research Foundation of Korea grant no. 2020R1I1A1A01066423. A. G. 
is supported by grant no. 2019/35/B/ST2/03531 of the Polish National Science Centre.

\clearpage

\appendix

\section{Bottom-baryon flavor wave functions }
In the bottom sector, we consider the {$ \bf  \bar 3$$\rm _F$-plet}  and  the  {$ \bf 6 \rm _F$-plet} representation of the flavor wave functions. \label{flavorcb}
In the following subsections, we give the flavor wave functions of a bottom baryon $A$ and its isospin quantum numbers  $|A,I,M_I\rangle$.

\subsubsection{$ \bf  \bar 3$$\rm _F$-plet}
\begin{eqnarray}
|\Xi^{-}_b,1/2,-1/2 \rangle:&=&\frac{1}{\sqrt{2}}(|dsb\rangle-|sdb\rangle)\\
|\Xi^{0}_b,1/2,1/2 \rangle:&=&\frac{1}{\sqrt{2}}(|usb\rangle-|sub\rangle)\\
|\Lambda^0_b,0,0\rangle:&=&\frac{1}{\sqrt{2}}(|udb\rangle-|dub\rangle)
\end{eqnarray}

\subsubsection{$ \bf 6 \rm _F$-plet}

\begin{eqnarray}
|\Omega_b^{-},0,0 \rangle:&=&|ssb\rangle\\
|\Xi^{\prime-}_b,1/2,-1/2 \rangle:&=&\frac{1}{\sqrt{2}}(|dsb\rangle+|sdb\rangle)\\
|\Xi^{\prime0}_b,1/2,1/2 \rangle:&=&\frac{1}{\sqrt{2}}(|usb\rangle+|sub\rangle)\\
|\Sigma^{+}_b,1,1 \rangle:&=&|uub\rangle\\
|\Sigma^-_b,1,-1 \rangle:&=&|ddb\rangle\\
|\Sigma^0_b,1, 0\rangle:&=&\frac{1}{\sqrt{2}}(|udb\rangle+|dub\rangle)
\end{eqnarray}

For the flavor-wave functions of light baryons, we follow the convention of Ref. \cite{Garcia-Tecocoatzi:2022zrf}.

\section{Meson flavor wave functions}

In the following, we give the flavor wave functions of a $C$   meson used in the calculation of the  strong  partial decay width. Here, we used  the isospin quantum numbers  $|C,I,M_I\rangle$.
\label{appme}

In the case of bottom-$B$ mesons, the flavor-wave functions are the same for the pseudoscalar  and  vector states. We use the following:

\ba
\begin{array}{lllcc}
|B^{0 *}_s,0,0\rangle&=&|B^{0 }_s,0,0\rangle&=&|b\bar{s}\rangle\nonumber\\
|B^{0 *},1/2,1/2\rangle&=&|B^{0 },1/2,1/2\rangle&=& |b\bar{d}\rangle\\
|B^{- *},1/2,-1/2\rangle&=&|B^{- },1/2,-1/2\rangle&=& |b\bar{u}\rangle\nonumber
\end{array}
\ea
The convention used in the calculation of the strong-decay  width when the final state has a  light meson is found in Ref. \cite{Garcia-Tecocoatzi:2022zrf}.

\section{Flavor coupling}
\label{flavor}
In the following subsections, we give the flavor coefficients $\mathcal{F}_{A\rightarrow BC}$ used to calculate the transition amplitudes. 
We compute 
$\mathcal{F}_{A\rightarrow BC}=\langle\phi_B \phi_C|\phi_0 \phi_A  \rangle$ where $\phi_{(A,B,C)}$ refers to the initial flavor wave function of a bottom baryon $\phi_{A}$, final baryon $\phi_B$, and  final meson $\phi_C$, respectively;  $\phi^{45}_{0}=(u\bar u +d\bar d +s \bar s)/\sqrt3$ is the flavor singlet-wave function of $SU_f(3)$. In addition, we compute the flavor decay coefficients of the isospin channels, since we assume that the isospin symmetry holds even though it is slightly broken. The corresponding charge channels are obtained by multiplying our $\mathcal{F}_{A\rightarrow BC}$ by the corresponding Clebsch-Gordan coefficient in the isospin space, using the convention of the isospin quantum numbers of the baryon and meson  flavor wave functions found in \ref{flavorcb} and \ref{appme},  for the light baryons we use the convention of Ref.\cite{Garcia-Tecocoatzi:2022zrf}. Thus, the flavor charge channel  for a specific projection  $(I,M_I)$ in the isospin space is obtained   as follows: 

\ba
\mathcal{F}_{A (I_A, M_{I_A})\rightarrow B(I_B, M_{I_B})C(I_C, M_{I_C})} =\nonumber\\ \langle\phi_B,I_B, M_{I_B}, \phi_C,I_C, M_{I_C}|\phi_0,0,0,  \phi_A ,I_A, M_{I_A}\rangle_F \nonumber \\= \langle I_B, M_{I_B},I_C, M_{I_C}|I_A, M_{I_A}\rangle\mathcal{F}_{A\rightarrow BC},
\ea
where $\langle I_B, M_{I_B},I_C, M_{I_C}|I_A, M_{I_A}\rangle$ is a Clebsch-Gordan coefficient and the flavor functions $\phi_i$ of each baryon and meson have a specific isospin projection $M_{I}$.

\subsection{Bottom baryons and pseudoscalar mesons}
We give the squared flavor-coupling coefficients, $\mathcal{F}^2_{A\rightarrow BC}$, when the final states have a pseudoscalar light  meson. Here, $A$ and $B$ are bottom baryons, and  the subindexes   $ \mathbf{\bar 3}_{\rm F} $ and ${\bf 6}_{\rm F}$  refer to the anti-triplet and the sextet baryon multiples. The $C$ is a pseudoscalar meson and the subindexes $\bf 8 $$\rm _F$ and $\bf 1$$\rm _F$ refer to the octet and singlet meson multiplets, respectively. 
\begin{itemize}

\item $ A_{\mathbf{\bar 3}_{\rm F}} \rightarrow  B_{\mathbf{6}_{\rm F}}+C_{\bf 8_{\rm F}}$ 

\begin{eqnarray}
\left( \begin{array}{c}  \Lambda_b \\ \\\Xi_b \end{array} \right) &\rightarrow &
\left( \begin{array}{cccc} 
\Xi'_b K & \Sigma_b \pi & \\ \\
\Sigma_b K & \Xi'_b \pi & \Xi'_b \eta  
\end{array} \right) 
\nonumber\\
&=& \left( \begin{array}{cccc} 
\frac{1 }{6} & \frac{1}{2} &  \\ \\
\frac{1 }{4} & \frac{1}{8} & \frac{1}{72} \\  
\end{array} \right) 
\end{eqnarray}

\item $ A_{\mathbf{\bar 3}_{\rm F}} \rightarrow  B_{\mathbf{6}_{\rm F}}+C_{\bf 1_{\rm F}}$ 

\begin{eqnarray}
\left( \begin{array}{c}   \Xi_b \end{array} \right) &\rightarrow &
\left( \begin{array}{c} 
 \Xi'_b \eta'  
\end{array} \right) 
=\Big( 
 \frac{1}{9} \Big)
\end{eqnarray}

\item $ A_{\mathbf{\bar 3}_{\rm F}} \rightarrow  B_{\mathbf{\bar 3}_{\rm F}}+C_{\bf 8_{\rm F}}$ 

\begin{eqnarray}
\left( \begin{array}{c}  \Lambda_b \\ \\\Xi_b \end{array} \right) &\rightarrow &
\left( \begin{array}{cccc} 
\Xi_b K & \Lambda_b \eta & \\ \\
\Lambda_b K & \Xi_b \pi & \Xi_b \eta  
\end{array} \right) 
\nonumber\\
&=& \left( \begin{array}{cccc} 
\frac{1 }{6} & \frac{1}{18} &  \\ \\
\frac{1 }{12} & \frac{1}{8} & \frac{1}{72} \\  
\end{array} \right) 
\end{eqnarray}

\item $ A_{\mathbf{\bar 3}_{\rm F}} \rightarrow  B_{\mathbf{\bar 3}_{\rm F}}+C_{\bf 1_{\rm F}}$ 

\begin{eqnarray}
\left( \begin{array}{c}  \Lambda_b \\ \\\Xi_b \end{array} \right) &\rightarrow &
\left( \begin{array}{c} 
 \Lambda_b \eta' \\ \\
 \Xi_b \eta'  
\end{array} \right) 
=\left( \begin{array}{c} 
\frac{1 }{9}  \\ \\
\frac{1 }{9}   
\end{array} \right) 
\end{eqnarray}

 \item $ A_{\mathbf{6}_{\rm F}} \rightarrow  B_{\mathbf{6}_{\rm F}}+C_{\bf 8_{\rm F}}$ 

\begin{eqnarray}
\left( \begin{array}{c} \Omega_b \\ \\ \Sigma_b \\ \\\Xi'_b \end{array} \right) &\rightarrow &
\left( \begin{array}{cccc} &\Xi'_b K  & \Omega_b \eta\\ \\
\Xi'_b K & \Sigma_b \pi & \Sigma_b \eta \\ \\
\Sigma_b K & \Xi'_b \pi & \Xi'_b \eta  
\end{array} \right) 
\nonumber\\
&=& \left( \begin{array}{cccc} & \frac{1}{3} & \frac{2}{9}\\  \\
\frac{1 }{6} & \frac{1}{3} & \frac{1}{18} \\ \\
\frac{1 }{4} & \frac{1}{8} & \frac{1}{72} \\  
\end{array} \right) 
\end{eqnarray}

\item $ A_{\mathbf{6}_{\rm F}} \rightarrow  B_{\mathbf{6}_{\rm F}}+C_{\bf 1_{\rm F}}$ 

\begin{eqnarray}
\left( \begin{array}{c} \Omega_b \\ \\ \Sigma_b \\ \\\Xi'_b \end{array} \right) \rightarrow 
\left( \begin{array}{c} \Omega_b \eta' \\ \\ \Sigma_b \eta' \\ \\ \Xi'_b \eta' \\  \end{array} \right) = 
 \left( \begin{array}{c} \frac{1}{3} \\ \\ \frac{1}{9} \\  \\ \frac{1}{9}  \end{array} \right) 
\label{eqn:10101_1}
\end{eqnarray}

  \item $ A_{\mathbf{6}_{\rm F}} \rightarrow  B_{ \mathbf{\bar3}_{\rm F}}+C_{\bf 8_{\rm F}}$ 
  
\begin{eqnarray}
\left( \begin{array}{c} \Omega_b \\ \\ \Sigma_b \\ \\\Xi'_b \end{array} \right) &\rightarrow &
\left( \begin{array}{cccc} &\Xi_b K  & \\ \\
\Xi_b K & \Lambda_b \pi &  \\ \\
\Lambda_b K & \Xi_b \pi & \Xi_b \eta  
\end{array} \right) 
\nonumber\\
&=& \left( \begin{array}{cccc} & \frac{1}{3} & \\  \\
\frac{1 }{6} & \frac{1}{2} &  \\ \\
\frac{1 }{12} & \frac{1}{8} & \frac{1}{72} \\  
\end{array} \right) 
\end{eqnarray}

\item $ A_{\mathbf{6}_{\rm F}} \rightarrow  B_{\mathbf{\bar3}_{\rm F}}+C_{\bf 1_{\rm F}} $

\begin{eqnarray}
\left(   \Xi'_b  \right) \rightarrow 
\left(    \Xi_b \eta'  \right) = 
 \left(  \frac{1}{9}  \right) 
\label{eqn:10101_2}
\end{eqnarray}

\end{itemize}
\subsection{Bottom baryons and vector mesons}
We give the squared flavor-coupling coefficients, $\mathcal{F}^2_{A\rightarrow BC}$ when the final states have a vector-light  meson. Here $A$ and $B$ are bottom baryons, and  the subindexes $ \mathbf{\bar 3}_{\rm F} $ and ${\bf 6}_{\rm F}$  refer to the anti-triplet and the sextet baryon multiplets. The $C$ is a vector meson  and the subindexes $\bf 8_{\rm F} $ and $\bf 1_{\rm F}$ refer to the octet and singlet meson multiplets, respectively. 
\begin{itemize}

\item $ A_{\mathbf{\bar 3}_{\rm F}} \rightarrow  B_{\mathbf{6}_{\rm F}}+C_{\bf 8_{\rm F}}$ 

\begin{eqnarray}
\left( \begin{array}{c}  \Lambda_b \\ \\\Xi_b \end{array} \right) &\rightarrow &
\left( \begin{array}{cccc} 
\Xi'_b K^* & \Sigma_b \rho & \\ \\
\Sigma_b K^* & \Xi'_b \rho & \Xi'_b \omega  
\end{array} \right) 
\nonumber\\
&=& \left( \begin{array}{cccc} 
\frac{1 }{6} & \frac{1}{2} &  \\ \\
\frac{1 }{4} & \frac{1}{8} & \frac{1}{24} \\  
\end{array} \right) 
\end{eqnarray}

\item $ A_{\mathbf{\bar 3}_{\rm F}} \rightarrow  B_{\mathbf{6}_{\rm F}}+C_{\bf 1_{\rm F}}$ 

\begin{eqnarray}
\left( \begin{array}{c}   \Xi_b \end{array} \right) &\rightarrow &
\left( \begin{array}{c} 
 \Xi'_c \phi  
\end{array} \right) 
=\Big( 
 \frac{1}{12} \Big)
\end{eqnarray}

\item $ A_{\mathbf{\bar 3}_{\rm F}} \rightarrow  B_{\mathbf{\bar 3}_{\rm F}}+C_{\bf 8_{\rm F}}$ 

\begin{eqnarray}
\left( \begin{array}{c}  \Lambda_b \\ \\\Xi_b \end{array} \right) &\rightarrow &
\left( \begin{array}{cccc} 
\Xi_b K^* & \Lambda_b \omega & \\ \\
\Lambda_b K^* & \Xi_b \rho & \Xi_b \omega 
\end{array} \right) 
\nonumber\\
&=& \left( \begin{array}{cccc} 
\frac{1 }{6} & \frac{1}{6} &  \\ \\
\frac{1 }{12} & \frac{1}{8} & \frac{1}{24} \\  
\end{array} \right) 
\end{eqnarray}

\item $ A_{\mathbf{\bar 3}_{\rm F}} \rightarrow  B_{\mathbf{\bar 3}_{\rm F}}+C_{\bf 1_{\rm F}}$ 

\begin{eqnarray}
\left( \begin{array}{c}  \Lambda_b \\ \\\Xi_b \end{array} \right) &\rightarrow &
\left( \begin{array}{c} 
 \Lambda_b \phi \\ \\
 \Xi_b \phi  
\end{array} \right) 
=\left( \begin{array}{c} 
0  \\ \\
\frac{1 }{12}   
\end{array} \right) 
\end{eqnarray}

 \item $ A_{\mathbf{6}_{\rm F}} \rightarrow  B_{\mathbf{6}_{\rm F}}+C_{\bf 8_{\rm F}}$ 

\begin{eqnarray}
\left( \begin{array}{c} \Omega_b \\ \\ \Sigma_b \\ \\\Xi'_b \end{array} \right) &\rightarrow &
\left( \begin{array}{cccc} &\Xi'_b K^*  & \\ \\
\Xi'_b K^* & \Sigma_b \rho & \Sigma_b \omega \\ \\
\Sigma_b K^* & \Xi'_b \rho & \Xi'_b \omega 
\end{array} \right) 
\nonumber\\
&=& \left( \begin{array}{cccc} & \frac{1}{3} & \\  \\
\frac{1 }{6} & \frac{1}{3} & \frac{1}{6} \\ \\
\frac{1 }{4} & \frac{1}{8} & \frac{1}{24} \\  
\end{array} \right) 
\end{eqnarray}

\item $ A_{\mathbf{6}_{\rm F}} \rightarrow  B_{\mathbf{6}_{\rm F}}+C_{\bf 1_{\rm F}}$ 

\begin{eqnarray}
\left( \begin{array}{c} \Omega_b \\ \\ \Sigma_b \\ \\\Xi'_b \end{array} \right) \rightarrow 
\left( \begin{array}{c} \Omega_b \phi \\ \\ \Sigma_b \phi \\ \\ \Xi'_b \phi \\  \end{array} \right) = 
 \left( \begin{array}{c} \frac{1}{9} \\ \\ 0\\  \\ \frac{1}{2}  \end{array} \right) 
\label{eqn:10101_3}
\end{eqnarray}

  \item $ A_{\mathbf{6}_{\rm F}} \rightarrow  B_{ \mathbf{\bar3}_{\rm F}}+C_{\bf 8_{\rm F}}$ 
  
\begin{eqnarray}
\left( \begin{array}{c} \Omega_b \\ \\ \Sigma_b \\ \\\Xi'_b \end{array} \right) &\rightarrow &
\left( \begin{array}{cccc} &\Xi_b K ^* & \\ \\
\Xi_b K^* & \Lambda_b \rho &  \\ \\
\Lambda_b K^* & \Xi_b \rho & \Xi_b \omega 
\end{array} \right) 
\nonumber\\
&=& \left( \begin{array}{cccc} & \frac{1}{3} & \\  \\
\frac{1 }{6} & \frac{1}{3} &  \\ \\
\frac{1 }{12} & \frac{1}{8} & \frac{1}{24} \\  
\end{array} \right) 
\end{eqnarray}

\item $ A_{\mathbf{6}_{\rm F}} \rightarrow  B_{\mathbf{\bar3}_{\rm F}}+C_{\bf 1_{\rm F}}$ 

\begin{eqnarray}
\left(   \Xi'_b  \right) \rightarrow 
\left(    \Xi_b \phi  \right) = 
 \left(  \frac{1}{12}  \right) 
\label{eqn:10101_4}
\end{eqnarray}

\end{itemize}

\subsection{Light baryons and bottom-(pseudoscalar/vector) mesons}
We give the $\mathcal{F}^2_{A\rightarrow BC}$ when the final states have a light baryon and a charm-(pseudoscalar/vector) meson. Since the mesons $D^0$ and $D^+$ form an isospin doublet, both are treated as $D$ in the tables; whereas $D_s$ is separated by the strangeness content. The subindexes   $ \mathbf{\bar 3}_{\rm F} $ and ${\bf 6}_{\rm F}$  refer to the anti-triplet and the sextet baryon multiples for the initial bottom baryon $A$, whereas the final $B$ baryons can have subindexes $\bf 8$ or $\bf 10$, according to whether the final light baryon belongs to the octet or decuplet baryon multiplets. 
Additionally, owing to the symmetry of the wave functions of the octet-light baryons, we can have only $\rho$ or $\lambda$ contributions in the final states, as indicated by a superindex.

\begin{itemize}

     \item $A_{\mathbf{\bar 3}_{\rm F}} \rightarrow  B_{\bf 8_{\rm F}}+C$ 
    
    \begin{eqnarray}\left( \begin{array}{c}
 \Lambda_b \\ \\ \Xi_b \end{array} \right) &\rightarrow &
\left( \begin{array}{cccc} 
N^\rho B & \Lambda^\rho_8B_s &  \\ \\
\Sigma^\rho_8 B & \Xi^\rho_8 B_s & \Lambda^\rho_8 B
\end{array} \right) 
\nonumber\\
&=& \left( \begin{array}{cccc} 
\frac{2 }{3} & \frac{2}{9} & \\ \\
\frac{1 }{2} & \frac{1}{3} & \frac{1}{18} \\  
\end{array} \right) 
\end{eqnarray}

    \item $A_{\mathbf{6}_{\rm F}} \rightarrow  B_{\bf 10_{\rm F}}+C$ 
    
    \begin{eqnarray}
\left( \begin{array}{c} \Omega_b \\ \\ \Sigma_b \\ \\ \Xi'_b \end{array} \right) &\rightarrow &
\left( \begin{array}{cccc} \Xi^*_{10}B&\Omega_{10} B_s  \\ \\
\Delta B & \Sigma^*_{10}B_s  \\ \\
\Sigma^*_{10} B & \Xi^*_{10} B_s 
\end{array} \right) 
= \left( \begin{array}{cccc} \frac{2}{9}& \frac{1}{3}  \\  \\
\frac{4 }{9} & \frac{1}{9} \\ \\
\frac{1 }{3} & \frac{2}{9}  \\  
\end{array} \right) 
\end{eqnarray}
    
       \item $A_{\mathbf{6}_{\rm F}} \rightarrow  B_{\bf 8_{\rm F}}+C$ 
    
    \begin{eqnarray}
\left( \begin{array}{c} \Omega_b \\ \\ \Sigma_b \\ \\ \Xi'_b \end{array} \right) &\rightarrow &
\left( \begin{array}{cccc} &\Xi^\lambda_8 B \\ \\
N^\lambda B & \Sigma^\lambda_8B_s  \\ \\
\Sigma^\lambda_8 B & \Xi^\lambda_8 B_s 
\end{array} \right) 
= \left( \begin{array}{cccc} & \frac{4}{9} & \\  \\
\frac{2 }{9} & \frac{2}{9} & \\ \\
\frac{1 }{6} & \frac{1}{9} & \\  
\end{array} \right) 
\end{eqnarray}
\end{itemize}

\section{The strong partial decay widths}
\label{partialdw}
The strong partial decay widths, $\Gamma_{\rm Strong}(A \rightarrow BC)$, of an initial baryon $A$ decaying into a final baryon $B$ plus a meson $C$, in all the open-flavor channels, are shown in Tables~\ref{tab:part_dec_lambdas}-\ref{tab:part_dec_Omega}. Here, we give the contribution of the isospin channels. The charge channel decay width for a baryon  $A$ with isospin $I_A$  and isospin projection $M_{I_A}$, $|A,I_A,M_{I_A}\rangle$, decaying 
into a baryon  $B$ with isospin $I_B$  and isospin projection $M_{I_B}$, $|B,I_B,M_{I_B}\rangle$, and a meson   $C$ with isospin $I_C$  and isospin projection $M_{I_C}$, $|C,I_C,M_{I_C}\rangle$, can be obtained as follows 
\begin{align}
 &  \Gamma_{\rm Strong}^{M_{I_A},M_{I_B},M_{I_C}}(A \rightarrow BC)=&\nonumber\\ 
 &\langle I_B, M_{I_B},I_C, M_{I_C}|I_A, M_{I_A}\rangle^2\Gamma_{\rm Strong}(A \rightarrow BC), &
\end{align}
where  $\langle I_B, M_{I_B},I_C, M_{I_C}|I_A, M_{I_A}\rangle$ is the Clebsch-Gordan coefficient in the isospin space, and the partial decay width   $\Gamma_{\rm Strong}(A \rightarrow BC)$ can be extracted from Tables~\ref{tab:part_dec_lambdas}-\ref{tab:part_dec_Omega}.

\begin{turnpage}
\begin{table*}[htbp]
\caption{$\Lambda_b(nnb)$ strong  partial  decay widths.
The flavor multiplet is denoted by the symbol $\mathcal{F}$. The first column reports the baryon name with its predicted  mass, calculated using the three-quark model Hamiltonian given by Eqs.~\ref{MassFormula} and \ref{eq:Hho}. The second column displays $\bf J^{\rm P}$, the third column shows the three-quark model state, $\left| l_{\lambda},l_{\rho}, k_{\lambda},k_{\rho}\right\rangle$, where $l_{\lambda,\rho}$ represent the orbital angular momenta and $k_{\lambda,\rho}$ denote the number of nodes of the $\lambda$ and $\rho$ oscillators.
The fourth column presents the spectroscopic notation $^{2S+1}L_J$.  The value of $N=n_\rho+n_\lambda$ distinguishes the $N=0,1,2$ energy bands. Starting from the fifth column, we provide the strong partial decay widths calculated using Eq.~\ref{gamma}. Each column corresponds to an open flavor strong decay channel, and the specific decay channels are indicated at the top of each column. The masses of the decay products are given in Table \ref {tab:exp_dec}. The values for the strong decay widths are given in MeV. The decay widths denoted by 0  are forbidden by phase space, while the decay widths denoted by $-$ are forbidden by selection rules.
Finally, the last column represents the sum of the strong partial  decay widths over all the decay channels.}
{\scriptsize
\begingroup
\setlength{\tabcolsep}{1.75pt} 
\renewcommand{\arraystretch}{1.35} 

\begin{tabular}{c c c c |p{0.58cm}p{0.58cm}p{0.58cm}p{0.58cm}p{0.58cm}p{0.58cm}p{0.58cm}p{0.58cm}p{0.58cm}p{0.58cm}p{0.58cm}p{0.58cm}p{0.58cm}p{0.58cm}p{0.58cm}p{0.58cm}p{0.58cm}p{0.75cm}} \hline \hline
$\mathcal{F}={\bf {\bar{3}}}_{\rm F}$  &    &    &    & $\Sigma_{b} \pi$  & $\Sigma^{*}_{b} \pi$  & $\Lambda_{b} \eta$  & $\Sigma_{b}\rho$  & $\Sigma_{b}^{*}\rho$  & $\Lambda_{b}\eta'$  & $\Lambda_{b}\omega$  & $\Xi_{b} K$  & $\Xi'_{b} K$  & $\Xi^{*}_{b} K$  & $\Xi_{b} K^{*}$  & $\Xi'_{b} K^{*}$  & $\Xi^{*}_{b} K^{*}$  & $N B$  & $\Gamma^{\rm Strong}$  \\
$\Lambda_b(nnb)$  & $ {\bf J}^P $  & $\vert l_{\lambda}, l_{\rho}, k_{\lambda}, k_{\rho} \rangle$  & $^{2S+1}L_{J}$  & MeV  & MeV  & MeV  & MeV  & MeV  & MeV  & MeV  & MeV  & MeV  & MeV  & MeV  & MeV  & MeV  & MeV  & \\
\hline
 $N=0$  &  &  &  &  &  \\
$\Lambda_b(5613)$  & ${\bf \frac{1}{2}}^+$ & $\vert \,0\,,\,0\,,\,0\,,\,0 \,\rangle $ &$^{2}S_{1/2}$&$0$   &$0$   &$0$   &$0$   &$0$   &$0$   &$0$   &$0$   &$0$   &$0$   &$0$   &$0$   &$0$   &$0$   &$0$  \\
\hline
 $N=1$  &  &  &  &  &  \\
$\Lambda_b(5918)$  & ${\bf \frac{1}{2}}^-$ & $\vert \,1\,,\,0\,,\,0\,,\,0 \,\rangle $ &$^{2}P_{1/2}$&$0$   &$0$   &$0$   &$0$   &$0$   &$0$   &$0$   &$0$   &$0$   &$0$   &$0$   &$0$   &$0$   &$0$   &$0$  \\
$\Lambda_b(5924)$  & ${\bf \frac{3}{2}}^-$ & $\vert \,1\,,\,0\,,\,0\,,\,0 \,\rangle $ &$^{2}P_{3/2}$&$0$   &$0$   &$0$   &$0$   &$0$   &$0$   &$0$   &$0$   &$0$   &$0$   &$0$   &$0$   &$0$   &$0$   &$0$  \\
$\Lambda_b(6114)$  & ${\bf \frac{1}{2}}^-$ & $\vert \,0\,,\,1\,,\,0\,,\,0 \,\rangle $ &$^{2}P_{1/2}$&9.3   &57.5   &$0$   &$0$   &$0$   &$0$   &$0$   &$0$   &$0$   &$0$   &$0$   &$0$   &$0$   &$0$   &66.8  \\
$\Lambda_b(6137)$  & ${\bf \frac{1}{2}}^-$ & $\vert \,0\,,\,1\,,\,0\,,\,0 \,\rangle $ &$^{4}P_{1/2}$&4.2   &31.3   &$0$   &$0$   &$0$   &$0$   &$0$   &$0$   &$0$   &$0$   &$0$   &$0$   &$0$   &$0$   &35.5  \\
$\Lambda_b(6121)$  & ${\bf \frac{3}{2}}^-$ & $\vert \,0\,,\,1\,,\,0\,,\,0 \,\rangle $ &$^{2}P_{3/2}$&76.9   &7.8   &$0$   &$0$   &$0$   &$0$   &$0$   &$0$   &$0$   &$0$   &$0$   &$0$   &$0$   &$0$   &84.7  \\
$\Lambda_b(6143)$  & ${\bf \frac{3}{2}}^-$ & $\vert \,0\,,\,1\,,\,0\,,\,0 \,\rangle $ &$^{4}P_{3/2}$&4.2   &123.6   &$0$   &$0$   &$0$   &$0$   &$0$   &$0$   &$0$   &$0$   &$0$   &$0$   &$0$   &$0$   &127.8  \\
$\Lambda_b(6153)$  & ${\bf \frac{5}{2}}^-$ & $\vert \,0\,,\,1\,,\,0\,,\,0 \,\rangle $ &$^{4}P_{5/2}$&26.4   &47.9   &$0$   &$0$   &$0$   &$0$   &$0$   &$0$   &$0$   &$0$   &$0$   &$0$   &$0$   &$0$   &74.3  \\
\hline
 $N=2$  &  &  &  &  &  \\
$\Lambda_b(6225)$  & ${\bf \frac{3}{2}}^+$ & $\vert \,2\,,\,0\,,\,0\,,\,0 \,\rangle $ &$^{2}D_{3/2}$&1.4   &8.3   &-   &$0$   &$0$   &$0$   &$0$   &$0$   &$0$   &$0$   &$0$   &$0$   &$0$   &3.3   &13.0  \\
$\Lambda_b(6235)$  & ${\bf \frac{5}{2}}^+$ & $\vert \,2\,,\,0\,,\,0\,,\,0 \,\rangle $ &$^{2}D_{5/2}$&3.1   &1.2   &-   &$0$   &$0$   &$0$   &$0$   &$0$   &$0$   &$0$   &$0$   &$0$   &$0$   &13.2   &17.5  \\
$\Lambda_b(6231)$  & ${\bf \frac{1}{2}}^+$ & $\vert \,0\,,\,0\,,\,1\,,\,0 \,\rangle $ &$^{2}S_{1/2}$&0.2   &0.4   &-   &$0$   &$0$   &$0$   &$0$   &$0$   &$0$   &$0$   &$0$   &$0$   &$0$   &0.5   &1.1  \\
$\Lambda_b(6624)$  & ${\bf \frac{1}{2}}^+$ & $\vert \,0\,,\,0\,,\,0\,,\,1 \,\rangle $ &$^{2}S_{1/2}$&0.3   &0.8   &-   &0.9   &0.1   &-   &1.6   &-   &0.4   &0.2   &$0$   &$0$   &$0$   &-   &4.3  \\
$\Lambda_b(6421)$  & ${\bf \frac{3}{2}}^+$ & $\vert \,1\,,\,1\,,\,0\,,\,0 \,\rangle $ &$^{2}D_{3/2}$&11.4   &48.7   &1.3   &$0$   &$0$   &$0$   &3.2   &2.5   &$0$   &$0$   &$0$   &$0$   &$0$   &-   &67.1  \\
$\Lambda_b(6431)$  & ${\bf \frac{5}{2}}^+$ & $\vert \,1\,,\,1\,,\,0\,,\,0 \,\rangle $ &$^{2}D_{5/2}$&81.4   &8.0   &8.3   &$0$   &$0$   &$0$   &0.7   &9.4   &0.3   &$0$   &$0$   &$0$   &$0$   &-   &108.1  \\
$\Lambda_b(6438)$  & ${\bf \frac{1}{2}}^+$ & $\vert \,1\,,\,1\,,\,0\,,\,0 \,\rangle $ &$^{4}D_{1/2}$&2.2   &23.5   &1.1   &$0$   &$0$   &$0$   &3.3   &4.3   &0.1   &$0$   &$0$   &$0$   &$0$   &-   &34.5  \\
$\Lambda_b(6444)$  & ${\bf \frac{3}{2}}^+$ & $\vert \,1\,,\,1\,,\,0\,,\,0 \,\rangle $ &$^{4}D_{3/2}$&2.8   &75.5   &1.2   &$0$   &$0$   &$0$   &12.5   &2.9   &0.1   &$0$   &$0$   &$0$   &$0$   &-   &95.0  \\
$\Lambda_b(6454)$  & ${\bf \frac{5}{2}}^+$ & $\vert \,1\,,\,1\,,\,0\,,\,0 \,\rangle $ &$^{4}D_{5/2}$&9.0   &95.1   &3.7   &$0$   &$0$   &$0$   &15.6   &4.5   &0.1   &$0$   &$0$   &$0$   &$0$   &-   &128.0  \\
$\Lambda_b(6468)$  & ${\bf \frac{7}{2}}^+$ & $\vert \,1\,,\,1\,,\,0\,,\,0 \,\rangle $ &$^{4}D_{7/2}$&29.1   &59.4   &12.1   &$0$   &$0$   &$0$   &4.8   &16.3   &0.7   &$0$   &$0$   &$0$   &$0$   &-   &122.4  \\
$\Lambda_b(6423)$  & ${\bf \frac{1}{2}}^-$ & $\vert \,1\,,\,1\,,\,0\,,\,0 \,\rangle $ &$^{2}P_{1/2}$&-   &0.5   &-   &$0$   &$0$   &$0$   &-   &-   &$0$   &$0$   &$0$   &$0$   &$0$   &-   &0.5  \\
$\Lambda_b(6429)$  & ${\bf \frac{3}{2}}^-$ & $\vert \,1\,,\,1\,,\,0\,,\,0 \,\rangle $ &$^{2}P_{3/2}$&1.2   &0.3   &0.1   &$0$   &$0$   &$0$   &-   &0.1   &$0$   &$0$   &$0$   &$0$   &$0$   &-   &1.7  \\
$\Lambda_b(6446)$  & ${\bf \frac{1}{2}}^-$ & $\vert \,1\,,\,1\,,\,0\,,\,0 \,\rangle $ &$^{4}P_{1/2}$&-   &0.3   &-   &$0$   &$0$   &$0$   &-   &-   &-   &$0$   &$0$   &$0$   &$0$   &-   &0.3  \\
$\Lambda_b(6452)$  & ${\bf \frac{3}{2}}^-$ & $\vert \,1\,,\,1\,,\,0\,,\,0 \,\rangle $ &$^{4}P_{3/2}$&0.1   &1.0   &-   &$0$   &$0$   &$0$   &0.1   &-   &-   &$0$   &$0$   &$0$   &$0$   &-   &1.2  \\
$\Lambda_b(6462)$  & ${\bf \frac{5}{2}}^-$ & $\vert \,1\,,\,1\,,\,0\,,\,0 \,\rangle $ &$^{4}P_{5/2}$&0.4   &1.3   &0.2   &$0$   &$0$   &$0$   &0.2   &0.2   &-   &$0$   &$0$   &$0$   &$0$   &-   &2.3  \\
$\Lambda_b(6456)$  & ${\bf \frac{3}{2}}^+$ & $\vert \,1\,,\,1\,,\,0\,,\,0 \,\rangle $ &$^{4}S_{3/2}$&2.3   &14.1   &1.1   &$0$   &$0$   &$0$   &8.2   &5.8   &0.3   &$0$   &$0$   &$0$   &$0$   &-   &31.8  \\
$\Lambda_b(6427)$  & ${\bf \frac{1}{2}}^+$ & $\vert \,1\,,\,1\,,\,0\,,\,0 \,\rangle $ &$^{2}S_{1/2}$&12.2   &7.0   &1.5   &$0$   &$0$   &$0$   &3.5   &5.2   &$0$   &$0$   &$0$   &$0$   &$0$   &-   &29.4  \\
$\Lambda_b(6618)$  & ${\bf \frac{3}{2}}^+$ & $\vert \,0\,,\,2\,,\,0\,,\,0 \,\rangle $ &$^{2}D_{3/2}$&21.5   &53.1   &-   &9.3   &0.1   &-   &40.7   &-   &3.1   &3.6   &$0$   &$0$   &$0$   &-   &131.4  \\
$\Lambda_b(6628)$  & ${\bf \frac{5}{2}}^+$ & $\vert \,0\,,\,2\,,\,0\,,\,0 \,\rangle $ &$^{2}D_{5/2}$&52.9   &93.0   &-   &1.0   &1.1   &-   &29.2   &-   &7.5   &0.5   &$0$   &$0$   &$0$   &-   &185.2  \\
\hline \hline
\end{tabular}

\endgroup
}
\label{tab:part_dec_lambdas}
\end{table*}
\end{turnpage}

\begin{turnpage}
\begin{table*}[htbp]
\caption{Same as \ref{tab:part_dec_lambdas}, but for $\Xi_b(snb)$ states. The order of the states is the same as in Table \ref{tab:All_mass_Xi}. The predicted masses, reported in Table \ref{tab:All_mass_Xi}, are obtained by the three-quark model Hamiltonian of Eqs. \ref{MassFormula} and \ref{eq:Hho}. }
{\scriptsize
\begingroup
\setlength{\tabcolsep}{1.75pt} 
\renewcommand{\arraystretch}{1.35} 

\begin{tabular}{c c c c |p{0.58cm}p{0.58cm}p{0.58cm}p{0.58cm}p{0.58cm}p{0.58cm}p{0.58cm}p{0.58cm}p{0.58cm}p{0.58cm}p{0.58cm}p{0.58cm}p{0.58cm}p{0.58cm}p{0.58cm}p{0.58cm}p{0.58cm}p{0.58cm}p{0.58cm}p{0.58cm}p{0.58cm}p{0.58cm}p{0.58cm}p{0.58cm}p{0.58cm}p{0.58cm}p{0.58cm}p{0.58cm}p{0.58cm}p{0.58cm}p{0.58cm}p{0.75cm}} \hline \hline
$\mathcal{F}={\bf {\bar{3}}}_{\rm F}$  &    &    &    & $\Lambda_{b} K$  & $\Xi_{b} \pi$  & $\Xi'_{b} \pi$  & $\Xi^{*}_{b} \pi$  & $\Sigma_{b} K$  & $\Sigma^{*}_{b} K$  & $\Xi_{b} \eta$  & $\Lambda_{b} K^{*}$  & $\Xi_{b} \rho$  & $\Xi'_{b} \rho$  & $\Xi^{*}_{b} \rho$  & $\Sigma_{b} K^{*}$  & $\Sigma^{*}_{b} K^{*}$  & $\Xi'_{b} \eta$  & $\Xi^{*}_{b} \eta$  & $\Xi_{b} \eta'$  & $\Xi'_{b} \eta'$  & $\Xi^{*}_{b} \eta'$  & $\Xi_{b} \omega$  & $\Xi'_{b} \omega$  & $\Xi^{*}_{b} \omega$  & $\Xi_{b} \phi$  & $\Xi'_{b} \phi$  & $\Xi^{*}_{b} \phi$  & $\Lambda_{8} B$  & $\Lambda_{8} B^{*}$  & $\Sigma_{8} B$  & $\Lambda_{8}^{*} B$  & $\Gamma^{\rm Strong}$   \\
$\Xi_b(snb)$  & $ {\bf J}^P $  & $\vert l_{\lambda}, l_{\rho}, k_{\lambda}, k_{\rho} \rangle$  & $^{2S+1}L_{J}$  & MeV  & MeV  & MeV  & MeV  & MeV  & MeV  & MeV  & MeV  & MeV  & MeV  & MeV  & MeV  & MeV  & MeV  & MeV  & MeV  & MeV  & MeV  & MeV  & MeV  & MeV  & MeV  & MeV  & MeV  & MeV  & MeV  & MeV  & MeV  & \\
\hline
 $N=0$  &  &  &  &  &  \\
$\Xi_b(5806)$  & ${\bf \frac{1}{2}}^+$ & $\vert \,0\,,\,0\,,\,0\,,\,0 \,\rangle $ &$^{2}S_{1/2}$&$0$   &$0$   &$0$   &$0$   &$0$   &$0$   &$0$   &$0$   &$0$   &$0$   &$0$   &$0$   &$0$   &$0$   &$0$   &$0$   &$0$   &$0$   &$0$   &$0$   &$0$   &$0$   &$0$   &$0$   &$0$   &$0$   &$0$   &$0$   &$0$  \\
\hline
 $N=1$  &  &  &  &  &  \\
$\Xi_b(6079)$  & ${\bf \frac{1}{2}}^-$ & $\vert \,1\,,\,0\,,\,0\,,\,0 \,\rangle $ &$^{2}P_{1/2}$&$0$   &-   &0.2   &$0$   &$0$   &$0$   &$0$   &$0$   &$0$   &$0$   &$0$   &$0$   &$0$   &$0$   &$0$   &$0$   &$0$   &$0$   &$0$   &$0$   &$0$   &$0$   &$0$   &$0$   &$0$   &$0$   &$0$   &$0$   &0.2  \\
$\Xi_b(6085)$  & ${\bf \frac{3}{2}}^-$ & $\vert \,1\,,\,0\,,\,0\,,\,0 \,\rangle $ &$^{2}P_{3/2}$&$0$   &-   &1.1   &$0$   &$0$   &$0$   &$0$   &$0$   &$0$   &$0$   &$0$   &$0$   &$0$   &$0$   &$0$   &$0$   &$0$   &$0$   &$0$   &$0$   &$0$   &$0$   &$0$   &$0$   &$0$   &$0$   &$0$   &$0$   &1.1  \\
$\Xi_b(6248)$  & ${\bf \frac{1}{2}}^-$ & $\vert \,0\,,\,1\,,\,0\,,\,0 \,\rangle $ &$^{2}P_{1/2}$&1.0   &0.4   &2.5   &4.8   &$0$   &$0$   &$0$   &$0$   &$0$   &$0$   &$0$   &$0$   &$0$   &$0$   &$0$   &$0$   &$0$   &$0$   &$0$   &$0$   &$0$   &$0$   &$0$   &$0$   &$0$   &$0$   &$0$   &$0$   &8.7  \\
$\Xi_b(6271)$  & ${\bf \frac{1}{2}}^-$ & $\vert \,0\,,\,1\,,\,0\,,\,0 \,\rangle $ &$^{4}P_{1/2}$&1.1   &0.3   &1.1   &3.5   &$0$   &$0$   &$0$   &$0$   &$0$   &$0$   &$0$   &$0$   &$0$   &$0$   &$0$   &$0$   &$0$   &$0$   &$0$   &$0$   &$0$   &$0$   &$0$   &$0$   &$0$   &$0$   &$0$   &$0$   &6.0  \\
$\Xi_b(6255)$  & ${\bf \frac{3}{2}}^-$ & $\vert \,0\,,\,1\,,\,0\,,\,0 \,\rangle $ &$^{2}P_{3/2}$&21.1   &23.7   &20.5   &0.6   &$0$   &$0$   &$0$   &$0$   &$0$   &$0$   &$0$   &$0$   &$0$   &$0$   &$0$   &$0$   &$0$   &$0$   &$0$   &$0$   &$0$   &$0$   &$0$   &$0$   &$0$   &$0$   &$0$   &$0$   &65.9  \\
$\Xi_b(6277)$  & ${\bf \frac{3}{2}}^-$ & $\vert \,0\,,\,1\,,\,0\,,\,0 \,\rangle $ &$^{4}P_{3/2}$&4.4   &5.0   &1.1   &15.6   &$0$   &$0$   &$0$   &$0$   &$0$   &$0$   &$0$   &$0$   &$0$   &$0$   &$0$   &$0$   &$0$   &$0$   &$0$   &$0$   &$0$   &$0$   &$0$   &$0$   &$0$   &$0$   &$0$   &$0$   &26.1  \\
$\Xi_b(6287)$  & ${\bf \frac{5}{2}}^-$ & $\vert \,0\,,\,1\,,\,0\,,\,0 \,\rangle $ &$^{4}P_{5/2}$&26.9   &30.5   &7.0   &4.1   &$0$   &$0$   &$0$   &$0$   &$0$   &$0$   &$0$   &$0$   &$0$   &$0$   &$0$   &$0$   &$0$   &$0$   &$0$   &$0$   &$0$   &$0$   &$0$   &$0$   &$0$   &$0$   &$0$   &$0$   &68.5  \\
\hline
 $N=2$  &  &  &  &  &  \\
$\Xi_b(6354)$  & ${\bf \frac{3}{2}}^+$ & $\vert \,2\,,\,0\,,\,0\,,\,0 \,\rangle $ &$^{2}D_{3/2}$&-   &-   &0.3   &0.5   &0.2   &0.8   &-   &$0$   &$0$   &$0$   &$0$   &$0$   &$0$   &$0$   &$0$   &$0$   &$0$   &$0$   &$0$   &$0$   &$0$   &$0$   &$0$   &$0$   &$0$   &$0$   &$0$   &$0$   &1.8  \\
$\Xi_b(6364)$  & ${\bf \frac{5}{2}}^+$ & $\vert \,2\,,\,0\,,\,0\,,\,0 \,\rangle $ &$^{2}D_{5/2}$&-   &-   &0.8   &0.1   &0.6   &0.1   &-   &$0$   &$0$   &$0$   &$0$   &$0$   &$0$   &$0$   &$0$   &$0$   &$0$   &$0$   &$0$   &$0$   &$0$   &$0$   &$0$   &$0$   &$0$   &$0$   &$0$   &$0$   &1.6  \\
$\Xi_b(6360)$  & ${\bf \frac{1}{2}}^+$ & $\vert \,0\,,\,0\,,\,1\,,\,0 \,\rangle $ &$^{2}S_{1/2}$&-   &-   &0.1   &-   &-   &-   &-   &$0$   &$0$   &$0$   &$0$   &$0$   &$0$   &$0$   &$0$   &$0$   &$0$   &$0$   &$0$   &$0$   &$0$   &$0$   &$0$   &$0$   &$0$   &$0$   &$0$   &$0$   &0.1  \\
$\Xi_b(6699)$  & ${\bf \frac{1}{2}}^+$ & $\vert \,0\,,\,0\,,\,0\,,\,1 \,\rangle $ &$^{2}S_{1/2}$&-   &-   &0.2   &0.7   &0.6   &1.4   &-   &1.1   &1.3   &$0$   &$0$   &$0$   &$0$   &-   &-   &$0$   &$0$   &$0$   &0.4   &$0$   &$0$   &$0$   &$0$   &$0$   &-   &-   &-   &$0$   &5.7  \\
$\Xi_b(6524)$  & ${\bf \frac{3}{2}}^+$ & $\vert \,1\,,\,1\,,\,0\,,\,0 \,\rangle $ &$^{2}D_{3/2}$&1.6   &2.1   &3.1   &6.4   &7.7   &23.6   &0.3   &1.0   &$0$   &$0$   &$0$   &$0$   &$0$   &0.1   &$0$   &$0$   &$0$   &$0$   &$0$   &$0$   &$0$   &$0$   &$0$   &$0$   &-   &-   &-   &$0$   &45.9  \\
$\Xi_b(6534)$  & ${\bf \frac{5}{2}}^+$ & $\vert \,1\,,\,1\,,\,0\,,\,0 \,\rangle $ &$^{2}D_{5/2}$&21.2   &24.4   &19.3   &0.6   &37.6   &2.3   &1.4   &0.3   &$0$   &$0$   &$0$   &$0$   &$0$   &0.3   &$0$   &$0$   &$0$   &$0$   &$0$   &$0$   &$0$   &$0$   &$0$   &$0$   &-   &-   &-   &$0$   &107.4  \\
$\Xi_b(6540)$  & ${\bf \frac{1}{2}}^+$ & $\vert \,1\,,\,1\,,\,0\,,\,0 \,\rangle $ &$^{4}D_{1/2}$&0.1   &0.5   &0.9   &3.1   &2.7   &11.1   &0.5   &1.3   &$0$   &$0$   &$0$   &$0$   &$0$   &0.1   &$0$   &$0$   &$0$   &$0$   &$0$   &$0$   &$0$   &$0$   &$0$   &$0$   &-   &-   &-   &$0$   &20.3  \\
$\Xi_b(6546)$  & ${\bf \frac{3}{2}}^+$ & $\vert \,1\,,\,1\,,\,0\,,\,0 \,\rangle $ &$^{4}D_{3/2}$&1.6   &2.0   &0.8   &12.2   &2.0   &42.7   &0.3   &5.3   &$0$   &$0$   &$0$   &$0$   &$0$   &-   &$0$   &$0$   &$0$   &$0$   &$0$   &$0$   &$0$   &$0$   &$0$   &$0$   &-   &-   &-   &$0$   &66.9  \\
$\Xi_b(6556)$  & ${\bf \frac{5}{2}}^+$ & $\vert \,1\,,\,1\,,\,0\,,\,0 \,\rangle $ &$^{4}D_{5/2}$&8.9   &10.2   &2.0   &14.8   &4.0   &52.3   &0.6   &7.3   &$0$   &$0$   &$0$   &$0$   &$0$   &-   &$0$   &$0$   &$0$   &$0$   &$0$   &$0$   &$0$   &$0$   &$0$   &$0$   &-   &-   &-   &$0$   &100.1  \\
$\Xi_b(6570)$  & ${\bf \frac{7}{2}}^+$ & $\vert \,1\,,\,1\,,\,0\,,\,0 \,\rangle $ &$^{4}D_{7/2}$&29.8   &34.2   &7.0   &4.3   &14.4   &19.5   &2.2   &2.2   &$0$   &$0$   &$0$   &$0$   &$0$   &0.2   &$0$   &$0$   &$0$   &$0$   &$0$   &$0$   &$0$   &$0$   &$0$   &$0$   &-   &-   &-   &$0$   &113.8  \\
$\Xi_b(6526)$  & ${\bf \frac{1}{2}}^-$ & $\vert \,1\,,\,1\,,\,0\,,\,0 \,\rangle $ &$^{2}P_{1/2}$&-   &-   &-   &0.1   &-   &0.2   &-   &-   &$0$   &$0$   &$0$   &$0$   &$0$   &-   &$0$   &$0$   &$0$   &$0$   &$0$   &$0$   &$0$   &$0$   &$0$   &$0$   &-   &-   &-   &$0$   &0.3  \\
$\Xi_b(6532)$  & ${\bf \frac{3}{2}}^-$ & $\vert \,1\,,\,1\,,\,0\,,\,0 \,\rangle $ &$^{2}P_{3/2}$&0.4   &0.5   &0.3   &-   &0.6   &0.1   &-   &-   &$0$   &$0$   &$0$   &$0$   &$0$   &-   &$0$   &$0$   &$0$   &$0$   &$0$   &$0$   &$0$   &$0$   &$0$   &$0$   &-   &-   &-   &$0$   &1.9  \\
$\Xi_b(6548)$  & ${\bf \frac{1}{2}}^-$ & $\vert \,1\,,\,1\,,\,0\,,\,0 \,\rangle $ &$^{4}P_{1/2}$&-   &-   &-   &-   &-   &0.1   &-   &-   &$0$   &$0$   &$0$   &$0$   &$0$   &-   &$0$   &$0$   &$0$   &$0$   &$0$   &$0$   &$0$   &$0$   &$0$   &$0$   &-   &-   &-   &$0$   &0.1  \\
$\Xi_b(6554)$  & ${\bf \frac{3}{2}}^-$ & $\vert \,1\,,\,1\,,\,0\,,\,0 \,\rangle $ &$^{4}P_{3/2}$&0.1   &0.1   &-   &0.1   &-   &0.5   &-   &0.1   &$0$   &$0$   &$0$   &$0$   &$0$   &-   &$0$   &$0$   &$0$   &$0$   &$0$   &$0$   &$0$   &$0$   &$0$   &$0$   &-   &-   &-   &$0$   &0.9  \\
$\Xi_b(6564)$  & ${\bf \frac{5}{2}}^-$ & $\vert \,1\,,\,1\,,\,0\,,\,0 \,\rangle $ &$^{4}P_{5/2}$&0.5   &0.6   &0.1   &0.2   &0.2   &0.7   &-   &0.1   &$0$   &$0$   &$0$   &$0$   &$0$   &-   &$0$   &$0$   &$0$   &$0$   &$0$   &$0$   &$0$   &$0$   &$0$   &$0$   &-   &-   &-   &$0$   &2.4  \\
$\Xi_b(6558)$  & ${\bf \frac{3}{2}}^+$ & $\vert \,1\,,\,1\,,\,0\,,\,0 \,\rangle $ &$^{4}S_{3/2}$&-   &0.3   &1.1   &6.1   &3.3   &17.2   &0.6   &4.0   &$0$   &$0$   &$0$   &$0$   &$0$   &0.1   &$0$   &$0$   &$0$   &$0$   &$0$   &$0$   &$0$   &$0$   &$0$   &$0$   &-   &-   &-   &$0$   &32.7  \\
$\Xi_b(6530)$  & ${\bf \frac{1}{2}}^+$ & $\vert \,1\,,\,1\,,\,0\,,\,0 \,\rangle $ &$^{2}S_{1/2}$&0.2   &0.9   &4.8   &2.2   &13.9   &7.0   &0.6   &1.3   &$0$   &$0$   &$0$   &$0$   &$0$   &0.2   &$0$   &$0$   &$0$   &$0$   &$0$   &$0$   &$0$   &$0$   &$0$   &$0$   &-   &-   &-   &$0$   &31.1  \\
$\Xi_b(6693)$  & ${\bf \frac{3}{2}}^+$ & $\vert \,0\,,\,2\,,\,0\,,\,0 \,\rangle $ &$^{2}D_{3/2}$&-   &-   &4.2   &17.3   &9.1   &42.9   &-   &22.8   &22.9   &$0$   &$0$   &$0$   &$0$   &0.3   &0.6   &$0$   &$0$   &$0$   &7.3   &$0$   &$0$   &$0$   &$0$   &$0$   &-   &-   &-   &$0$   &127.4  \\
$\Xi_b(6703)$  & ${\bf \frac{5}{2}}^+$ & $\vert \,0\,,\,2\,,\,0\,,\,0 \,\rangle $ &$^{2}D_{5/2}$&-   &-   &11.3   &6.3   &24.9   &30.4   &-   &12.6   &8.4   &$0$   &$0$   &$0$   &$0$   &0.8   &0.1   &$0$   &$0$   &$0$   &2.6   &$0$   &$0$   &$0$   &$0$   &$0$   &-   &-   &-   &$0$   &97.4  \\
\hline \hline
\end{tabular}

\endgroup
}
\label{tab:part_dec_cascades_anti3}
\end{table*}
\end{turnpage}

\begin{turnpage}
\begin{table*}[htbp]
\caption{Same as \ref{tab:part_dec_lambdas}, but for $\Sigma_b(nnb)$ states. The order of the states is the same as in Table \ref{tab:All_mass_Sigma}. The predicted masses, reported in Table \ref{tab:All_mass_Sigma}, are obtained by the three-quark model Hamiltonian of Eqs. \ref{MassFormula} and \ref{eq:Hho}.  $N^*_1$, $N^*_2$, $N^*_3$, and $N^*_4$ represent $N(1520)$, $N(1535)$, $N(1680)$, and $N(1720)$, respectively.}
{\scriptsize
\begingroup
\setlength{\tabcolsep}{1.75pt} 
\renewcommand{\arraystretch}{1.35} 

\begin{tabular}{c c c c |p{0.58cm}p{0.58cm}p{0.58cm}p{0.58cm}p{0.58cm}p{0.58cm}p{0.58cm}p{0.58cm}p{0.58cm}p{0.58cm}p{0.58cm}p{0.58cm}p{0.58cm}p{0.58cm}p{0.58cm}p{0.58cm}p{0.58cm}p{0.58cm}p{0.58cm}p{0.58cm}p{0.58cm}p{0.58cm}p{0.58cm}p{0.58cm}p{0.58cm}p{0.58cm}p{0.58cm}p{0.58cm}p{0.58cm}p{0.75cm}} \hline \hline
$\mathcal{F}={\bf {6}}_{\rm F}$   &    &    &    & $\Sigma_{b} \pi$  & $\Sigma^{*}_{b} \pi$  & $\Lambda_{b} \pi$  & $\Sigma_{b} \eta$  & $\Xi_{b} K$  & $\Sigma_{b}\rho$  & $\Sigma^{*}_{b}\rho$  & $\Lambda_{b}\rho$  & $\Sigma^{*}_{b}\eta$  & $\Sigma_{b}\eta'$  & $\Sigma^{*}_{b}\eta'$  & $\Xi'_{b}K$  & $\Xi^{*}_{b}K$  & $\Xi_{b} K^{*}$  & $\Xi'_{b} K^{*}$  & $\Xi^{*}_{b} K^{*}$  & $\Sigma_{b}\omega$  & $\Sigma^{*}_{b}\omega$  & $N B$  & $\Sigma_{8} B_{s}$  & $N B^{*}$  & $\Delta B$  & $N^{*}_{1} B$  & $N^{*}_{2} B$  & $N^{*}_{3} B$  & $N^{*}_{4} B$  & $\Gamma^{\rm Strong}$   \\
$\Sigma_b(nnb)$  & $ {\bf J}^P $  & $\vert l_{\lambda}, l_{\rho}, k_{\lambda}, k_{\rho} \rangle$  & $^{2S+1}L_{J}$  & MeV  & MeV  & MeV  & MeV  & MeV  & MeV  & MeV  & MeV  & MeV  & MeV  & MeV  & MeV  & MeV  & MeV  & MeV  & MeV  & MeV  & MeV  & MeV  & MeV  & MeV  & MeV  & MeV  & MeV  & MeV  & MeV  & \\
\hline
 $N=0$  &  &  &  &  &  \\
$\Sigma_b(5804)$  & ${\bf \frac{1}{2}}^+$ & $\vert \,0\,,\,0\,,\,0\,,\,0 \,\rangle $ &$^{2}S_{1/2}$&$0$   &$0$   &3.9   &$0$   &$0$   &$0$   &$0$   &$0$   &$0$   &$0$   &$0$   &$0$   &$0$   &$0$   &$0$   &$0$   &$0$   &$0$   &$0$   &$0$   &$0$   &$0$   &$0$   &$0$   &$0$   &$0$   &3.9  \\
$\Sigma_b(5832)$  & ${\bf \frac{3}{2}}^+$ & $\vert \,0\,,\,0\,,\,0\,,\,0 \,\rangle $ &$^{4}S_{3/2}$&$0$   &$0$   &10.0   &$0$   &$0$   &$0$   &$0$   &$0$   &$0$   &$0$   &$0$   &$0$   &$0$   &$0$   &$0$   &$0$   &$0$   &$0$   &$0$   &$0$   &$0$   &$0$   &$0$   &$0$   &$0$   &$0$   &10.0  \\
\hline
 $N=1$  &  &  &  &  &  \\
$\Sigma_b(6108)$  & ${\bf \frac{1}{2}}^-$ & $\vert \,1\,,\,0\,,\,0\,,\,0 \,\rangle $ &$^{2}P_{1/2}$&3.4   &20.1   &-   &$0$   &$0$   &$0$   &$0$   &$0$   &$0$   &$0$   &$0$   &$0$   &$0$   &$0$   &$0$   &$0$   &$0$   &$0$   &$0$   &$0$   &$0$   &$0$   &$0$   &$0$   &$0$   &$0$   &23.5  \\
$\Sigma_b(6131)$  & ${\bf \frac{1}{2}}^-$ & $\vert \,1\,,\,0\,,\,0\,,\,0 \,\rangle $ &$^{4}P_{1/2}$&1.6   &11.1   &0.5   &$0$   &$0$   &$0$   &$0$   &$0$   &$0$   &$0$   &$0$   &$0$   &$0$   &$0$   &$0$   &$0$   &$0$   &$0$   &$0$   &$0$   &$0$   &$0$   &$0$   &$0$   &$0$   &$0$   &13.2  \\
$\Sigma_b(6114)$  & ${\bf \frac{3}{2}}^-$ & $\vert \,1\,,\,0\,,\,0\,,\,0 \,\rangle $ &$^{2}P_{3/2}$&27.2   &2.7   &54.0   &$0$   &$0$   &$0$   &$0$   &$0$   &$0$   &$0$   &$0$   &$0$   &$0$   &$0$   &$0$   &$0$   &$0$   &$0$   &$0$   &$0$   &$0$   &$0$   &$0$   &$0$   &$0$   &$0$   &83.9  \\
$\Sigma_b(6137)$  & ${\bf \frac{3}{2}}^-$ & $\vert \,1\,,\,0\,,\,0\,,\,0 \,\rangle $ &$^{4}P_{3/2}$&1.5   &44.1   &11.3   &$0$   &$0$   &$0$   &$0$   &$0$   &$0$   &$0$   &$0$   &$0$   &$0$   &$0$   &$0$   &$0$   &$0$   &$0$   &$0$   &$0$   &$0$   &$0$   &$0$   &$0$   &$0$   &$0$   &56.9  \\
$\Sigma_b(6147)$  & ${\bf \frac{5}{2}}^-$ & $\vert \,1\,,\,0\,,\,0\,,\,0 \,\rangle $ &$^{4}P_{5/2}$&9.4   &16.7   &69.8   &$0$   &$0$   &$0$   &$0$   &$0$   &$0$   &$0$   &$0$   &$0$   &$0$   &$0$   &$0$   &$0$   &$0$   &$0$   &$0$   &$0$   &$0$   &$0$   &$0$   &$0$   &$0$   &$0$   &95.9  \\
$\Sigma_b(6304)$  & ${\bf \frac{1}{2}}^-$ & $\vert \,0\,,\,1\,,\,0\,,\,0 \,\rangle $ &$^{2}P_{1/2}$&0.1   &134.1   &-   &$0$   &-   &$0$   &$0$   &$0$   &$0$   &$0$   &$0$   &$0$   &$0$   &$0$   &$0$   &$0$   &$0$   &$0$   &-   &$0$   &-   &$0$   &$0$   &$0$   &$0$   &$0$   &134.2  \\
$\Sigma_b(6311)$  & ${\bf \frac{3}{2}}^-$ & $\vert \,0\,,\,1\,,\,0\,,\,0 \,\rangle $ &$^{2}P_{3/2}$&67.1   &62.2   &-   &$0$   &-   &$0$   &$0$   &$0$   &$0$   &$0$   &$0$   &$0$   &$0$   &$0$   &$0$   &$0$   &$0$   &$0$   &-   &$0$   &-   &$0$   &$0$   &$0$   &$0$   &$0$   &129.3  \\
\hline
 $N=2$  &  &  &  &  &  \\
$\Sigma_b(6415)$  & ${\bf \frac{3}{2}}^+$ & $\vert \,2\,,\,0\,,\,0\,,\,0 \,\rangle $ &$^{2}D_{3/2}$&2.7   &4.2   &6.0   &0.1   &0.6   &$0$   &$0$   &1.0   &0.1   &$0$   &$0$   &$0$   &$0$   &$0$   &$0$   &$0$   &$0$   &$0$   &4.3   &$0$   &38.6   &$0$   &$0$   &$0$   &$0$   &$0$   &57.6  \\
$\Sigma_b(6425)$  & ${\bf \frac{5}{2}}^+$ & $\vert \,2\,,\,0\,,\,0\,,\,0 \,\rangle $ &$^{2}D_{5/2}$&7.3   &1.5   &15.1   &0.3   &1.3   &$0$   &$0$   &0.1   &-   &$0$   &$0$   &$0$   &$0$   &$0$   &$0$   &$0$   &$0$   &$0$   &11.0   &$0$   &93.3   &$0$   &$0$   &$0$   &$0$   &$0$   &129.9  \\
$\Sigma_b(6431)$  & ${\bf \frac{1}{2}}^+$ & $\vert \,2\,,\,0\,,\,0\,,\,0 \,\rangle $ &$^{4}D_{1/2}$&0.1   &6.7   &2.2   &-   &0.5   &$0$   &$0$   &2.3   &0.2   &$0$   &$0$   &-   &$0$   &$0$   &$0$   &$0$   &$0$   &$0$   &4.5   &$0$   &61.3   &$0$   &$0$   &$0$   &$0$   &$0$   &77.8  \\
$\Sigma_b(6437)$  & ${\bf \frac{3}{2}}^+$ & $\vert \,2\,,\,0\,,\,0\,,\,0 \,\rangle $ &$^{4}D_{3/2}$&0.7   &7.3   &6.6   &-   &0.7   &$0$   &$0$   &5.0   &0.4   &$0$   &$0$   &-   &$0$   &$0$   &$0$   &$0$   &$0$   &$0$   &18.7   &$0$   &66.5   &$0$   &$0$   &$0$   &$0$   &$0$   &105.9  \\
$\Sigma_b(6448)$  & ${\bf \frac{5}{2}}^+$ & $\vert \,2\,,\,0\,,\,0\,,\,0 \,\rangle $ &$^{4}D_{5/2}$&1.6   &6.6   &12.5   &0.1   &1.2   &$0$   &$0$   &4.1   &0.4   &$0$   &$0$   &-   &$0$   &$0$   &$0$   &$0$   &$0$   &$0$   &39.1   &$0$   &67.1   &$0$   &$0$   &$0$   &$0$   &$0$   &132.7  \\
$\Sigma_b(6462)$  & ${\bf \frac{7}{2}}^+$ & $\vert \,2\,,\,0\,,\,0\,,\,0 \,\rangle $ &$^{4}D_{7/2}$&2.2   &7.8   &18.5   &0.1   &2.0   &$0$   &$0$   &2.5   &0.2   &$0$   &$0$   &0.1   &$0$   &$0$   &$0$   &$0$   &$0$   &$0$   &55.8   &$0$   &56.0   &$0$   &$0$   &$0$   &$0$   &$0$   &145.2  \\
$\Sigma_b(6421)$  & ${\bf \frac{1}{2}}^+$ & $\vert \,0\,,\,0\,,\,1\,,\,0 \,\rangle $ &$^{2}S_{1/2}$&0.3   &0.1   &0.1   &-   &0.1   &$0$   &$0$   &0.1   &-   &$0$   &$0$   &$0$   &$0$   &$0$   &$0$   &$0$   &$0$   &$0$   &0.3   &$0$   &2.9   &$0$   &$0$   &$0$   &$0$   &$0$   &3.9  \\
$\Sigma_b(6450)$  & ${\bf \frac{3}{2}}^+$ & $\vert \,0\,,\,0\,,\,1\,,\,0 \,\rangle $ &$^{4}S_{3/2}$&0.1   &0.3   &0.1   &-   &0.1   &$0$   &$0$   &0.3   &-   &$0$   &$0$   &-   &$0$   &$0$   &$0$   &$0$   &$0$   &$0$   &0.9   &$0$   &2.1   &$0$   &$0$   &$0$   &$0$   &$0$   &3.9  \\
$\Sigma_b(6813)$  & ${\bf \frac{1}{2}}^+$ & $\vert \,0\,,\,0\,,\,0\,,\,1 \,\rangle $ &$^{2}S_{1/2}$&0.2   &0.1   &2.7   &-   &-   &11.6   &0.2   &0.1   &-   &0.2   &0.1   &0.4   &0.3   &1.2   &$0$   &$0$   &6.0   &0.1   &-   &-   &-   &-   &-   &-   &$0$   &$0$   &23.2  \\
$\Sigma_b(6842)$  & ${\bf \frac{3}{2}}^+$ & $\vert \,0\,,\,0\,,\,0\,,\,1 \,\rangle $ &$^{4}S_{3/2}$&0.1   &0.3   &3.3   &-   &-   &0.6   &16.1   &-   &-   &0.1   &0.4   &0.1   &0.7   &1.2   &0.1   &$0$   &0.3   &8.3   &-   &-   &-   &-   &-   &-   &$0$   &$0$   &31.6  \\
$\Sigma_b(6611)$  & ${\bf \frac{3}{2}}^+$ & $\vert \,1\,,\,1\,,\,0\,,\,0 \,\rangle $ &$^{2}D_{3/2}$&7.0   &117.8   &-   &1.3   &-   &9.0   &0.1   &209.2   &20.1   &$0$   &$0$   &3.3   &5.1   &$0$   &$0$   &$0$   &3.3   &$0$   &-   &-   &-   &-   &$0$   &$0$   &$0$   &$0$   &376.2  \\
$\Sigma_b(6621)$  & ${\bf \frac{5}{2}}^+$ & $\vert \,1\,,\,1\,,\,0\,,\,0 \,\rangle $ &$^{2}D_{5/2}$&77.8   &77.0   &-   &8.3   &-   &1.7   &1.1   &66.6   &2.9   &$0$   &$0$   &14.6   &0.5   &$0$   &$0$   &$0$   &0.7   &0.4   &-   &-   &-   &-   &$0$   &$0$   &$0$   &$0$   &251.6  \\
$\Sigma_b(6613)$  & ${\bf \frac{1}{2}}^-$ & $\vert \,1\,,\,1\,,\,0\,,\,0 \,\rangle $ &$^{2}P_{1/2}$&-   &2.2   &-   &-   &-   &-   &-   &1.8   &0.2   &$0$   &$0$   &-   &-   &$0$   &$0$   &$0$   &-   &-   &-   &-   &-   &-   &$0$   &$0$   &$0$   &$0$   &4.2  \\
$\Sigma_b(6619)$  & ${\bf \frac{3}{2}}^-$ & $\vert \,1\,,\,1\,,\,0\,,\,0 \,\rangle $ &$^{2}P_{3/2}$&1.2   &1.1   &-   &0.1   &-   &-   &-   &1.9   &0.1   &$0$   &$0$   &0.2   &-   &$0$   &$0$   &$0$   &-   &-   &-   &-   &-   &-   &$0$   &$0$   &$0$   &$0$   &4.6  \\
$\Sigma_b(6617)$  & ${\bf \frac{1}{2}}^+$ & $\vert \,1\,,\,1\,,\,0\,,\,0 \,\rangle $ &$^{2}S_{1/2}$&2.0   &1.5   &-   &1.5   &-   &9.0   &1.0   &26.2   &3.5   &$0$   &$0$   &6.1   &2.9   &$0$   &$0$   &$0$   &3.5   &0.3   &-   &-   &-   &-   &$0$   &$0$   &$0$   &$0$   &57.5  \\
$\Sigma_b(6807)$  & ${\bf \frac{3}{2}}^+$ & $\vert \,0\,,\,2\,,\,0\,,\,0 \,\rangle $ &$^{2}D_{3/2}$&49.8   &3.7   &120.5   &4.3   &11.8   &213.3   &0.4   &2.2   &2.1   &1.4   &1.2   &8.0   &8.4   &13.6   &$0$   &$0$   &108.3   &0.2   &-   &-   &-   &-   &-   &$0$   &$0$   &$0$   &549.2  \\
$\Sigma_b(6817)$  & ${\bf \frac{5}{2}}^+$ & $\vert \,0\,,\,2\,,\,0\,,\,0 \,\rangle $ &$^{2}D_{5/2}$&85.6   &54.0   &160.6   &10.1   &25.4   &116.2   &1.4   &73.8   &4.7   &3.6   &0.2   &21.8   &2.4   &0.9   &$0$   &$0$   &54.9   &0.7   &-   &-   &-   &-   &-   &-   &$0$   &$0$   &616.3  \\
$\Sigma_b(6824)$  & ${\bf \frac{1}{2}}^+$ & $\vert \,0\,,\,2\,,\,0\,,\,0 \,\rangle $ &$^{4}D_{1/2}$&17.2   &15.6   &213.5   &0.7   &10.6   &13.2   &676.3   &13.9   &5.0   &0.4   &2.8   &0.2   &13.4   &18.5   &$0$   &$0$   &6.6   &340.9   &-   &-   &-   &-   &-   &-   &$0$   &$0$   &1348.8  \\
$\Sigma_b(6830)$  & ${\bf \frac{3}{2}}^+$ & $\vert \,0\,,\,2\,,\,0\,,\,0 \,\rangle $ &$^{4}D_{3/2}$&14.0   &4.2   &130.3   &1.2   &13.5   &25.2   &291.0   &38.8   &2.4   &0.6   &4.9   &2.2   &15.4   &35.7   &$0$   &$0$   &12.7   &148.4   &-   &-   &-   &-   &-   &-   &$0$   &$0$   &740.5  \\
$\Sigma_b(6840)$  & ${\bf \frac{5}{2}}^+$ & $\vert \,0\,,\,2\,,\,0\,,\,0 \,\rangle $ &$^{4}D_{5/2}$&13.6   &75.2   &77.8   &2.0   &19.0   &15.4   &29.5   &66.0   &6.9   &1.0   &5.1   &4.7   &13.4   &22.1   &0.7   &$0$   &7.7   &15.6   &-   &-   &-   &-   &-   &-   &$0$   &$0$   &375.7  \\
$\Sigma_b(6854)$  & ${\bf \frac{7}{2}}^+$ & $\vert \,0\,,\,2\,,\,0\,,\,0 \,\rangle $ &$^{4}D_{7/2}$&27.8   &173.6   &211.6   &3.2   &32.3   &32.3   &235.6   &267.6   &18.6   &1.9   &2.5   &6.6   &14.1   &24.8   &0.6   &$0$   &15.6   &109.8   &-   &-   &-   &-   &-   &-   &$0$   &$0$   &1178.5  \\
\hline \hline
\end{tabular}

\endgroup
}
\label{tab:part_dec_Sigma}
\end{table*}
\end{turnpage}

\begin{turnpage}
\begin{table*}[htbp]
\caption{Same as \ref{tab:part_dec_lambdas}, but for $\Xi'_b(snb)$ states.  The order of the states is the same as in Table \ref{tab:All_mass_Xiprime}. The predicted masses, reported in Table \ref{tab:All_mass_Xiprime},  are obtained by the three-quark model Hamiltonian of Eqs. \ref{MassFormula} and \ref{eq:Hho}.}
{\scriptsize
\begingroup
\setlength{\tabcolsep}{1.75pt} 
\renewcommand{\arraystretch}{1.35} 

\begin{tabular}{c c c c |p{0.58cm}p{0.58cm}p{0.58cm}p{0.58cm}p{0.58cm}p{0.58cm}p{0.58cm}p{0.58cm}p{0.58cm}p{0.58cm}p{0.58cm}p{0.58cm}p{0.58cm}p{0.58cm}p{0.58cm}p{0.58cm}p{0.58cm}p{0.58cm}p{0.58cm}p{0.58cm}p{0.58cm}p{0.58cm}p{0.58cm}p{0.58cm}p{0.58cm}p{0.58cm}p{0.58cm}p{0.58cm}p{0.58cm}p{0.58cm}p{0.58cm}p{0.75cm}} \hline \hline
$\mathcal{F}={\bf {6}}_{\rm F}$   &    &    &    & $\Lambda_{b} K$  & $\Xi_{b} \pi$  & $\Xi'_{b} \pi$  & $\Xi^{*}_{b} \pi$  & $\Sigma_{b} K$  & $\Sigma^{*}_{b} K$  & $\Xi_{b} \eta$  & $\Lambda_{b} K^{*}$  & $\Xi_{b} \rho$  & $\Xi'_{b} \rho$  & $\Xi^{*}_{b} \rho$  & $\Sigma_{b} K^{*}$  & $\Sigma^{*}_{b} K^{*}$  & $\Xi'_{b} \eta$  & $\Xi^{*}_{b} \eta$  & $\Xi_{b} \eta'$  & $\Xi'_{b} \eta'$  & $\Xi^{*}_{b} \eta'$  & $\Xi_{b} \omega$  & $\Xi'_{b} \omega$  & $\Xi^{*}_{b} \omega$  & $\Xi_{b} \phi$  & $\Xi'_{b} \phi$  & $\Xi^{*}_{b} \phi$  & $\Sigma_{8} B$  & $\Xi_{8} B_{s}$  & $\Sigma_{8} B^{*}$  & $\Sigma_{10} B$  & $\Gamma^{\rm Strong}$   \\
$\Xi'_b(snb)$  & $ {\bf J}^P $  & $\vert l_{\lambda}, l_{\rho}, k_{\lambda}, k_{\rho} \rangle$  & $^{2S+1}L_{J}$  & MeV  & MeV  & MeV  & MeV  & MeV  & MeV  & MeV  & MeV  & MeV  & MeV  & MeV  & MeV  & MeV  & MeV  & MeV  & MeV  & MeV  & MeV  & MeV  & MeV  & MeV  & MeV  & MeV  & MeV  & MeV  & MeV  & MeV  & MeV  & \\
\hline
 $N=0$  &  &  &  &  &  \\
$\Xi'_b(5925)$  & ${\bf \frac{1}{2}}^+$ & $\vert \,0\,,\,0\,,\,0\,,\,0 \,\rangle $ &$^{2}S_{1/2}$&$0$   &$0$   &$0$   &$0$   &$0$   &$0$   &$0$   &$0$   &$0$   &$0$   &$0$   &$0$   &$0$   &$0$   &$0$   &$0$   &$0$   &$0$   &$0$   &$0$   &$0$   &$0$   &$0$   &$0$   &$0$   &$0$   &$0$   &$0$   &$0$  \\
$\Xi'_b(5953)$  & ${\bf \frac{3}{2}}^+$ & $\vert \,0\,,\,0\,,\,0\,,\,0 \,\rangle $ &$^{4}S_{3/2}$&$0$   &0.2   &$0$   &$0$   &$0$   &$0$   &$0$   &$0$   &$0$   &$0$   &$0$   &$0$   &$0$   &$0$   &$0$   &$0$   &$0$   &$0$   &$0$   &$0$   &$0$   &$0$   &$0$   &$0$   &$0$   &$0$   &$0$   &$0$   &0.2  \\
\hline
 $N=1$  &  &  &  &  &  \\
$\Xi'_b(6198)$  & ${\bf \frac{1}{2}}^-$ & $\vert \,1\,,\,0\,,\,0\,,\,0 \,\rangle $ &$^{2}P_{1/2}$&1.1   &0.5   &1.4   &$0$   &$0$   &$0$   &$0$   &$0$   &$0$   &$0$   &$0$   &$0$   &$0$   &$0$   &$0$   &$0$   &$0$   &$0$   &$0$   &$0$   &$0$   &$0$   &$0$   &$0$   &$0$   &$0$   &$0$   &$0$   &3.0  \\
$\Xi'_b(6220)$  & ${\bf \frac{1}{2}}^-$ & $\vert \,1\,,\,0\,,\,0\,,\,0 \,\rangle $ &$^{4}P_{1/2}$&1.8   &0.8   &0.7   &0.4   &$0$   &$0$   &$0$   &$0$   &$0$   &$0$   &$0$   &$0$   &$0$   &$0$   &$0$   &$0$   &$0$   &$0$   &$0$   &$0$   &$0$   &$0$   &$0$   &$0$   &$0$   &$0$   &$0$   &$0$   &3.7  \\
$\Xi'_b(6204)$  & ${\bf \frac{3}{2}}^-$ & $\vert \,1\,,\,0\,,\,0\,,\,0 \,\rangle $ &$^{2}P_{3/2}$&9.9   &11.2   &8.4   &$0$   &$0$   &$0$   &$0$   &$0$   &$0$   &$0$   &$0$   &$0$   &$0$   &$0$   &$0$   &$0$   &$0$   &$0$   &$0$   &$0$   &$0$   &$0$   &$0$   &$0$   &$0$   &$0$   &$0$   &$0$   &29.5  \\
$\Xi'_b(6226)$  & ${\bf \frac{3}{2}}^-$ & $\vert \,1\,,\,0\,,\,0\,,\,0 \,\rangle $ &$^{4}P_{3/2}$&2.1   &2.4   &0.5   &2.6   &$0$   &$0$   &$0$   &$0$   &$0$   &$0$   &$0$   &$0$   &$0$   &$0$   &$0$   &$0$   &$0$   &$0$   &$0$   &$0$   &$0$   &$0$   &$0$   &$0$   &$0$   &$0$   &$0$   &$0$   &7.6  \\
$\Xi'_b(6237)$  & ${\bf \frac{5}{2}}^-$ & $\vert \,1\,,\,0\,,\,0\,,\,0 \,\rangle $ &$^{4}P_{5/2}$&13.0   &14.6   &3.0   &0.8   &$0$   &$0$   &$0$   &$0$   &$0$   &$0$   &$0$   &$0$   &$0$   &$0$   &$0$   &$0$   &$0$   &$0$   &$0$   &$0$   &$0$   &$0$   &$0$   &$0$   &$0$   &$0$   &$0$   &$0$   &31.4  \\
$\Xi'_b(6367)$  & ${\bf \frac{1}{2}}^-$ & $\vert \,0\,,\,1\,,\,0\,,\,0 \,\rangle $ &$^{2}P_{1/2}$&-   &-   &0.6   &44.8   &7.2   &144.0   &-   &$0$   &$0$   &$0$   &$0$   &$0$   &$0$   &$0$   &$0$   &$0$   &$0$   &$0$   &$0$   &$0$   &$0$   &$0$   &$0$   &$0$   &$0$   &$0$   &$0$   &$0$   &196.6  \\
$\Xi'_b(6374)$  & ${\bf \frac{3}{2}}^-$ & $\vert \,0\,,\,1\,,\,0\,,\,0 \,\rangle $ &$^{2}P_{3/2}$&-   &-   &22.8   &6.1   &50.1   &18.2   &-   &$0$   &$0$   &$0$   &$0$   &$0$   &$0$   &$0$   &$0$   &$0$   &$0$   &$0$   &$0$   &$0$   &$0$   &$0$   &$0$   &$0$   &$0$   &$0$   &$0$   &$0$   &97.2  \\
\hline
 $N=2$  &  &  &  &  &  \\
$\Xi'_b(6473)$  & ${\bf \frac{3}{2}}^+$ & $\vert \,2\,,\,0\,,\,0\,,\,0 \,\rangle $ &$^{2}D_{3/2}$&0.8   &0.9   &0.8   &0.7   &1.6   &2.4   &0.1   &$0$   &$0$   &$0$   &$0$   &$0$   &$0$   &$0$   &$0$   &$0$   &$0$   &$0$   &$0$   &$0$   &$0$   &$0$   &$0$   &$0$   &1.6   &$0$   &4.8   &$0$   &13.7  \\
$\Xi'_b(6483)$  & ${\bf \frac{5}{2}}^+$ & $\vert \,2\,,\,0\,,\,0\,,\,0 \,\rangle $ &$^{2}D_{5/2}$&2.2   &2.5   &2.0   &0.1   &3.6   &0.4   &0.1   &$0$   &$0$   &$0$   &$0$   &$0$   &$0$   &-   &$0$   &$0$   &$0$   &$0$   &$0$   &$0$   &$0$   &$0$   &$0$   &$0$   &3.6   &$0$   &15.7   &$0$   &30.2  \\
$\Xi'_b(6489)$  & ${\bf \frac{1}{2}}^+$ & $\vert \,2\,,\,0\,,\,0\,,\,0 \,\rangle $ &$^{4}D_{1/2}$&-   &-   &0.1   &1.0   &0.3   &3.6   &-   &$0$   &$0$   &$0$   &$0$   &$0$   &$0$   &-   &$0$   &$0$   &$0$   &$0$   &$0$   &$0$   &$0$   &$0$   &$0$   &$0$   &6.2   &$0$   &13.5   &$0$   &24.7  \\
$\Xi'_b(6495)$  & ${\bf \frac{3}{2}}^+$ & $\vert \,2\,,\,0\,,\,0\,,\,0 \,\rangle $ &$^{4}D_{3/2}$&0.9   &1.0   &0.2   &1.4   &0.5   &5.1   &0.1   &$0$   &$0$   &$0$   &$0$   &$0$   &$0$   &-   &$0$   &$0$   &$0$   &$0$   &$0$   &$0$   &$0$   &$0$   &$0$   &$0$   &8.5   &$0$   &17.7   &$0$   &35.4  \\
$\Xi'_b(6506)$  & ${\bf \frac{5}{2}}^+$ & $\vert \,2\,,\,0\,,\,0\,,\,0 \,\rangle $ &$^{4}D_{5/2}$&1.9   &2.2   &0.4   &1.3   &0.8   &4.5   &0.1   &$0$   &$0$   &$0$   &$0$   &$0$   &$0$   &-   &$0$   &$0$   &$0$   &$0$   &$0$   &$0$   &$0$   &$0$   &$0$   &$0$   &14.2   &$0$   &20.9   &$0$   &46.3  \\
$\Xi'_b(6520)$  & ${\bf \frac{7}{2}}^+$ & $\vert \,2\,,\,0\,,\,0\,,\,0 \,\rangle $ &$^{4}D_{7/2}$&2.6   &3.0   &0.6   &0.6   &1.3   &2.4   &0.2   &0.1   &$0$   &$0$   &$0$   &$0$   &$0$   &-   &$0$   &$0$   &$0$   &$0$   &$0$   &$0$   &$0$   &$0$   &$0$   &$0$   &24.6   &$0$   &11.4   &$0$   &46.8  \\
$\Xi'_b(6479)$  & ${\bf \frac{1}{2}}^+$ & $\vert \,0\,,\,0\,,\,1\,,\,0 \,\rangle $ &$^{2}S_{1/2}$&0.1   &0.1   &0.1   &-   &0.3   &0.1   &-   &$0$   &$0$   &$0$   &$0$   &$0$   &$0$   &$0$   &$0$   &$0$   &$0$   &$0$   &$0$   &$0$   &$0$   &$0$   &$0$   &$0$   &0.2   &$0$   &0.7   &$0$   &1.6  \\
$\Xi'_b(6508)$  & ${\bf \frac{3}{2}}^+$ & $\vert \,0\,,\,0\,,\,1\,,\,0 \,\rangle $ &$^{4}S_{3/2}$&0.1   &0.1   &-   &0.1   &0.1   &0.3   &-   &$0$   &$0$   &$0$   &$0$   &$0$   &$0$   &-   &$0$   &$0$   &$0$   &$0$   &$0$   &$0$   &$0$   &$0$   &$0$   &$0$   &0.9   &$0$   &1.0   &$0$   &2.6  \\
$\Xi'_b(6818)$  & ${\bf \frac{1}{2}}^+$ & $\vert \,0\,,\,0\,,\,0\,,\,1 \,\rangle $ &$^{2}S_{1/2}$&0.1   &-   &0.1   &0.2   &0.2   &0.2   &-   &0.4   &1.0   &4.3   &$0$   &10.7   &0.1   &0.1   &-   &0.3   &$0$   &$0$   &0.3   &1.4   &$0$   &-   &$0$   &$0$   &-   &-   &-   &-   &19.4  \\
$\Xi'_b(6847)$  & ${\bf \frac{3}{2}}^+$ & $\vert \,0\,,\,0\,,\,0\,,\,1 \,\rangle $ &$^{4}S_{3/2}$&0.1   &-   &-   &0.4   &-   &0.3   &-   &0.3   &0.9   &0.3   &$0$   &0.7   &16.3   &-   &0.1   &0.4   &$0$   &$0$   &0.3   &0.1   &$0$   &0.2   &$0$   &$0$   &-   &-   &-   &-   &20.4  \\
$\Xi'_b(6642)$  & ${\bf \frac{3}{2}}^+$ & $\vert \,1\,,\,1\,,\,0\,,\,0 \,\rangle $ &$^{2}D_{3/2}$&-   &-   &2.1   &35.8   &5.4   &113.8   &-   &43.4   &25.7   &$0$   &$0$   &$0$   &$0$   &0.3   &0.2   &$0$   &$0$   &$0$   &7.5   &$0$   &$0$   &$0$   &$0$   &$0$   &-   &$0$   &-   &$0$   &234.2  \\
$\Xi'_b(6653)$  & ${\bf \frac{5}{2}}^+$ & $\vert \,1\,,\,1\,,\,0\,,\,0 \,\rangle $ &$^{2}D_{5/2}$&-   &-   &22.9   &4.4   &49.4   &21.7   &-   &9.1   &5.3   &$0$   &$0$   &$0$   &$0$   &1.2   &-   &$0$   &$0$   &$0$   &1.6   &$0$   &$0$   &$0$   &$0$   &$0$   &-   &$0$   &-   &$0$   &115.6  \\
$\Xi'_b(6644)$  & ${\bf \frac{1}{2}}^-$ & $\vert \,1\,,\,1\,,\,0\,,\,0 \,\rangle $ &$^{2}P_{1/2}$&-   &-   &-   &0.4   &-   &1.6   &-   &0.3   &0.2   &$0$   &$0$   &$0$   &$0$   &-   &-   &$0$   &$0$   &$0$   &0.1   &$0$   &$0$   &$0$   &$0$   &$0$   &-   &$0$   &-   &$0$   &2.6  \\
$\Xi'_b(6651)$  & ${\bf \frac{3}{2}}^-$ & $\vert \,1\,,\,1\,,\,0\,,\,0 \,\rangle $ &$^{2}P_{3/2}$&-   &-   &0.4   &0.2   &0.9   &0.8   &-   &0.4   &0.2   &$0$   &$0$   &$0$   &$0$   &-   &-   &$0$   &$0$   &$0$   &0.1   &$0$   &$0$   &$0$   &$0$   &$0$   &-   &$0$   &-   &$0$   &3.0  \\
$\Xi'_b(6649)$  & ${\bf \frac{1}{2}}^+$ & $\vert \,1\,,\,1\,,\,0\,,\,0 \,\rangle $ &$^{2}S_{1/2}$&-   &-   &1.3   &7.7   &4.9   &12.6   &-   &14.5   &13.2   &$0$   &$0$   &$0$   &$0$   &0.6   &0.1   &$0$   &$0$   &$0$   &4.0   &$0$   &$0$   &$0$   &$0$   &$0$   &-   &$0$   &-   &$0$   &58.9  \\
$\Xi'_b(6812)$  & ${\bf \frac{3}{2}}^+$ & $\vert \,0\,,\,2\,,\,0\,,\,0 \,\rangle $ &$^{2}D_{3/2}$&13.5   &13.7   &9.3   &5.9   &20.4   &10.0   &0.8   &4.1   &10.0   &57.7   &$0$   &145.2   &0.1   &0.6   &0.7   &1.5   &$0$   &$0$   &3.4   &18.1   &$0$   &$0$   &$0$   &$0$   &-   &-   &-   &-   &315.0  \\
$\Xi'_b(6822)$  & ${\bf \frac{5}{2}}^+$ & $\vert \,0\,,\,2\,,\,0\,,\,0 \,\rangle $ &$^{2}D_{5/2}$&22.0   &24.8   &21.8   &5.0   &48.6   &23.0   &2.0   &6.9   &2.8   &11.1   &$0$   &29.8   &1.0   &1.7   &0.1   &3.7   &$0$   &$0$   &0.9   &3.4   &$0$   &0.1   &$0$   &$0$   &-   &-   &-   &-   &208.7  \\
$\Xi'_b(6828)$  & ${\bf \frac{1}{2}}^+$ & $\vert \,0\,,\,2\,,\,0\,,\,0 \,\rangle $ &$^{4}D_{1/2}$&20.2   &17.4   &1.5   &11.0   &3.1   &23.6   &0.4   &9.1   &15.7   &3.5   &$0$   &8.7   &404.9   &-   &1.0   &1.6   &$0$   &$0$   &5.3   &1.1   &$0$   &1.1   &$0$   &$0$   &-   &-   &-   &-   &529.2  \\
$\Xi'_b(6834)$  & ${\bf \frac{3}{2}}^+$ & $\vert \,0\,,\,2\,,\,0\,,\,0 \,\rangle $ &$^{4}D_{3/2}$&15.2   &15.4   &2.6   &8.7   &5.8   &11.3   &0.9   &18.5   &30.4   &6.9   &$0$   &17.0   &212.1   &0.2   &1.3   &2.2   &$0$   &$0$   &10.2   &2.2   &$0$   &2.7   &$0$   &$0$   &-   &-   &-   &-   &363.6  \\
$\Xi'_b(6845)$  & ${\bf \frac{5}{2}}^+$ & $\vert \,0\,,\,2\,,\,0\,,\,0 \,\rangle $ &$^{4}D_{5/2}$&13.2   &16.4   &4.3   &10.7   &9.6   &33.0   &1.6   &13.8   &18.5   &4.5   &$0$   &10.9   &41.8   &0.4   &1.1   &3.8   &$0$   &$0$   &6.2   &1.4   &$0$   &2.6   &$0$   &$0$   &-   &-   &-   &-   &193.8  \\
$\Xi'_b(6859)$  & ${\bf \frac{7}{2}}^+$ & $\vert \,0\,,\,2\,,\,0\,,\,0 \,\rangle $ &$^{4}D_{7/2}$&28.8   &32.0   &6.7   &22.3   &15.1   &89.3   &2.5   &46.1   &38.1   &3.7   &0.2   &10.2   &31.2   &0.5   &0.8   &6.7   &$0$   &$0$   &12.3   &1.1   &$0$   &1.8   &$0$   &$0$   &-   &-   &-   &-   &349.4  \\
\hline \hline
\end{tabular}

\endgroup
}
\label{tab:part_dec_cascades}
\end{table*}
\end{turnpage}

\begin{turnpage}
\begin{table*}[htbp]
\caption{Same as \ref{tab:part_dec_lambdas}, but for $\Omega_b(snb)$ states.  The order of the states is the same as in Table \ref{tab:All_mass_Omega}. The predicted masses, reported in Table \ref{tab:All_mass_Omega}, are obtained by the three-quark model Hamiltonian of Eqs. \ref{MassFormula} and \ref{eq:Hho}.}
{\scriptsize
\begingroup
\setlength{\tabcolsep}{1.75pt} 
\renewcommand{\arraystretch}{1.35} 

\begin{tabular}{c c c c |p{0.58cm}p{0.58cm}p{0.58cm}p{0.58cm}p{0.58cm}p{0.58cm}p{0.58cm}p{0.58cm}p{0.58cm}p{0.58cm}p{0.58cm}p{0.58cm}p{0.58cm}p{0.58cm}p{0.58cm}p{0.58cm}p{0.58cm}p{0.75cm}} \hline \hline
$\mathcal{F}={\bf {6}}_{\rm F}$   &    &    &    & $\Xi_{b} K$  & $\Xi'_{b} K$  & $\Xi^{*}_{b} K$  & $\Xi_{b} K^{*}$  & $\Xi'_{b}K^{*}$  & $\Xi^{*}_{b} K^{*}$  & $\Omega_{b} \eta$  & $\Omega^{*}_{b} \eta$  & $\Omega_{b} \phi$  & $\Omega^{*}_{b} \phi$  & $\Omega_{b} \eta'$  & $\Omega^{*}_{b} \eta'$  & $\Xi_{8} B$  & $\Xi_{10} B$  & $\Gamma^{\rm Strong}$  \\
$\Omega_b(ssb)$  & $ {\bf J}^P $  & $\vert l_{\lambda}, l_{\rho}, k_{\lambda}, k_{\rho} \rangle$  & $^{2S+1}L_{J}$  & MeV  & MeV  & MeV  & MeV  & MeV  & MeV  & MeV  & MeV  & MeV  & MeV  & MeV  & MeV  & MeV  & MeV  & \\
\hline
 $N=0$  &  &  &  &  &  \\
$\Omega_b(6064)$  & ${\bf \frac{1}{2}}^+$ & $\vert \,0\,,\,0\,,\,0\,,\,0 \,\rangle $ &$^{2}S_{1/2}$&$0$   &$0$   &$0$   &$0$   &$0$   &$0$   &$0$   &$0$   &$0$   &$0$   &$0$   &$0$   &$0$   &$0$   &$0$  \\
$\Omega_b(6093)$  & ${\bf \frac{3}{2}}^+$ & $\vert \,0\,,\,0\,,\,0\,,\,0 \,\rangle $ &$^{4}S_{3/2}$&$0$   &$0$   &$0$   &$0$   &$0$   &$0$   &$0$   &$0$   &$0$   &$0$   &$0$   &$0$   &$0$   &$0$   &$0$  \\
\hline
 $N=1$  &  &  &  &  &  \\
$\Omega_b(6315)$  & ${\bf \frac{1}{2}}^-$ & $\vert \,1\,,\,0\,,\,0\,,\,0 \,\rangle $ &$^{2}P_{1/2}$&4.6   &$0$   &$0$   &$0$   &$0$   &$0$   &$0$   &$0$   &$0$   &$0$   &$0$   &$0$   &$0$   &$0$   &4.6  \\
$\Omega_b(6337)$  & ${\bf \frac{1}{2}}^-$ & $\vert \,1\,,\,0\,,\,0\,,\,0 \,\rangle $ &$^{4}P_{1/2}$&10.7   &$0$   &$0$   &$0$   &$0$   &$0$   &$0$   &$0$   &$0$   &$0$   &$0$   &$0$   &$0$   &$0$   &10.7  \\
$\Omega_b(6321)$  & ${\bf \frac{3}{2}}^-$ & $\vert \,1\,,\,0\,,\,0\,,\,0 \,\rangle $ &$^{2}P_{3/2}$&24.0   &$0$   &$0$   &$0$   &$0$   &$0$   &$0$   &$0$   &$0$   &$0$   &$0$   &$0$   &$0$   &$0$   &24.0  \\
$\Omega_b(6343)$  & ${\bf \frac{3}{2}}^-$ & $\vert \,1\,,\,0\,,\,0\,,\,0 \,\rangle $ &$^{4}P_{3/2}$&6.3   &$0$   &$0$   &$0$   &$0$   &$0$   &$0$   &$0$   &$0$   &$0$   &$0$   &$0$   &$0$   &$0$   &6.3  \\
$\Omega_b(6353)$  & ${\bf \frac{5}{2}}^-$ & $\vert \,1\,,\,0\,,\,0\,,\,0 \,\rangle $ &$^{4}P_{5/2}$&40.5   &$0$   &$0$   &$0$   &$0$   &$0$   &$0$   &$0$   &$0$   &$0$   &$0$   &$0$   &$0$   &$0$   &40.5  \\
$\Omega_b(6465)$  & ${\bf \frac{1}{2}}^-$ & $\vert \,0\,,\,1\,,\,0\,,\,0 \,\rangle $ &$^{2}P_{1/2}$&-   &9.8   &$0$   &$0$   &$0$   &$0$   &$0$   &$0$   &$0$   &$0$   &$0$   &$0$   &$0$   &$0$   &9.8  \\
$\Omega_b(6471)$  & ${\bf \frac{3}{2}}^-$ & $\vert \,0\,,\,1\,,\,0\,,\,0 \,\rangle $ &$^{2}P_{3/2}$&-   &53.5   &$0$   &$0$   &$0$   &$0$   &$0$   &$0$   &$0$   &$0$   &$0$   &$0$   &$0$   &$0$   &53.5  \\
\hline
 $N=2$  &  &  &  &  &  \\
$\Omega_b(6568)$  & ${\bf \frac{3}{2}}^+$ & $\vert \,2\,,\,0\,,\,0\,,\,0 \,\rangle $ &$^{2}D_{3/2}$&2.4   &1.6   &$0$   &$0$   &$0$   &$0$   &$0$   &$0$   &$0$   &$0$   &$0$   &$0$   &$0$   &$0$   &4.0  \\
$\Omega_b(6578)$  & ${\bf \frac{5}{2}}^+$ & $\vert \,2\,,\,0\,,\,0\,,\,0 \,\rangle $ &$^{2}D_{5/2}$&6.2   &3.5   &-   &$0$   &$0$   &$0$   &$0$   &$0$   &$0$   &$0$   &$0$   &$0$   &$0$   &$0$   &9.7  \\
$\Omega_b(6584)$  & ${\bf \frac{1}{2}}^+$ & $\vert \,2\,,\,0\,,\,0\,,\,0 \,\rangle $ &$^{4}D_{1/2}$&0.5   &0.3   &0.1   &$0$   &$0$   &$0$   &$0$   &$0$   &$0$   &$0$   &$0$   &$0$   &$0$   &$0$   &0.9  \\
$\Omega_b(6590)$  & ${\bf \frac{3}{2}}^+$ & $\vert \,2\,,\,0\,,\,0\,,\,0 \,\rangle $ &$^{4}D_{3/2}$&2.6   &0.5   &0.3   &$0$   &$0$   &$0$   &$0$   &$0$   &$0$   &$0$   &$0$   &$0$   &$0$   &$0$   &3.4  \\
$\Omega_b(6600)$  & ${\bf \frac{5}{2}}^+$ & $\vert \,2\,,\,0\,,\,0\,,\,0 \,\rangle $ &$^{4}D_{5/2}$&5.5   &0.8   &0.4   &$0$   &$0$   &$0$   &-   &$0$   &$0$   &$0$   &$0$   &$0$   &1.0   &$0$   &7.7  \\
$\Omega_b(6614)$  & ${\bf \frac{7}{2}}^+$ & $\vert \,2\,,\,0\,,\,0\,,\,0 \,\rangle $ &$^{4}D_{7/2}$&7.7   &1.3   &0.2   &$0$   &$0$   &$0$   &0.1   &$0$   &$0$   &$0$   &$0$   &$0$   &8.2   &$0$   &17.5  \\
$\Omega_b(6574)$  & ${\bf \frac{1}{2}}^+$ & $\vert \,0\,,\,0\,,\,1\,,\,0 \,\rangle $ &$^{2}S_{1/2}$&0.4   &0.3   &-   &$0$   &$0$   &$0$   &$0$   &$0$   &$0$   &$0$   &$0$   &$0$   &$0$   &$0$   &0.7  \\
$\Omega_b(6602)$  & ${\bf \frac{3}{2}}^+$ & $\vert \,0\,,\,0\,,\,1\,,\,0 \,\rangle $ &$^{4}S_{3/2}$&0.4   &0.1   &-   &$0$   &$0$   &$0$   &-   &$0$   &$0$   &$0$   &$0$   &$0$   &0.1   &$0$   &0.6  \\
$\Omega_b(6874)$  & ${\bf \frac{1}{2}}^+$ & $\vert \,0\,,\,0\,,\,0\,,\,1 \,\rangle $ &$^{2}S_{1/2}$&0.2   &1.2   &0.9   &3.5   &5.5   &$0$   &1.2   &0.6   &$0$   &$0$   &$0$   &$0$   &-   &-   &13.1  \\
$\Omega_b(6902)$  & ${\bf \frac{3}{2}}^+$ & $\vert \,0\,,\,0\,,\,0\,,\,1 \,\rangle $ &$^{4}S_{3/2}$&0.1   &0.2   &2.2   &3.5   &0.6   &$0$   &0.3   &1.5   &$0$   &$0$   &$0$   &$0$   &-   &-   &8.4  \\
$\Omega_b(6718)$  & ${\bf \frac{3}{2}}^+$ & $\vert \,1\,,\,1\,,\,0\,,\,0 \,\rangle $ &$^{2}D_{3/2}$&-   &8.5   &61.6   &22.3   &$0$   &$0$   &3.7   &19.8   &$0$   &$0$   &$0$   &$0$   &-   &$0$   &115.9  \\
$\Omega_b(6728)$  & ${\bf \frac{5}{2}}^+$ & $\vert \,1\,,\,1\,,\,0\,,\,0 \,\rangle $ &$^{2}D_{5/2}$&-   &55.3   &5.4   &6.0   &$0$   &$0$   &13.8   &1.8   &$0$   &$0$   &$0$   &$0$   &-   &$0$   &82.3  \\
$\Omega_b(6720)$  & ${\bf \frac{1}{2}}^-$ & $\vert \,1\,,\,1\,,\,0\,,\,0 \,\rangle $ &$^{2}P_{1/2}$&-   &-   &0.7   &0.2   &$0$   &$0$   &-   &0.2   &$0$   &$0$   &$0$   &$0$   &-   &$0$   &1.1  \\
$\Omega_b(6726)$  & ${\bf \frac{3}{2}}^-$ & $\vert \,1\,,\,1\,,\,0\,,\,0 \,\rangle $ &$^{2}P_{3/2}$&-   &1.1   &0.4   &0.2   &$0$   &$0$   &0.2   &0.1   &$0$   &$0$   &$0$   &$0$   &-   &$0$   &2.0  \\
$\Omega_b(6724)$  & ${\bf \frac{1}{2}}^+$ & $\vert \,1\,,\,1\,,\,0\,,\,0 \,\rangle $ &$^{2}S_{1/2}$&-   &13.7   &24.7   &16.0   &$0$   &$0$   &8.2   &9.9   &$0$   &$0$   &$0$   &$0$   &-   &$0$   &72.5  \\
$\Omega_b(6868)$  & ${\bf \frac{3}{2}}^+$ & $\vert \,0\,,\,2\,,\,0\,,\,0 \,\rangle $ &$^{2}D_{3/2}$&27.3   &19.1   &20.1   &33.1   &59.4   &$0$   &8.5   &12.4   &$0$   &$0$   &$0$   &$0$   &-   &-   &179.9  \\
$\Omega_b(6878)$  & ${\bf \frac{5}{2}}^+$ & $\vert \,0\,,\,2\,,\,0\,,\,0 \,\rangle $ &$^{2}D_{5/2}$&58.4   &51.2   &6.8   &3.1   &12.1   &$0$   &22.0   &3.0   &$0$   &$0$   &$0$   &$0$   &-   &-   &156.6  \\
$\Omega_b(6884)$  & ${\bf \frac{1}{2}}^+$ & $\vert \,0\,,\,2\,,\,0\,,\,0 \,\rangle $ &$^{4}D_{1/2}$&23.1   &0.6   &32.1   &46.2   &4.7   &$0$   &0.3   &19.1   &$0$   &$0$   &$0$   &$0$   &-   &-   &126.1  \\
$\Omega_b(6890)$  & ${\bf \frac{3}{2}}^+$ & $\vert \,0\,,\,2\,,\,0\,,\,0 \,\rangle $ &$^{4}D_{3/2}$&30.7   &5.2   &35.2   &88.4   &9.8   &$0$   &2.3   &23.4   &$0$   &$0$   &$0$   &$0$   &-   &-   &195.0  \\
$\Omega_b(6900)$  & ${\bf \frac{5}{2}}^+$ & $\vert \,0\,,\,2\,,\,0\,,\,0 \,\rangle $ &$^{4}D_{5/2}$&44.2   &11.0   &31.1   &53.0   &7.4   &$0$   &4.9   &19.9   &$0$   &$0$   &$0$   &$0$   &-   &-   &171.5  \\
$\Omega_b(6914)$  & ${\bf \frac{7}{2}}^+$ & $\vert \,0\,,\,2\,,\,0\,,\,0 \,\rangle $ &$^{4}D_{7/2}$&74.1   &15.6   &37.0   &74.7   &4.6   &$0$   &6.8   &17.7   &$0$   &$0$   &$0$   &$0$   &-   &-   &230.5  \\
\hline \hline
\end{tabular}

\endgroup
}
\label{tab:part_dec_Omega}
\end{table*}
\end{turnpage}

\section{Decay products}
\label{app2}
\begin{table}[h!]
\caption{Masses of final state baryons and mesons used  in the calculation of the decay widths as from PDG \cite{Workman:2022ynf}.}
\begin{tabular}{c | l }\hline \hline
                 & Mass in GeV  \\ \hline
$m_{\pi}$       & $0.13725 \pm 0.00295$ \\
$m_{K}$          & $0.49564 \pm 0.00279$ \\
$m_{\eta}$       & $0.54786 \pm 0.00002$ \\
$m_{\eta'}$       & $0.95778 \pm 0.00006$ \\
$m_{\rho}$       & $0.77518 \pm 0.00045$ \\
$m_{K^*}$          & $0.89555 \pm 0.00100$ \\
$m_{\omega}$       & $0.78266 \pm 0.00002$ \\
$m_{\phi}$       & $1.01946 \pm 0.00002$ \\
$m_{B}$          & $5.27966 \pm 0.00012$ \\
$m_{B_s}$          & $5.36692 \pm 0.00010$ \\
$m_{B^*}$          & $5.32471 \pm 0.00021$ \\
$m_{N}$          & $0.93891 \pm 0.00091$ \\
$m_{N(1520)}$    & $ 1.51500\pm 0.00500$ \\
$m_{N(1535)}$        & $1.53000 \pm 0.01500$ \\
$m_{N(1680)}$        & $1.68500\pm 0.00500$ \\
$m_{N(1720)}$        & $1.72000 \pm 0.03500$ \\
$m_{\Delta}$       & $1.23200 \pm 0.00200$ \\
$m_{\Lambda}$    & $1.11568 \pm 0.00001$ \\
$m_{\Lambda(1520)}$    & $1.51900 \pm 0.00010$ \\
$m_{\Xi_8}$        & $1.31820\pm 0.00360$ \\
$m_{\Xi_{10}}$       & $1.53370 \pm 0.00250$ \\
$m_{\Sigma_8}$     & $1.11932 \pm 0.00340$ \\
$m_{\Sigma_{10}}$ & $1.38460 \pm 0.00460$ \\
$m_{\Lambda_b}$    & $5.61960 \pm 0.00010$ \\
$m_{\Xi_b}$        & $5.79700 \pm 0.00060$ \\
$m_{\Xi'_b}$      & $5.93502 \pm 0.00005$ \\
$m_{\Xi^*_b}$      & $6.07800 \pm 0.00006$ \\
$m_{\Sigma_b}$     & $5.81056 \pm 0.00025$ \\
$m_{\Sigma^*_b}$ & $5.83032 \pm 0.00030$ \\
$m_{\Omega_b}$ & $6.04520 \pm 0.00120$ \\
$m_{\Omega^*_b}$ & $6.09300 \pm 0.00060$ \\
\hline\hline
\end{tabular}
\label{tab:exp_dec}
\end{table}



\clearpage

\begin{thebibliography}{10}
\providecommand{\url}[1]{{#1}}
\providecommand{\urlprefix}{URL }
\expandafter\ifx\csname urlstyle\endcsname\relax
  \providecommand{\doi}[1]{DOI \discretionary{}{}{}#1}\else
  \providecommand{\doi}{DOI \discretionary{}{}{}\begingroup
  \urlstyle{rm}\Url}\fi



\bibitem{LHCb:2012kxf}
R.~Aaij \textit{et al.} [LHCb],
Phys. Rev. Lett. \textbf{109} (2012), 172003
doi:10.1103/PhysRevLett.109.172003
[arXiv:1205.3452 [hep-ex]].

\bibitem{CDF:2013pvu}
T.~A.~Aaltonen \textit{et al.} [CDF],
Phys. Rev. D \textbf{88} (2013) no.7, 071101
doi:10.1103/PhysRevD.88.071101
[arXiv:1308.1760 [hep-ex]].

\bibitem{LHCb:2018vuc}
R.~Aaij \textit{et al.} [LHCb],
Phys. Rev. Lett. \textbf{121} (2018) no.7, 072002
doi:10.1103/PhysRevLett.121.072002
[arXiv:1805.09418 [hep-ex]].

\bibitem{LHCb:2018haf}
R.~Aaij \textit{et al.} [LHCb],
Phys. Rev. Lett. \textbf{122} (2019) no.1, 012001
doi:10.1103/PhysRevLett.122.012001
[arXiv:1809.07752 [hep-ex]].

\bibitem{LHCb:2019soc}
R.~Aaij \textit{et al.} [LHCb],
Phys. Rev. Lett. \textbf{123} (2019) no.15, 152001
doi:10.1103/PhysRevLett.123.152001
[arXiv:1907.13598 [hep-ex]].

\bibitem{LHCb:2020tqd}
R.~Aaij \textit{et al.} [LHCb],
Phys. Rev. Lett. \textbf{124} (2020) no.8, 082002
doi:10.1103/PhysRevLett.124.082002
[arXiv:2001.00851 [hep-ex]].

\bibitem{LHCb:2021ssn}
R.~Aaij \textit{et al.} [LHCb],
Phys. Rev. Lett. \textbf{128} (2022) no.16, 162001
doi:10.1103/PhysRevLett.128.162001
[arXiv:2110.04497 [hep-ex]].

\bibitem{LHCb:2020lzx}
R.~Aaij \textit{et al.} [LHCb],
JHEP \textbf{06} (2020), 136
doi:10.1007/JHEP06(2020)136
[arXiv:2002.05112 [hep-ex]].

\bibitem{CMS:2020zzv}
A.~M.~Sirunyan \textit{et al.} [CMS],
Phys. Lett. B \textbf{803} (2020), 135345
doi:10.1016/j.physletb.2020.135345
[arXiv:2001.06533 [hep-ex]].


\bibitem{Capstick:1986bm}
S.~Capstick and N.~Isgur,
AIP Conf. Proc. \textbf{132} (1985), 267-271
doi:10.1063/1.35361

\bibitem{Ebert:2007nw}
D.~Ebert, R.~N.~Faustov and V.~O.~Galkin,
Phys. Lett. B \textbf{659} (2008), 612-620
doi:10.1016/j.physletb.2007.11.037
[arXiv:0705.2957 [hep-ph]].

\bibitem{Ebert:2011kk}
D.~Ebert, R.~N.~Faustov and V.~O.~Galkin,
Phys. Rev. D \textbf{84} (2011), 014025
doi:10.1103/PhysRevD.84.014025
[arXiv:1105.0583 [hep-ph]].

\bibitem{Roberts:2007ni}
W.~Roberts and M.~Pervin,
Int. J. Mod. Phys. A \textbf{23} (2008), 2817-2860
doi:10.1142/S0217751X08041219
[arXiv:0711.2492 [nucl-th]].

\bibitem{Yoshida:2015tia}
T.~Yoshida, E.~Hiyama, A.~Hosaka, M.~Oka and K.~Sadato,
Phys. Rev. D \textbf{92} (2015) no.11, 114029
doi:10.1103/PhysRevD.92.114029
[arXiv:1510.01067 [hep-ph]].

\bibitem{Chen:2016phw}
H.~X.~Chen, Q.~Mao, A.~Hosaka, X.~Liu and S.~L.~Zhu,
Phys. Rev. D \textbf{94} (2016) no.11, 114016
doi:10.1103/PhysRevD.94.114016
[arXiv:1611.02677 [hep-ph]].

\bibitem{Bagan:1992tp}
E.~Bagan, M.~Chabab, H.~G.~Dosch and S.~Narison,
Phys. Lett. B \textbf{287} (1992), 176-178
doi:10.1016/0370-2693(92)91896-H

\bibitem{Gutierrez-Guerrero:2019uwa}
L.~X.~Guti\'errez-Guerrero, A.~Bashir, M.~A.~Bedolla and E.~Santopinto,
Phys. Rev. D \textbf{100} (2019) no.11, 114032
doi:10.1103/PhysRevD.100.114032
[arXiv:1911.09213 [nucl-th]].

\bibitem{Garcilazo:2007eh}
H.~Garcilazo, J.~Vijande and A.~Valcarce,
J. Phys. G \textbf{34} (2007), 961-976
doi:10.1088/0954-3899/34/5/014
[arXiv:hep-ph/0703257 [hep-ph]].

\bibitem{Hasenfratz:1980ka}
P.~Hasenfratz, R.~R.~Horgan, J.~Kuti and J.~M.~Richard,
Phys. Lett. B \textbf{94} (1980), 401-404
doi:10.1016/0370-2693(80)90906-5

\bibitem{Kim:2020imk}
Y.~Kim, E.~Hiyama, M.~Oka and K.~Suzuki,
Phys. Rev. D \textbf{102} (2020) no.1, 014004
doi:10.1103/PhysRevD.102.014004
[arXiv:2003.03525 [hep-ph]].

\bibitem{Kim:2021ywp}
Y.~Kim, Y.~R.~Liu, M.~Oka and K.~Suzuki,
Phys. Rev. D \textbf{104} (2021) no.5, 054012
doi:10.1103/PhysRevD.104.054012
[arXiv:2105.09087 [hep-ph]].

\bibitem{Vijande:2014uma}
J.~Vijande, A.~Valcarce and H.~Garcilazo,
Phys. Rev. D \textbf{90} (2014) no.9, 094004
doi:10.1103/PhysRevD.90.094004
[arXiv:1507.03736 [hep-ph]].

\bibitem{Korner:1994nh}
J.~G.~Korner, M.~Kramer and D.~Pirjol,
Prog. Part. Nucl. Phys. \textbf{33} (1994), 787-868
doi:10.1016/0146-6410(94)90053-1
[arXiv:hep-ph/9406359 [hep-ph]].

\bibitem{Chen:2016spr}
H.~X.~Chen, W.~Chen, X.~Liu, Y.~R.~Liu and S.~L.~Zhu,
Rept. Prog. Phys. \textbf{80} (2017) no.7, 076201
doi:10.1088/1361-6633/aa6420
[arXiv:1609.08928 [hep-ph]].

\bibitem{Crede:2013kia}
V.~Crede and W.~Roberts,
Rept. Prog. Phys. \textbf{76} (2013), 076301
doi:10.1088/0034-4885/76/7/076301
[arXiv:1302.7299 [nucl-ex]].

\bibitem{Amhis:2019ckw}
Y.~S.~Amhis \textit{et al.} [HFLAV],
Eur. Phys. J. C \textbf{81} (2021) no.3, 226
doi:10.1140/epjc/s10052-020-8156-7
[arXiv:1909.12524 [hep-ex]].

\bibitem{Workman:2022ynf}
R.~L.~Workman \textit{et al.} [Particle Data Group],
PTEP \textbf{2022} (2022), 083C01
doi:10.1093/ptep/ptac097

\bibitem{Wang:2017kfr}
K.~L.~Wang, Y.~X.~Yao, X.~H.~Zhong and Q.~Zhao,
Phys. Rev. D \textbf{96} (2017) no.11, 116016
doi:10.1103/PhysRevD.96.116016
[arXiv:1709.04268 [hep-ph]].



\bibitem{Yao:2018jmc}
Y.~X.~Yao, K.~L.~Wang and X.~H.~Zhong,
Phys. Rev. D \textbf{98} (2018) no.7, 076015
doi:10.1103/PhysRevD.98.076015
[arXiv:1803.00364 [hep-ph]].

\bibitem{Nagahiro:2016nsx}
H.~Nagahiro, S.~Yasui, A.~Hosaka, M.~Oka and H.~Noumi,
Phys. Rev. D \textbf{95} (2017) no.1, 014023
doi:10.1103/PhysRevD.95.014023
[arXiv:1609.01085 [hep-ph]].

\bibitem{Liang:2020hbo}
W.~Liang and Q.~F.~L\"u,
Eur. Phys. J. C \textbf{80} (2020) no.3, 198
doi:10.1140/epjc/s10052-020-7759-3
[arXiv:2001.02221 [hep-ph]].

\bibitem{He:2021xrh}
H.~Z.~He, W.~Liang, Q.~F.~L\"u and Y.~B.~Dong,
Sci. China Phys. Mech. Astron. \textbf{64} (2021) no.6, 261012
doi:10.1007/s11433-021-1704-x
[arXiv:2102.07391 [hep-ph]].


\bibitem{Tawfiq:1999cf}
  S.~Tawfiq, J.~G.~Korner and P.~J.~O'Donnell,
  Electromagnetic transitions of heavy baryons in the SU$(2N_f)\times$ O(3) symmetry,
  Phys.\ Rev.\ D {\bf 63}, 034005 (2001).

\bibitem{Chow:1995nw}
  C.~K.~Chow,
  Radiative decays of excited $\Lambda_Q$ baryons in the bound state picture,
  Phys.\ Rev.\ D {\bf 54}, 3374 (1996).

\bibitem{Gamermann:2010ga}
  D.~Gamermann, C.~E.~Jimenez-Tejero and A.~Ramos,
  Radiative decays of dynamically generated charmed baryons,
  Phys.\ Rev.\ D {\bf 83}, 074018 (2011).

\bibitem{Ivanov:1998wj}
  M.~A.~Ivanov, J.~G.~Korner and V.~E.~Lyubovitskij,
  One photon transitions between heavy baryons in a relativistic three quark model,
  Phys.\ Lett.\ B {\bf 448}, 143 (1999).



\bibitem{Zhu:1998ih}
  S.~L.~Zhu and Y.~B.~Dai,
  Radiative decays of heavy hadrons from light cone QCD sum rules in the leading order of HQET,
  Phys.\ Rev.\ D {\bf 59}, 114015 (1999).

\bibitem{Wang:2009ic}
  Z.~G.~Wang,
  Analysis of the vertexes $\Xi_Q^*\Xi_Q'V$, $\Sigma_Q^*\Sigma_QV$ and radiative decays $\Xi_Q^* \to \Xi_Q'\gamma$, $\Sigma_Q^*\to \Sigma_Q\gamma$,
  Eur.\ Phys.\ J.\ A {\bf 44}, 105 (2010).

\bibitem{Wang:2009cd}
  Z.~G.~Wang,
  Analysis of the vertexes $\Omega_Q^* \Omega_Q\phi$ and radiative decays $\Omega_Q^* \to \Omega_Q \gamma$,''
  Phys.\ Rev.\ D {\bf 81}, 036002 (2010).

\bibitem{Aliev:2014bma}
  T.~M.~Aliev, K.~Azizi and H.~Sundu,
  Radiative $\Omega _{Q}^{*}\rightarrow \Omega _{Q}\gamma $ and $\Xi _{Q}^{*}\rightarrow \Xi ^{\prime}_Q \gamma $ transitions in light cone QCD,
  Eur.\ Phys.\ J.\ C {\bf 75}, 14 (2015).

\bibitem{Aliev:2009jt}
  T.~M.~Aliev, K.~Azizi and A.~Ozpineci,
  Radiative decays of the heavy flavored baryons in light cone QCD sum rules,
  Phys.\ Rev.\ D {\bf 79}, 056005 (2009).

\bibitem{Aliev:2016xvq}
  T.~M.~Aliev, T.~Barakat and M.~Savc{\i},
  Analysis of the radiative decays $\Sigma_Q \to \Lambda_Q \gamma$ and $\Xi^\prime_Q \to \Xi_Q \gamma$ in light cone sum rules,
  Phys.\ Rev.\ D {\bf 93}, no. 5, 056007 (2016).

\bibitem{Aliev:2011bm}
  T.~M.~Aliev, M.~Savc{\i} and V.~S.~Zamiralov,
  Vector meson dominance and radiative decays of heavy spin-3/2 baryons to heavy spin-1/2 baryons,
  Mod.\ Phys.\ Lett.\ A {\bf 27}, 1250054 (2012).

\bibitem{Banuls:1999br}
  M.~C.~Banuls, A.~Pich and I.~Scimemi,
  Electromagnetic decays of heavy baryons,
  Phys.\ Rev.\ D {\bf 61}, 094009 (2000).

\bibitem{Cheng:1992xi}
  H.~Y.~Cheng, C.~Y.~Cheung, G.~L.~Lin, Y.~C.~Lin, T.~M.~Yan and H.~L.~Yu,
  Chiral Lagrangians for radiative decays of heavy hadrons,''
  Phys.\ Rev.\ D {\bf 47}, 1030 (1993).

\bibitem{Jiang:2015xqa}
  N.~Jiang, X.~L.~Chen and S.~L.~Zhu,
  Electromagnetic decays of the charmed and bottom baryons in chiral perturbation theory,
  Phys.\ Rev.\ D {\bf 92}, 054017 (2015).

\bibitem{Bernotas:2013eia}
  A.~Bernotas and V.~\v{S}imonis,
  Radiative M1 transitions of heavy baryons in the bag model,
  Phys.\ Rev.\ D {\bf 87}, 074016 (2013).




\bibitem{Santopinto:2018ljf}
E.~Santopinto, A.~Giachino, J.~Ferretti, H.~Garc\'\i{}a-Tecocoatzi, M.~A.~Bedolla, R.~Bijker and E.~Ortiz-Pacheco,
Eur. Phys. J. C \textbf{79} (2019) no.12, 1012
doi:10.1140/epjc/s10052-019-7527-4
[arXiv:1811.01799 [hep-ph]].


\bibitem{LHCb:2017uwr}
R.~Aaij \textit{et al.} [LHCb],
Phys. Rev. Lett. \textbf{118} (2017) no.18, 182001
doi:10.1103/PhysRevLett.118.182001
[arXiv:1703.04639 [hep-ex]].

\bibitem{Bijker:2020tns}
R.~Bijker, H.~Garc\'\i{}a-Tecocoatzi, A.~Giachino, E.~Ortiz-Pacheco and E.~Santopinto,
Phys. Rev. D \textbf{105} (2022) no.7, 074029
doi:10.1103/PhysRevD.105.074029
[arXiv:2010.12437 [hep-ph]].

\bibitem{Garcia-Tecocoatzi:2022zrf}
H.~Garcia-Tecocoatzi, A.~Giachino, J.~Li, A.~Ramirez-Morales and E.~Santopinto,
Phys. Rev. D \textbf{107} (2023) no.3, 034031
doi:10.1103/PhysRevD.107.034031
[arXiv:2205.07049 [hep-ph]].

\bibitem{Ferretti:2015rsa}
J.~Ferretti and E.~Santopinto,
Phys. Rev. D \textbf{97}, no.11, 114020 (2018)
doi:10.1103/PhysRevD.97.114020
[arXiv:1506.04415 [hep-ph]].

\bibitem{Efron1994}
B.~Efron, R.J. Tibshirani, \emph{{An introduction to the bootstrap}}.
\newblock Mono. Stat. Appl. Probab. (Chapman and Hall, London, 1993).
\newblock \urlprefix\url{https://cds.cern.ch/record/526679}

\bibitem{Santopinto:2004hw}
E.~Santopinto,
Phys. Rev. C \textbf{72}, 022201 (2005)
doi:10.1103/PhysRevC.72.022201
[arXiv:hep-ph/0412319 [hep-ph]].

\bibitem{Micu:1968mk}
L.~Micu,
Nucl. Phys. B \textbf{10} (1969), 521-526
doi:10.1016/0550-3213(69)90039-X

\bibitem{LeYaouanc:1972vsx}
A.~Le Yaouanc, L.~Oliver, O.~Pene and J.~C.~Raynal,
Phys. Rev. D \textbf{8} (1973), 2223-2234
doi:10.1103/PhysRevD.8.2223

\bibitem{Bijker:2015gyk}
R.~Bijker, J.~Ferretti, G.~Galat\`a, H.~Garc\'\i{}a-Tecocoatzi and E.~Santopinto,
Phys. Rev. D \textbf{94} (2016) no.7, 074040
doi:10.1103/PhysRevD.94.074040
[arXiv:1506.07469 [hep-ph]].

\bibitem{Chen:2007xf}
C.~Chen, X.~L.~Chen, X.~Liu, W.~Z.~Deng and S.~L.~Zhu,
Phys. Rev. D \textbf{75} (2007), 094017
doi:10.1103/PhysRevD.75.094017
[arXiv:0704.0075 [hep-ph]].

\bibitem{Blundell:1995ev}
H.~G.~Blundell and S.~Godfrey,
Phys. Rev. D \textbf{53} (1996), 3700-3711
doi:10.1103/PhysRevD.53.3700
[arXiv:hep-ph/9508264 [hep-ph]].

\bibitem{Molina:2020zao}
D.~Molina, M.~De Sanctis, C.~Fern\'andez-Ram\'\i{}rez and E.~Santopinto,
Eur. Phys. J. C \textbf{80} (2020) no.6, 526
doi:10.1140/epjc/s10052-020-8099-z
[arXiv:2001.05408 [hep-ph]].

\bibitem{JAMES1975343}
F.~James, M.~Roos, Computer Physics Communications \textbf{10}(6), 343  (1975).
\newblock \doi{https://doi.org/10.1016/0010-4655(75)90039-9}.
\newblock
  \urlprefix\url{http://www.sciencedirect.com/science/article/pii/0010465575900399}


\bibitem{Harris2020}
C.R. Harris, K.J. Millman, S.J. van~der Walt, R.~Gommers, P.~Virtanen,
  D.~Cournapeau, E.~Wieser, J.~Taylor, S.~Berg, N.J. Smith, R.~Kern, M.~Picus,
  S.~Hoyer, M.H. van Kerkwijk, M.~Brett, A.~Haldane, J.F. del R{\'i}o,
  M.~Wiebe, P.~Peterson, P.~G{\'e}rard-Marchant, K.~Sheppard, T.~Reddy,
  W.~Weckesser, H.~Abbasi, C.~Gohlke, T.E. Oliphant, Nature \textbf{585}(7825),
  357 (2020).
\newblock \doi{10.1038/s41586-020-2649-2}.
\newblock \urlprefix\url{https://doi.org/10.1038/s41586-020-2649-2}




\bibitem{ParticleDataGroup:2018ovx}
M.~Tanabashi \textit{et al.} [Particle Data Group],
Phys. Rev. D \textbf{98} (2018) no.3, 030001
doi:10.1103/PhysRevD.98.030001

\bibitem{Liu:2007fg}
X.~Liu, H.~X.~Chen, Y.~R.~Liu, A.~Hosaka and S.~L.~Zhu,
Phys. Rev. D \textbf{77} (2008), 014031
doi:10.1103/PhysRevD.77.014031
[arXiv:0710.0123 [hep-ph]].

\bibitem{Mao:2015gya}
Q.~Mao, H.~X.~Chen, W.~Chen, A.~Hosaka, X.~Liu and S.~L.~Zhu,
Phys. Rev. D \textbf{92} (2015) no.11, 114007
doi:10.1103/PhysRevD.92.114007
[arXiv:1510.05267 [hep-ph]].

\bibitem{Mohanta:2019mxo}
P.~Mohanta and S.~Basak,
Phys. Rev. D \textbf{101} (2020) no.9, 094503
doi:10.1103/PhysRevD.101.094503
[arXiv:1911.03741 [hep-lat]].


\bibitem{Aaij:2020yyt}
R.~Aaij \textit{et al.} [LHCb],
Phys. Rev. Lett. \textbf{124} (2020) no.22, 222001
doi:10.1103/PhysRevLett.124.222001
[arXiv:2003.13649 [hep-ex]].

\bibitem{Gell-Mann:1962yej}
M.~Gell-Mann,
Phys. Rev. \textbf{125} (1962), 1067-1084
doi:10.1103/PhysRev.125.1067

\bibitem{Okubo:1961jc}
S.~Okubo,
Prog. Theor. Phys. \textbf{27} (1962), 949-966
doi:10.1143/PTP.27.949











\end{thebibliography}
\end{document}